\documentclass[rmp,preprintnumbers,amsmath,amssymb,superscriptaddress,aps]{revtex4}
\pdfoutput=1
\usepackage[font=footnotesize,labelfont=sf,bf,flushleft]{caption}
\usepackage{graphicx}
\usepackage[usenames,dvipsnames]{xcolor}
\usepackage{amsmath,bbm}
\usepackage{hyperref}
\setcitestyle{numbers,square}

\begin{document}

\newcommand{\NV}{{NV$^{-}$}}
\newcommand{\gs}{{$^{3}A_{2}$}}
\newcommand{\Es}{{$^{3}E$}}

\newcommand{\kB}{{k_{\rm B}}}
\newcommand{\eo}{{\epsilon_0}}
\newcommand{\er}{{\epsilon_{\rm r}}}
\newcommand{\dif}{{\rm d}}
\newcommand{\ldB}{\lambda_{\rm dB}}

\newcommand{\w}[1]{\omega_{\rm #1}} 
\newcommand{\ks}[1]{{\rm k}_{#1}} 
\newcommand{\Rad}{{R}}
\newcommand{\V}[1]{V_{\rm #1}} 
\newcommand{\pol}[1]{\alpha_{\rm #1}} 
\newcommand{\pabs}{{\sigma_{\rm abs}}}
\newcommand{\mass}{M} 
\newcommand{\Qm}{Q_{\rm m}} 
\newcommand{\acf}{\langle q(t)q(0)\rangle} 
\newcommand{\varq}{{\sigma_{q}^2(t)}}
\newcommand{\PSD}{{S_{qq}(\omega)}} 

\newcommand{\waist}{{\rm w}_0}
\newcommand{\Po}{P_{{\rm opt}}}
\newcommand{\Io}{I_{{\rm opt}}}
\newcommand{\wc}{\omega_{\rm cav}}
\newcommand{\kl}{k_{\rm L}}
\newcommand{\Upot}[1]{U_{\rm #1}}

\newcommand{\go}[1]{g_{\rm #1}} 
\newcommand{\Fmin}{F_{\rm min}} 
\newcommand{\fb}{u_{\rm fb}(t)} 
\newcommand{\torque}{N_{\theta}} 

\newcommand{\Gam}[1]{\Gamma_{\rm #1}}
\newcommand{\T}[1]{T_{\rm #1}}
\newcommand{\Pg}{P_{{\rm gas}}}
\newcommand{\acc}{{\alpha_{\rm C}}}
\newcommand{\Sf}[2]{S_{\rm #1}^{\rm #2}}
\newcommand{\mfp}{l_{{\rm gas}}}
\newcommand{\Kn}{{\rm Kn}}
\newcommand{\mg}{m_{\rm gas}}
\newcommand{\etag}{\eta_{\rm gas}}

\newcommand{\Det}{{\Delta}} 
\newcommand{\kap}[1]{\kappa_{\rm #1}} 
\newcommand{\F}{{\mathcal{F}}} 
\newcommand{\HPo}{\hat{P}_{\rm opt}} 
\newcommand{\Hq}{\hat{q}_{\rm opt}} 
\newcommand{\Vint}{U_{\rm int}} 
\newcommand{\nm}{n_{\rm m}} 
\newcommand{\Coop}{C_{\rm QM}} 
\newcommand{\nopt}{n_{\rm opt}} 
\newcommand{\env}{\mathcal{E}} 

\newcommand{\zpf}[1]{{#1_{\rm zpf}}} 
\newcommand{\rc}{r_{\rm CSL}} 
\newcommand{\lcsl}{\gamma_0^{\rm CSL}} 
\newcommand{\Lcsl}{\gamma_{\rm CSL}} 
\newcommand{\LDP}{\gamma_{\rm DP}} 
\newcommand{\Gdec}{{\tilde \Gamma}} 
\newcommand{\gdec}{{\tilde \gamma}} 
\newcommand{\adec}{{\tilde a}} 
\newcommand{\Ldec}{{\tilde \Lambda}} 
\newcommand{\LRdec}[1]{\Lambda_{\rm #1}} 

\title{Optomechanics with Levitated Particles.}
\date{\today}
\author{James Millen}
\email{james.millen@kcl.ac.uk}
\affiliation{ Department of Physics, King's College London, Strand, London, WC2R 2LS, UK.}
\author{Tania S. Monteiro}
\affiliation{ Department of Physics and Astronomy, University College London, Gower Street, London, WC1E 6BT, UK.}
\author{Robert Pettit}
\affiliation{ The Institute of Optics, University of Rochester, 480 Intercampus Drive, River Campus, Rochester, NY 14627, USA.}
\author{A. Nick Vamivakas}
\affiliation{ The Institute of Optics, University of Rochester, 480 Intercampus Drive, River Campus, Rochester, NY 14627, USA.}

\begin{abstract}
Optomechanics is concerned with the use of light to control mechanical objects. As a field, it has been hugely successful in the production of precise and novel sensors, the development of low-dissipation nanomechanical devices, and the manipulation of quantum signals. Micro- and nano-particles levitated in optical fields act as nanoscale oscillators, making them excellent low-dissipation optomechanical objects, with minimal thermal contact to the environment when operating in vacuum. Levitated optomechanics is seen as the most promising route for studying high-mass quantum physics, with the promise of creating macroscopically separated superposition states at masses of $10^6$\,amu and above. Optical feedback, both using active monitoring or the passive interaction with an optical cavity, can be used to cool the centre-of-mass of levitated nanoparticles well below 1\,mK, paving the way to operation in the quantum regime. In addition, trapped mesoscopic particles are the paradigmatic system for studying nanoscale stochastic processes, and have already demonstrated their utility in state-of-the-art force sensing.
\end{abstract}

\maketitle	
\newpage

\section*{Introduction}
\label{sec:intro}

\setcounter{footnote}{0}

It is a pleasant coincidence, that whilst writing this review the Nobel Prize in Physics 2018 was jointly awarded to the American scientist Arthur Ashkin, for his development of optical tweezers. By focusing a beam of light, small objects can be manipulated through radiation pressure and/or gradient forces. This technology is now available off-the-shelf due to its applicability in the bio- and medical-sciences, where it has found utility in studying cells and other microscopic entities.

The pleasant coincidences continue, when one notes that the 2017 Nobel Prize in Physics was awarded to Weiss, Thorne and Barish for their work on the LIGO gravitational wave detector. This amazingly precise experiment is, ultimately, an \emph{optomechanical device}, where the position of a mechanical oscillator is monitored via its coupling to an optical cavity. The field of optomechanics is in the ascendency \cite{CavOptReview}, showing great promise in the development of quantum technologies and force sensing. These applications are somewhat limited by unavoidable energy dissipation and thermal loading at the nanoscale \cite{Ekinci2005}, which despite impressive progress in soft-clamping technology \cite{Tsaturyan2017} means that these technologies will likely always operate in cryogenic environments.

Enter the work of Ashkin: in 1977 he showed that dielectric particles could be levitated and cooled under vacuum conditions \cite{Ashkin1977b}. By levitating particles at low pressures, they naturally decouple from the thermal environment, and since the mechanical mode is the centre-of-mass motion of a particle, energy dissipation via strain vanishes. The field of \emph{levitated optomechanics} really took off in 2010, when three independent proposals illustrated that levitated nanoparticles could be coupled to optical cavities \cite{Barker2010a, Chang2010, RomeroIsart2010}. This promises cooling to the quantum regime, and state engineering once you are there. This excited researchers interested in fundamental quantum physics, since it seemed realistic to perform interferometry with the centre-of-mass of a dropped particle to test the limits of the quantum superposition principle \cite{RomeroIsart2011a}\footnote{This has been proposed in a standard optomechanical system, but the experimental conditions required are daunting \cite{Marshall2003}.}.

Simultaneously, Li \emph{et al.} began pioneering studies into exploring nanoscale processes with levitated microparticles, explicitly observing ballistic Brownian motion for the first time \cite{Li2010, Li2013}. It had already been realized that trapped Brownian particles were paradigmatic for studying nano-thermodynamic processes \cite{SeifertReview}, but the ability to operate in low-pressure underdamped regimes, as well as to vary the coupling to the thermal environment (as offered in levitated systems) inspired a slew of works, including the first observation of the Kramers turnover \cite{Rondin2017} and the observation of photon recoil noise \cite{Jain2016}.

This review is structured as follows: in Sec.~\ref{sec:Basics} we outline the basic physics involved in levitating dielectric particles; in Sec.~\ref{sec:thermodynamics} we briefly review the study of nanothermodynamics with optically levitated particles; in Sec.~\ref{sec:det_fb} we discuss methods to use active feedback to cool the centre-of-mass motion; and in Sec.~\ref{sec:sensing} we illustrate the utility of the system in force sensing.

Moving onto quantum applications: in Sec.~\ref{sec:cav_cool} we introduce levitated cavity optomechanics; in Sec.~\ref{sec:quantum} we discuss potential tests of quantum physics using massive objects; and in Sec.~\ref{sec:NV} we consider coupling to spins \emph{within} levitated nanoparticles. In Sec.~\ref{sec:further} we cover some cutting-edge topics, before an outlook in the concluding Sec.~\ref{sec:conclusion}.

\tableofcontents
\newpage

\section{Optically trapped particles: the basics}
\label{sec:Basics}

\setcounter{footnote}{0}

To avoid limitations associated with mechanically tethered oscillators, levitation-based experiments have been developed \cite{Li2010, Chang2010, RomeroIsart2011a, Gieseler2012,Kiesel2013,Monteiro2013}. In these experiments, the mechanical object is typically held by an intense optical field rather than being tethered to the environment. In doing so, the primary dissipation comes from interactions with the surrounding gas, which can be minimized by working in vacuum, and noise in the optical field. In levitation-based experiments, the mechanical object is typically a micro- or nano-sized particle, with a geometry chosen to highlight a specific type of motion. For example, while spherical particles are ideal for monitoring centre-of-mass motion, ellipsoidal or cylindrical particles can be used to investigate rotation and libration \cite{Hoang2016, Kuhn2017a}. Particles can also be fabricated from birefringent materials, providing further means to influence motion \cite{Vogel2009, Arita2016}. 

\begin{figure}
\centering
\includegraphics[width=3.5in]{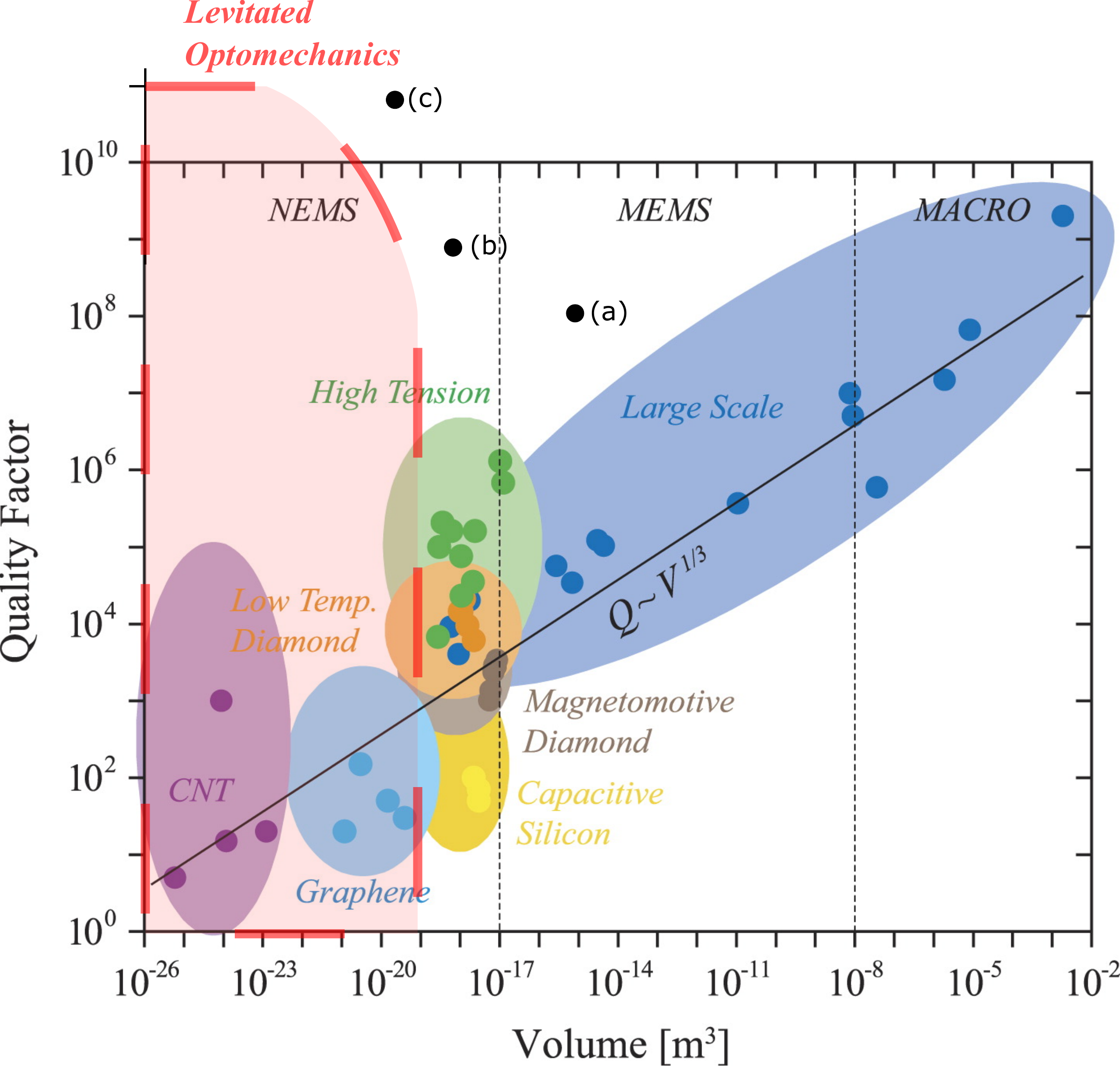}
\caption{\textbf{Scaling of mechanical quality factor with oscillator volume}, modified from \cite{Imboden2014}. The red shaded area represents the range in which levitated optomechanical experiments are predicted to operate. Black points represent the state-of-the-art tethered optomechanics experiments: (a) Ref.~\cite{Tsaturyan2017}, (b) Ref.~\cite{Ghadimi2018}, (c) Ref.~\cite{Maccabe2019}.}
\label{Qfactor_fig}
\end{figure}

An interesting comparison between levitated optomechanical resonators and traditional (tethered) opto- and electro-mechanical resonators is made in Fig.~\ref{Qfactor_fig}. There is a general scaling of the tethered resonator's quality factor proportional to the cube root of the resonator's volume. For levitated optomechanical systems this is not the case, and quality factors well above the general trend can be achieved, as illustrated in Fig.~\ref{Qfactor_fig}. The variation in quality factor with particle volume is due to the effects of gas-induced damping and photon recoil, see Sec.~\ref{sec:eom}. Oscillators of low mass and high quality factor are particularly useful for force sensing. Also included on this figure are the state-of-the-art tethered experiments, where engineering is used to minimize mechanical loss. A full comparison between the quality factors of tethered and levitated systems is reserved for Sec.~\ref{sec:conclusion}.

\subsection{Optical forces and trapping geometries}

We now describe a dielectric sphere of radius $\Rad\ll\lambda$, where $\lambda$ is the optical trapping wavelength, a regime where we can neglect the radiation pressure force \cite{Millen2018, Gieseler2018}. The interaction of such a sphere with a light field of frequency $\w{L} = 2\pi c/\lambda$ is governed by the complex polarizability $\pol{}(\w{L})$:

\begin{equation}
\label{eqn:polarisability}
\pol{}(\w{L}) = 4\pi\eo \Rad^3\frac{\er(\w{L})-1}{\er(\w{L})+2},
\end{equation}
\noindent
where the frequency dependent permittivity is related to the complex refractive index through $\er{}(\w{L}) = n(\w{L})^2$. While the real part $\pol{}'$ determines the optical potential, the imaginary part $\pol{}''$ determines optical absorption, with absorption cross-section $\pabs = \w{L}\pol{}''/(c\eo)$.

A particle can be confined by the optical potential formed by tightly focused light, a system which can be modelled as a harmonic oscillator in three spatial dimensions. The acting gradient force can be expressed as

\begin{equation}
\langle\mathbf{F}_{\rm grad}\rangle=\frac{\pol{}'}{2}\langle \bf{\nabla E}^{2}\rangle,
\end{equation}
\noindent
where $\bf{E}$ is the electric field of the light. If we make the simplifying assumption that the focused beam is Gaussian and assume that the particle occupies displacements small with respect to the beam waist and Rayleigh range, the gradient force acting on the particle is well-approximated by a linear restoring force

\begin{equation}
\label{eqn:linforce}
\langle \mathrm{F}_{{\rm grad}, q}\rangle=-\ks{q}q \qquad q\in\{x,y,z\},
\end{equation}
\noindent
where the $x$ and $y$ coordinates are taken to be the degrees-of-freedom transverse to the direction of propagation of the optical beam, and the $z$ coordinate is parallel to the direction of propagation. For a description when the force is non-linear, see Sec.~\ref{sec:thermodynamics}.

The spring constants $\ks{q}$ are different for each degree-of-freedom for a linearly polarized Gaussian beam

\begin{equation}
\ks{q} = \frac{4\pol{}'\Po(t)}{\pi c \eo {\rm w}_x {\rm w}_y {\rm w}_q^2},
\end{equation}
\noindent
where ${\rm w}_q$ is the beam waist along the $q$-direction, ${\rm w}_z$ is related to the Rayleigh range $z_0$ through ${\rm w}_z = \sqrt{2}z_0$, and $\Po(t)$ is the power contained within the optical beam\footnote{For a nanoparticle trapped in a standing-wave formed by counter-propagating beams of equal polarization, or the field of an optical cavity, the axial spring constant is $\ks{z} = 2\pol{}'\kl^2\Po(t)/\pi c \eo {\rm w}_x {\rm w}_y$, where $\kl$ is the wavenumber of the trapping light.}. Time dependence has been included in the power term to hint at the possibility of controlling particle dynamics through this variable; the ability to dynamically vary the spring constant is a key advantage of optically levitated oscillators.

\subsection{Equations of Motion}
\label{sec:eom}
The linear restoring force in eqn.~\eqref{eqn:linforce} indicates that we can construct the equation of motion for each degree-of-freedom $q$ of the particle's center-of-mass (c.o.m.) in the following way

\begin{equation}
\mass\ddot{q}(t)=-\mass\Gam{CM}\dot{q}(t)-\mass\w{q}^2q(t)+\sqrt{2\pi\Sf{ff}{}}\eta(t),
\label{eqn:EoM}
\end{equation} 
\noindent
where $\w{q}=\sqrt{\ks{q}/\mass}$ is the mechanical oscillation frequency of the trapped particle, $\Gam{CM}$ is the total momentum damping rate acting on the particle, and $\mass$ is its mass. $\Sf{ff}{}$ is the force spectral density associated with coupling to a bath at temperature $\T{env}$ at a rate $\Gam{CM}$, such that $2\pi\Sf{ff}{} = 2M\kB\T{env}\Gam{CM}$, and $\eta(t)$ encodes a white-noise process, such that $\langle\eta(t)\rangle = 0$, $\langle\eta(t)\eta(t')\rangle = \delta(t-t')$. This model holds as long as the dynamics are linear, i.e. the oscillation amplitude of the particle in the optical trap is small. Nonlinear contributions to the motion are negligible under the condition \cite{Gieseler2013, Millen2018}

\begin{equation}
\frac{3\kB\T{CM}}{{\rm w}_q^2\Gam{CM}\w{q}\mass} \ll 1,
\end{equation}
\noindent
where $\T{CM}$ is the temperature of the c.o.m. (which may differ from $\T{env}$).  When this condition is not met, the different motional degrees-of-freedom are no longer independent \cite{Gieseler2013}. There is a further discussion of nonlinearities in Sec.~\ref{sec:thermodynamics}.

The momentum damping rate of a levitated oscillator in ambient or low-vacuum conditions is dominated by collisions with the background gas $\Gam{CM}\approx\Gam{gas}$. For a spherical particle in a rarefied gas, the damping rate is \cite{Millen2018}

\begin{equation}
\label{Ggas}
\Gam{gas}=\frac{6\pi\etag \Rad}{m}\frac{0.619}{0.619+\Kn}\bigl(1+c_{\mathrm{K}}\bigr),
\end{equation}
\noindent
where $\etag$ is the dynamic viscosity of the background gas $\etag  = 18.27\times10^{-6}\,$kg\,(ms)$^{-1}$ for air, $\Kn=\mfp/\Rad$ is the Knudsen number, given by the ratio of the mean free path $\mfp$ of the background gas to the radius of the trapped particle, and $c_{\mathrm{K}}=0.31\Kn/(0.785+1.152\Kn+\Kn^{2})$. When $\Kn\gg1$, known as the Knudsen regime, the damping is linearly proportional to pressure 

\begin{equation} 
\Gam{gas}^{{\rm Kn}>1} = \frac{8}{3}\sqrt{\frac{2\mg}{\pi\kB\T{im}}}\Rad^2\Pg. {\rm \ \ \ \ (Low\ pressure)}
\end{equation}
\noindent
The transition to the Knudsen regime occurs at a pressure $\Pg \approx 54.4\,$mbar$/\Rad(\mu{\rm m})$. Mechanical quality factors $\Qm = \w{q}/\Gam{CM}$ range from $\sim10$ at 10\,mbar to $\sim10^8$ at $10^{-6}$\,mbar. 

Another stochastic force which an optically trapped particle experiences is that due to the discrete photon nature of light, known as photon shot noise. This has been recently measured \cite{Jain2016}, and leads to a damping rate \cite{Gordon1980, Millen2018}

\begin{equation}
\Gam{rad}=\frac{c_{\rm dp}P_{\rm scat}}{\mass c^2},
\label{eqn:photon}
\end{equation}
\noindent
where $c_{\rm dp}$ depends on the motion of the particle relative to the polarization of the light ($c_{\rm dp} = 2/5$ for motion parallel to the polarization, $c_{\rm dp} = 4/5$ for motion perpendicular to the polarization), and $P_{\rm scat}$ is the power of the light scattered by the particle. This depends upon the polarizability of the particle $\pol{}$, the wavevector $\kl$ and intensity $\Io$ of the trapping light: $P_{\rm scat} = \vert\pol{}\vert^2\kl^4\Io/6\pi\eo^2$. In general, $\Gam{rad} \ll \Gam{gas}$ until pressures below $\sim 10^{-6}\,$mbar are reached, at which point it becomes the dominant damping mechanism. The total damping rate $\Gam{CM}$ is the sum of all the different momentum damping rates. 

A nanoparticle exposed only to photon shot noise would reach an equilibrium temperature given by the photon energy \cite{Jain2016}, $\T{CM} = \hbar\w{L}/2\kB$. This temperature is in general very high, and necessitates continuous additional cooling (i.e. active feedback, Sec.~\ref{sec:det_fb} or passive cavity cooling, Sec.~\ref{sec:cav_cool}) to stabilize optically trapped nanoparticles at low pressures. Figure~\ref{fig:internal_temp} later in the manuscript illustrates the implication of the competing heating and damping mechanisms on the c.o.m. temperature.

\subsection{Autocorrelation function, Power Spectral Density and c.o.m. Temperature}
\label{sec:acf}

\begin{figure}
\centering
\includegraphics{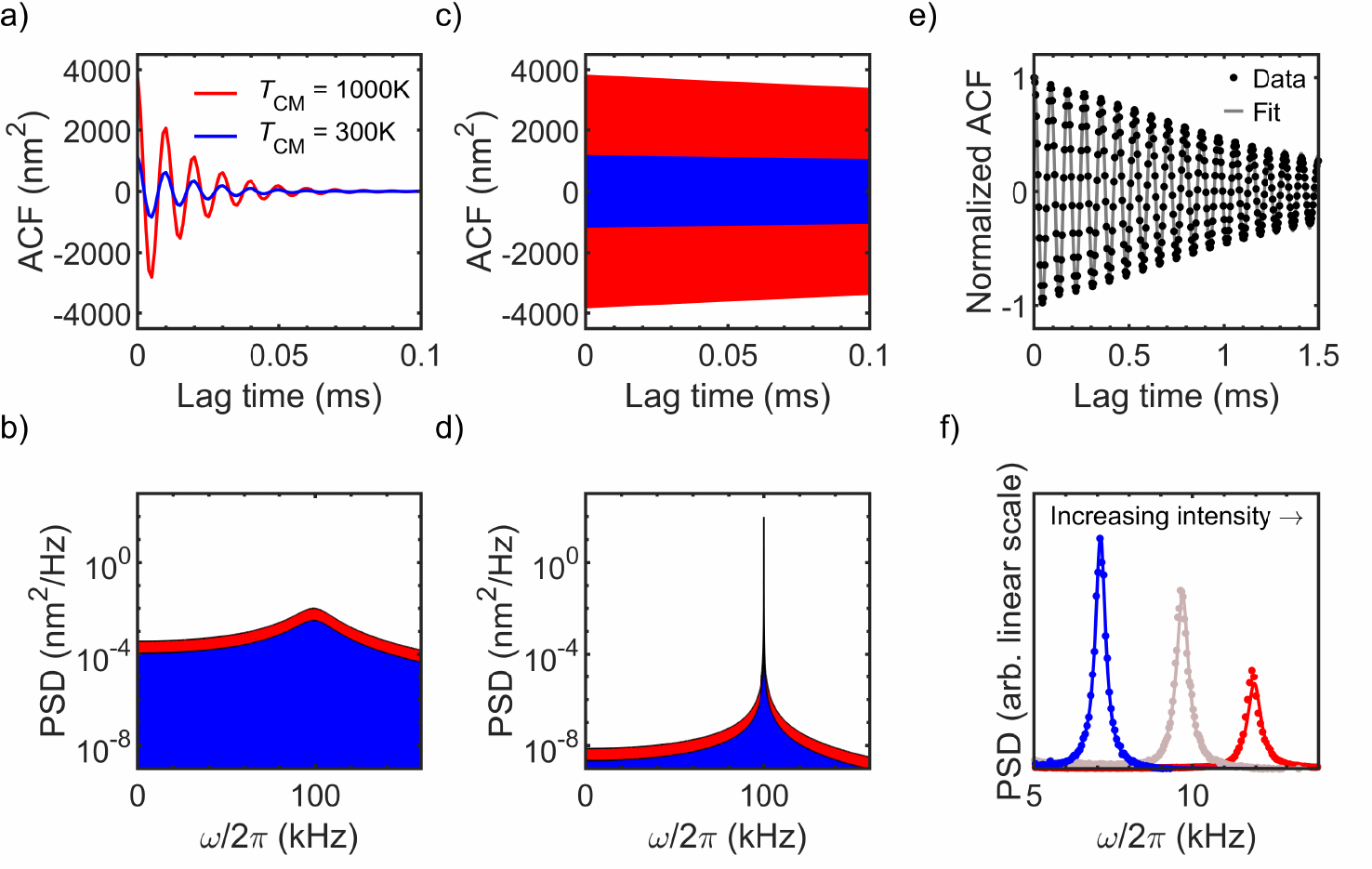}
\caption{\textbf{Autocorrelation functions (ACF) and Power Spectral Densities (PSD) in the underdamped regime} Simulated a) ACF and b) PSD at a gas pressure of 50\,mbar for $\T{CM} = 1000\,$K (red) and $\T{CM} = 300\,$K (blue). Simulated c) ACF and d) PSD at a gas pressure of $1 \times 10^{-3}$\,mbar with the same $\T{CM}$ as above (oscillations not resolved). Simulations are for an $\Rad = 100\,$nm silica sphere with a mechanical frequency $\w{x} = 2\pi \times 100\,$kHz. Example experimental e) ACF and f) PSD of a levitated $\Rad = 105\,$nm silica sphere at 1\,mbar. Points are data and solid lines are fits based on eqns.~\eqref{eqn:acf} \& \eqref{eqn:PSD}. In f), the trapping laser intensity sets $\T{CM}$, leading it to increase from left-to-right (along with an increase in trapping frequency). Reproduced using the data from \cite{Millen2014}.}
\label{fig:acf}
\end{figure}

It is not always straightforward to directly analyse eqn.~\eqref{eqn:EoM} due to the stochastic term. The first tool we will consider is the position autocorrelation function (ACF) $\acf$ for the position variable $q(t)$:

\begin{equation}
\acf = \frac{\kB\T{CM}}{\mass\w{q}^2} - \frac{1}{2}\varq.
\label{eqn:acf}
\end{equation}
\noindent
In the underdamped regime ($\Gam{CM} < \w{q}$), the position variance $\varq$ is given by

\begin{equation}
\varq = \frac{2\kB\T{CM}}{\mass\w{q}^2} \left [ 1 - e^{-\frac{1}{2}\Gam{CM}t} \left ( \cos(\w{q}t) + \frac{\Gam{CM}}{2\w{q}}\sin(\w{q}t) \right ) \right ].
\end{equation}
\noindent
For a discussion of the autocorrelation function in different damping regimes, and between different variables, see \cite{Millen2018}. The c.o.m. temperature of the particle can be extracted from the $t = 0$ value of $\acf$. Examples of $\acf$ for different pressures and temperatures are shown in Fig.~\ref{fig:acf}a), c).

The position ACF is the Fourier transform of the power spectral density (PSD) $\PSD = \frac{1}{2\pi}\int_{-\infty}^{\infty} \acf e^{-i\omega t}\dif t$, and is a convenient tool for analysing the response of the different degrees-of-freedom in frequency space. It is given by

\begin{equation}
\PSD = \frac{\Gam{CM}\kB\T{CM}/\pi\mass}{(\omega^2-\w{q}^2)^2 + \Gam{CM}^2\omega^2}.
\label{eqn:PSD}
\end{equation}
\noindent
Examples of the PSD at different pressures and $\T{CM}$ are shown in Figs.~\ref{fig:acf} b), d).  It is common to acquire $\PSD$ experimentally, and fit the data to eqn.~\eqref{eqn:PSD} to extract $\T{CM}$ \cite{Millen2014, Hebestreit2018}. Another method is to note that $\langle q(0)q(0)\rangle = \int_{-\infty}^{\infty} \PSD \dif \omega$, with  $\langle q(0)q(0)\rangle = \kB\T{CM}/\mass\w{q}^2$, i.e. by integrating over the PSD one can also extract the c.o.m. temperature.

This analysis assumes that the motion of the trapped particle is linear, whereas in the underdamped regime the particle may undergo non-linear dynamics, as discussed in Sec.~\ref{sec:thermodynamics}. It is still possible to extract the energy of the particle in this regime by analysing the PSD, see \cite{Gieseler2013, Millen2018}.

\section{Thermodynamics}

\label{sec:thermodynamics}

Trapped mesoscale objects are excellent test-beds for a range of thermodynamic phenomena. For a thorough discussion of using levitated nanoparticles to investigate thermodynamics, see recent reviews by Gieseler \& Millen \cite{Millen2018, Gieseler2018}.

A nano- or micro-particle levitated in an optical trap couples to the thermal bath provided by collisions with the surrounding gas. The motional energy of the particle is comparable to that of the thermal fluctuations of the bath: a $1\, \mu$m diameter silica sphere weighs $\sim 10^{-15}$\, kg, and has velocities in an optical tweezer of $\sim\,$mm\,s$^{-1}$, which at room temperature ($\T{env}=300\,$K) yields a kinetic energy of $\sim \kB \T{env}$. With optical trap depths $>10^4\,$K, this means that the motion of micron-sized particles is sensitive to thermal fluctuations, but not destructively so. Unlike macroscopic thermal systems, or microscopic systems with relevant internal degrees-of-freedom, considering only the centre-of-mass (c.o.m.) motion is sufficient to fully describe their behaviour\footnote{Recent work with levitated nanoparticles also considers rotational degrees of freedom \cite{Kuhn2017a}, see Sec.~\ref{sec:rotation}.}.

Working in a gaseous environment gives us the ability to dynamically vary the coupling to the bath by changing the pressure. Hence, one has access to \emph{underdamped} dynamics, as opposed to the overdamped dynamics always observed in a liquid. One way to define the transition between these regions is to compare the harmonic frequency $\omega_q$ of a trapped particle to the momentum damping rate $\Gam{CM}$, such that dynamics are underdamped when $\Gam{CM} << \omega_q$. As an example, with typical $\omega_q \sim 100\,$kHz, an $\Rad = 100$\,nm silica nanosphere in a room temperature gas experiences $\Gam{CM} \sim\,$MHz at atmospheric pressures, and $\Gam{CM} \sim\,$mHz at $10^{-6}\,$mbar pressures. This enables the study of equilibration processes on experimentally accessible timescales, and hence the verification of fluctuation relations \cite{Gieseler2014, Hoang2018}.

\subsubsection{Brownian Motion:}

Monitoring the Brownian motion of an object is an excellent window into an archetypal stochastic process. We consider a single coordinate $q(t)$. For a full discussion of the dynamics when the particle is harmonically trapped, see \cite{Li2013}. The Langevin equation for a free particle, where the dominant noise process is due to collisions with gas molecules, is

\begin{equation} 
\mass\ddot{q} + \mass\Gam{gas}\dot{q} = \sqrt{2\pi\Sf{ff}{gas}}\eta(t),
\end{equation} 
\noindent
with terms defined after eqn.~\eqref{eqn:EoM}. The mean squared displacement (MSD) for such a particle is

\begin{equation}
\label{eqn:msd}
\langle q(t)^2 \rangle = \frac{2\kB\T{env}}{\mass\Gam{gas}^2}(\Gam{gas}t - 1 + e^{-\Gam{gas}t}).
\end{equation} 
\noindent
To explore this result, we note that the relevant timescale is the momentum relaxation time $\tau = 1/\Gam{gas}$. On long timescales $t\gg\tau$, eqn.~\eqref{eqn:msd} approximates to $\langle q(t)^2 \rangle = \frac{2\kB\T{env}}{\mass\Gam{gas}}t$, which is the diffusive motion as predicted by Einstein. On short timescales $t\ll\tau$, eqn.~\eqref{eqn:msd} approximates to $\langle q(t)^2 \rangle = \frac{\kB\T{env}}{\mass}t^2$, which describes ballistic motion. The transition from diffusive to ballistic motion is set by the gas pressure (via $\tau = 1/\Gam{gas}$), with the particle motion being ballistic at low pressures. The first ever observation of the transition from diffusive to ballistic dynamics was made using a levitated particle by Li \emph{et al.} \cite{Li2010}.

Brownian motion can cause a trapped nanoparticle to explore nonlinear regions of the trapping potential, as observed by Gieseler \emph{et al.} \cite{Gieseler2013}. It is normally assumed that excursions of the oscillator are small, with the thermal amplitude of motion $a_q^{\rm th} = \sqrt{2\kB\T{CM}/(\mass\w{q}^2)} < \Rad$, and hence we can consider the potential to be harmonic. However, for a high-$\Qm$ oscillator, this no longer holds, and the different degrees of freedom $q = \{x,y,z\}$ are no longer decoupled. For an optical tweezer, in the directions transverse to the beam propagation direction, the dominant nonlinearity is the cubic ``Duffing'' term \cite{Gieseler2013, Gieseler2014b}, and the Langevin equation reads

\begin{equation} 
\mass\ddot{q_i} + \mass\Gam{gas}\dot{q_i} + \mass\w{i}^2 \left [ q_i + \left (\sum_{j=x,y,z}\xi_{j}q_j^2\right)q_i\right ]   = \sqrt{2\pi\Sf{ff}{gas}}\eta(t),
\end{equation}
\noindent
where $\xi_i$ is the nonlinear coefficient in the $i = (x,y,z)$ direction, which in an optical tweezer can be approximated as $\xi_i = -1/{\rm w}_i^2$. The consequence of this nonlinearity is that the mechanical frequency is not constant, and is red-shifted by an amount $\Delta\w{i} = \frac{3}{8}\w{i}\sum_j\xi_j a_j^2$, where $a_i$ is the instantaneous oscillation amplitude in the corresponding direction. This frequency shift broadens and skews the power spectral density \cite{Gieseler2013}. When $\Delta\w{i} \ll \Gam{gas}$ the nonlinear term can be neglected.

\subsubsection{Thermally activated escape:}

We have discussed the dynamics of a particle confined within a potential, and subject to fluctuating forces from the environment. Due to the stochastic nature of the imparted force, there is a probability that the particle will gain enough energy to escape the potential, even when it is confined by a potential much deeper than $\kB\T{env}$, in a process known as Kramers escape. It is often physically relevant in Nature to consider the stochastically driven transition between two states, for example the transition between different protein configurations. In the underdamped regime, the transition rate increases with \emph{increasing} friction, and in the overdamped regime the transition rate increases with \emph{decreasing} friction, with the transition region labelled the turnover. The Kramers turnover was first experimentally measured using a levitated nanoparticle hopping between two potential wells formed by focused laser beams \cite{Rondin2017, Ricci2017}, exploiting the fact that the friction rate can be varied over many orders of magnitude through a change in the gas pressure $\Pg$.

\subsubsection{Heat Engines:}
\label{sec:engines}
When considering a nano-scale engine, the work performed per duty cycle becomes comparable in scale to the thermal energy of the piston, and it is entirely possible for the engine to run in reverse for short times, due to the fluctuating nature of energy transfer with the heat bath. This is the scale at which biological systems operate, and a regime which levitated nano- and micro-particles have access to.

There have been many realizations of the overdamped heat engine \cite{MartinezReview}, where the construction of optimized work-extraction protocols is simplified, since the equations of motion are such that the position is independent of the velocity. An analytic solution to the optimization problem is not possible in the underdamped case, where the position and velocity variables cannot be separated \cite{GomezMarin2008, Dechant2017}, and numerical methods must be used. In both regimes, the optimum protocols call for instantaneous jumps in some control parameter \cite{GomezMarin2008}, such as the trap stiffness, which is easier to realize in the underdamped regime due to the rapid response of the particle.

It seems challenging to realize an underdamped (levitated) stochastic heat engine when by definition coupling to the heat bath (surrounding gas) is weak. Dechant \emph{et al.} propose a realization of an underdamped heat engine, based on an optically levitated nanoparticle inside an optical cavity \cite{Dechant2015}. In this case, the heat bath is provided through a combination of residual gas (1\,mbar) and the interaction with the optical cavity, which can cool the motion of the particle. 

\subsection{Internal temperature}
\label{sec:int_temp}

So far in this review, when we discuss temperature, we refer to the c.o.m. temperature $\T{CM}^{(q)} = \mass\w{q}^2\langle q^2\rangle$, which can be changed from the ambient temperature $\T{env}$ through feedback (Sec.~\ref{sec:det_fb}) or cavity cooling (Sec.~\ref{sec:cav_cool}). In this section we discuss the role of the \emph{internal}, or bulk, temperature of the particle $\T{int}$. For simplicity, we will consider the particle to have a uniform temperature, though Millen \emph{et al.}  \cite{Millen2014} observed an anisotropic temperature distribution across the surface of silica microspheres due to internal lensing.

It is well documented that the bulk temperature of a levitated particle affects its dynamics. Absorbing particles are repelled from optical intensity maxima through the photophoretic effect \cite{Lewittes1982}. In this process, the particle is anisotropically heated by absorbing light. When gas collides with the particle, it sticks for some time to the surface, and then leaves, converting some of the surface heat into momentum. The departing gas molecule imparts momentum to the particle, giving it a kick. Hence, there is a stronger kick away from the hot surface, and the particle moves away from the region of high light intensity. By employing complex optical beam geometries, the photophoretic effect can lead to stable trapping and manipulation \cite{Shvedov2011}.

The process by which a surface exchanges thermal energy with a gas is called accommodation, which is characterized by the energy accommodation coefficient

\begin{equation} 
\label{eqn:accomm}
 \acc = \frac{\T{em} - \T{im}}{\T{int} - \T{im}},
\end{equation} 
\noindent 
where $\T{im}$ is the temperature of the impinging gas molecules and $\T{em}$ the temperature of the gas molecules emitted from the surface after accommodation. Accommodation quantifies the fraction of the thermal energy that the colliding molecule removes from the surface, such that $\acc =1$ means the gas molecule fully thermalizes with the surface.

When in the Knudsen regime (see Sec.~\ref{sec:Basics}), the impinging gas molecules thermalize with the environment rather than the emitted gas, such that $\T{im} \equiv \T{env} (\neq \T{em})$. In this regime, the particle is subject to \emph{two independent} fluctuating thermal baths, one provided by the impinging gas molecules at $\T{im}$, and one by the emitted molecules at $\T{em}$. This non-equilibrium situation can be characterized by an effective c.o.m. temperature \cite{Millen2014} 

\begin{equation} 
\label{eqn:TCM}
\T{CM} = \frac{\T{im}\Gam{im} + \T{em}\Gam{em}}{\Gam{tot}},
\end{equation} 
\noindent
with a damping rate $\Gam{tot} = \Gam{im}+ \Gam{em}$. Generally, $\Gam{im}$ is given by eqn.~\eqref{Ggas} from the previous section. The damping rate due to the emitted gas is $\Gam{em}= \frac{\pi}{8}\Gam{im}\sqrt{\T{em}/\T{im}}$. Note that $\Gam{em}$ depends on $\T{int}$ through eqn.~\eqref{eqn:accomm}, as has been observed \cite{Millen2014}. 

Practically, when one analyses the motion of the particle, for example via the power spectral density, one will measure $\T{CM}$ and $\Gam{CM}$. Recent work has demonstrated a shift of a few percent in the trapping frequency $\w{q}$ with $\T{int}$, due to the dependence of the material density and refractive index upon temperature \cite{Hebestreit2018}.

\subsubsection{Absorption and emission:}
The bulk temperature of a levitated particle depends on several competing processes: heating through optical absorption of the trapping light and optical absorption of blackbody radiation, and cooling through blackbody emission and energy exchange with the background gas. The rate at which a sphere absorbs or emits blackbody energy\footnote{This assumes that the sphere is much smaller than typical blackbody radiation wavelengths, which is true for sub-micron particles.} is given by \cite{Chang2010}

\begin{equation}
\label{eqn:bbenergy}
\begin{aligned}
 \dot{E}_{\mathrm{bb,abs}} &= \frac{24 \xi(5)}{\pi^2\eo c^3 \hbar^4} \pol{bb}'' (\kB \T{env})^5\\
 \dot{E}_{\mathrm{bb,emis}} &= -\frac{24 \xi(5)}{\pi^2\eo c^3 \hbar^4} \pol{bb}'' (\kB \T{int})^5,
\end{aligned}
\end{equation} 
\noindent
where $\xi(5) \approx 1.04$ is the Riemann zeta function, and $\pol{bb}$ is averaged over the blackbody spectrum, such that for silica $\pol{bb}'' \approx 4\pi\eo \Rad^3 \times 0.1$ \cite{Chang2010}. Next we consider the cooling power due to collisions with gas molecules \cite{Chang2010}

\begin{equation}
\label{eqn:gascool}
\dot{E}_{\mathrm{gas}} = -\acc \sqrt{\frac{2}{3\pi}}(\pi \Rad^2)\Pg v_{\rm th} \frac{\gamma_{\rm sh} + 1}{\gamma_{\rm sh} - 1}\left (\frac{\T{int}}{\T{im}}-1 \right),
\end{equation}
\noindent
where $v_{\rm th}$ is the mean thermal velocity of the impinging gas molecules and $\gamma_{\rm sh} = 7/5$ is the specific heat ratio of a diatomic gas. This expression holds in the Knudsen regime. Combining all of these leads to a rate equation that describes $\T{int}$

\begin{equation}
\label{eqn:intemp}
\mass c_{\rm shc} \frac{\dif \T{int}}{\dif t} = \Io\pabs + \dot{E}_{\mathrm{gas}}(\T{int}) + \dot{E}_{\mathrm{bb,abs}} + \dot{E}_{\mathrm{bb,emis}}(\T{int}),
\end{equation} 
\noindent
where $c_{\rm shc}$ is the specific heat capacity for the particle material. Using eqn.~\eqref{eqn:intemp}, one can calculate the steady state temperature of a sphere levitated in vacuum. It is possible to reach extremely high temperatures: Silica levitated at 1\,mbar has been observed to reach its melting point at 1,873\,K, and gold nanoparticles reach 1000s\,K at atmospheric pressure \cite{Jauffred2015}\footnote{The absorption of gold is greatly enhanced by the presence of plasmonic resonances.}. 

The variation in $\T{int}$ and $\T{CM}$ with pressure for different sized particles is shown in Fig.~\ref{fig:internal_temp}.

\subsubsection{Practical considerations and particle instability} 

\begin{figure}[t]
	 {\includegraphics{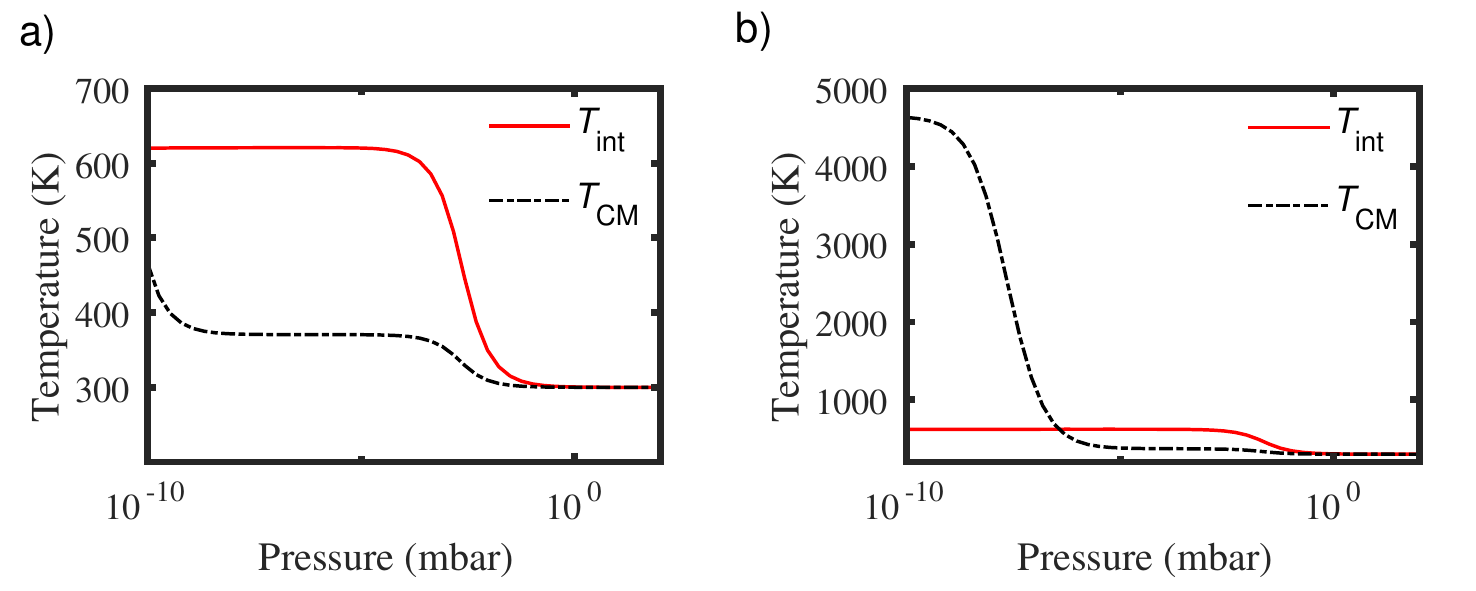}}	\centering
\caption{\label{fig:internal_temp} 
\textbf{Interaction of internal and centre-of-mass temperatures.} Variation in the bulk temperature $\T{int}$ (solid lines) and centre-of-mass temperature $\T{CM}$ (dot-dashed lines) with pressure for a) $\Rad = 10\,$nm, and b) $\Rad = 100\,$nm silica spheres, with $\T{im} \equiv \T{env} = 300\,$K. These dynamics are due to the balance between optical absorption, blackbody absorption and emission (eqn.~\eqref{eqn:bbenergy}), photon recoil heating (eqn.~\eqref{eqn:photon}), and cooling due to collisions with gas molecules (eqn.~\eqref{eqn:gascool}). This figure assumes a sphere trapped with a realistic laser intensity of $6\times10^{11}\,$W\,m$^{-2}$ at a wavelength of 1550\,nm. The optical trap depth is a) $U_0/\kB = 520\,$K and b) $U_0/\kB = 5\times10^5\,$K. For silica we use a complex refractive index $n = 1.45 +(2.5\times10^{-9})i$ \cite{Bateman2014}, material density $\rho = 2198\,$kg\,m$^{-3}$, and we assume the surrounding gas is $N_2$, with a corresponding surface accommodation coefficient $\acc = 0.65$.
}
\end{figure}

It is clear from eqn.~\eqref{eqn:intemp} that the material properties greatly effect the thermal behaviour of levitated particles, and from eqn.~\eqref{eqn:bbenergy} that both the optical absorption and emission rates depend on the absorption cross-section $\pabs = \w{L}\pol{}''/(c\eo)$. This means that low absorption materials also radiate their heat away slowly. This may be of consequence when working in ultra-high vacuum, or during protocols where the trapping light is switched off for periods of time.

One can measure the internal temperature $\T{int}$ by monitoring the c.o.m. dynamics, as discussed above. Another method is to use a material that emits light with a temperature dependent spectrum. For levitated nanoparticles, this method has been used to estimate the temperature of nanodiamonds through measurement of the NV$^-$ centre fluorescence \cite{Hoang2016b}, and the temperature of nanocrystals of YLF through measurement of the spectrum of Yb$^{3+}$ impurities \cite{Rahman2017}. There are a whole host of tracers and dyes that could be employed to do the same task \cite{Brites2012}. It is also, in principle, possible to directly measure the blackbody spectrum of a levitated nanocrystal \cite{Brites2012}.

At this point it is worth asking whether increases in internal temperature $\T{int}$, which affect the c.o.m. temperature $\T{CM}$, are enough to explain the widely reported problem of particle escape from optical traps at low pressures \cite{Millen2014, Ranjit2016}. This is not a simple question to answer, as $\T{CM}$ depends on a balance between blackbody absorption and emission (eqn.~\eqref{eqn:bbenergy}), cooling from the surrounding gas (eqn.~\eqref{eqn:gascool}), and further heating due to photon recoil \cite{Jain2016} (eqn.~\eqref{eqn:photon}), which in turn depend sensitively on the particle size and shape. In Fig.~\ref{fig:internal_temp} we show some indicative examples of the trade-off between these processes. We also note that recent work \cite{Svak2018} has shown that circularly polarized light can cause particles to undergo unstable orbits, which could play a role in regimes of low damping and imperfect polarization control.

\section{Detection and Feedback Control}
\label{sec:det_fb}

In this section, we consider methods for detecting the motion of optically trapped particles. This information can then be used to \emph{control} the motion of the particles via feedback. We focus on the use of feedback to extract energy from the levitated oscillator, but note that feedback can also be used to study non-linear processes, such as phonon lasing \cite{Pettit2019}.

\subsection{Detection and calibration}
\label{sec:det}

\begin{figure}[t]
\centering
\includegraphics{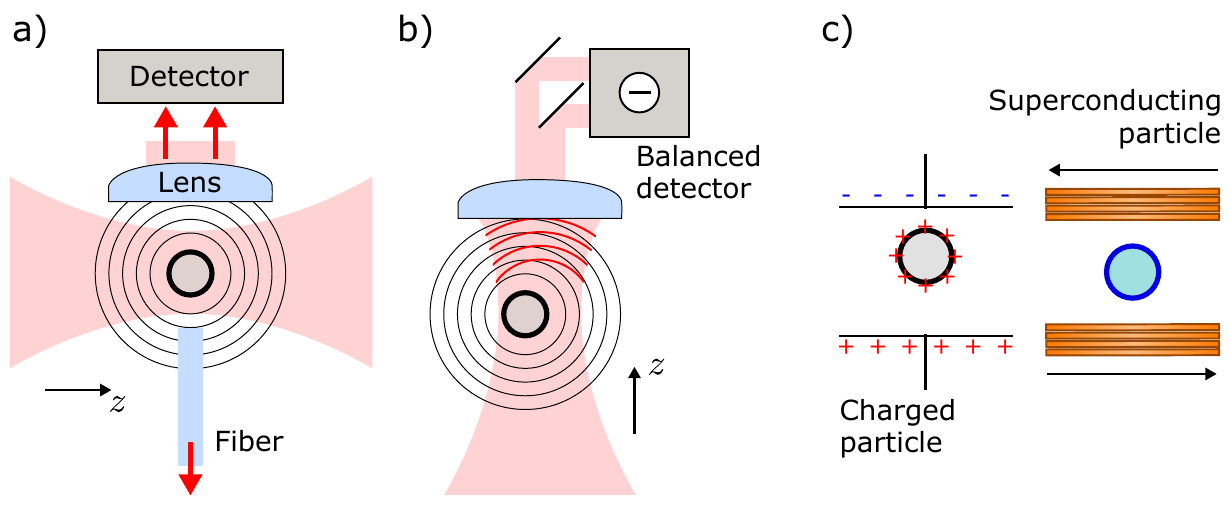}
\caption{\textbf{Particle detection} a) The light scattered from an optically trapped particle contains information about its position. This light can be collected by a lens, a multimode fiber, or an optical cavity (not shown). b) The interference between the scattered light and the trapping light acts as a homodyne measurement of the particle's position when measured with a balanced detector. c) Non-optically trapped particles can be detected by other means: charged particles induce a current in nearby electrodes (left) \cite{Goldwater2018}, and superconducting particles induce a current in anti-Helmholtz pick-up coils (right) \cite{RomeroIsart2012}. }
\label{fig:detection}
\end{figure}

The ease of detection of a levitated nanoparticle depends strongly on its size. As briefly mentioned in Sec.~\ref{sec:Basics}, the power of the light scattered by a sub-wavelength sphere within an optical field is

\begin{equation}
P_{\rm scat} = \vert\pol{}\vert^2\kl^4\Io/6\pi\eo^2.
\label{scattering}
\end{equation}
\noindent
Note that $P_{\rm scat}\propto \pol{}^2 \propto \Rad^6$, so the amount of light it is possible to detect rapidly drops with particle radius. The scattered light depends upon the local intensity $\Io$, and so varies as the particle moves through the spatially varying intensity profile of a focussed laser beam, yielding position sensitivity. This also means that position resolution is improved by using tightly focussed beams, and using a standing wave increases resolution along the $z$-direction (Fig.~\ref{fig:detection}a)). Anisotropic particles have polarizabilities that vary with their alignment relative to the polarization vector of the light field, meaning that $P_{\rm scat}$ is alignment-dependent \cite{Kuhn2015, Kuhn2017a}.

The scattered light can be collected using a lens and imaged onto a photodetector \cite{Millen2015}, or by placing a multi-mode optical fiber close to the trapping region \cite{Kuhn2015, Kuhn2017a}, see Fig.~\ref{fig:detection}a). The collected signal contains information about all degrees-of-freedom, which can be analysed separately in frequency space. To collect information about different degrees-of-freedom separately, the scattered light can be imaged onto a quadrant photodiode \cite{Millen2014, Ranjit2015a, Ranjit2016}, or onto a camera. The latter method is generally low bandwidth, though is suitable for low frequency oscillators and has favourable noise characteristics \cite{Slezak2018, Bullier2019}. It is possible to use a fast camera and stroboscopic illumination to achieve acquisition rates above 1\,MHz \cite{Arita2018}. 

A highly sensitive method for detecting nanoparticles is to make an interferometric measurement of position \cite{Rahman2018}. This method was pioneered with levitated particles by Gieseler \emph{et al.} \cite{Gieseler2012}, and has enabled $\sim 1\,$pm\,Hz$^{-1/2}$ position sensitivity. The light which the particle scatters interferes with the trapping light. Collecting this pattern with a lens produces an image of the momentum distribution of the particle. The trapping light acts to amplify the scattered-light signal, as familiar from other homodyne detection techniques. For small oscillations (i.e. in a linear optical potential), this technique produces a signal for motion in the $q$-direction proportional to $E_{\rm scat}E_{\rm ref}q/{\rm w}_q$, where ${\rm w}_q$ is the beam waist in the $q$-direction, as defined in Sec.~\ref{sec:Basics}. The fields are those incident upon the detector, with $E_{\rm scat}$ being the field scattered by the particle, and $E_{\rm ref}$ being the field due to the trapping light. For more details see Ref.~\cite{Gieseler2012}.

The total intensity of the pattern at a fixed plane is proportional to the $z$-position only, and to measure the $x,y$-positions one spatially splits the beam and makes a balanced detection, as illustrated in Fig.~\ref{fig:detection}b), which removes the intensity modulation due to the $z$-motion\footnote{particle rotation has also been detected using this method \cite{Ahn2018, Reimann2018}.}. Due to the dependence on the beam waist, optimal application of this method requires the use of high numerical aperture trapping optics. The limiting factors of this technique are the collection efficiency of the scattered light, and detector noise, see Ref.~\cite{Tebbenjohanns2019} for a thorough discussion. 

The collection efficiency could be improved by using optical microcavities. Recently, such microcavities \cite{Wachter2019} have been used to detect the motion of free nanoparticles \cite{Kuhn2017c}, and the near-field of a photonic crystal cavity has been used to detect the motion of a levitated nanoparticle \cite{Magrini2018}, achieving a position sensitivity of $\sim 3\,$pm\,Hz$^{-1/2}$. 

Macroscopic optical cavities can also be used to detect particle motion \cite{Asenbaum2013, Kiesel2013}, as also discussed in Sec.~\ref{sec:sens_cav}. This currently represents the state-of-the art detection method, providing position measurement at the $10^{-14}$\,m\,Hz$^{-1/2}$ level \cite{Delic2018, Windey2018}. When working with such cavities, their narrow bandwidth should be taken into consideration.

A final set of detection methods are non-optical, Fig.~\ref{fig:detection}c). Once could use inductive detection of charged nanoparticles \cite{Goldwater2018}, which is predicted to be able to resolve displacements below 1\,pm and detect sub-nm sized particles. This method also gives direct access to the velocity of the particle, which is useful for feedback cooling, as discussed below. It is also proposed to detect magnetically levitated superconducting spheres via pick-up coils \cite{RomeroIsart2012}.

\subsubsection{Calibration:}

Once the particle has been detected, the detected signal must be converted into position, i.e. the detector must be calibrated. This is a significant source of uncertainty in many experiments. 

The standard method is to analyse the PSD or the position variance, as discussed in Sec.~\ref{sec:acf}, and use the equipartition theorem to calibrate the detector. The limitations of this method are that it assumes the particle is in equilibrium with its environment, which is often not true \cite{Millen2014, Hebestreit2018}, and that the motion of the particle is purely harmonic, which is also often not true \cite{Gieseler2013} (though this can be compensated for \cite{Hebestreit2018b}). This method requires accurate knowledge of the local temperature, and the pressure in the vicinity of the particle (which can only be measured to 10\% accuracy or worse). The mass of the particle must be known, which in principle can be inferred by measuring the damping rate, but this requires confidence on the particle shape and density (and again the local pressure), and so represents at least another 20\% uncertainty. In addition, it has been found \cite{Hebestreit2018b} that detector calibration is pressure dependent. In all, there is a 20-30\% uncertainty on detector calibration via analysis of the thermal motion of a levitated particle, as discussed in more detail in \cite{Hebestreit2018b}.

It is possible to use another calibrated force to calibrate a detector, for example an electric force acting on a charged particle \cite{Hempston2017, Hebestreit2018b}, but this requires a precise knowledge of the applied force. One can exploit particles trapped in optical standing waves, since the well-defined optical wavelength acts as a kind of ruler to calibrate the motion of the particles as they cross nodes in the field \cite{Kuhn2015, Ranjit2016, Kuhn2017a}, but this requires a detection method with a wide field-of-view compared to the nm-level thermal motion of a trapped particle. Finally, it is possible to use a heterodyne measurement to extract various experimental parameters, such as the damping, with high precision \cite{Vovrosh2017, Dawson2019}.

\subsection{Feedback Cooling}

By detecting the motion of a levitated particle, feedback methods can be utilized to extract energy from each degree-of-freedom. Such cooling is of great interest in the quest to reach the quantum regime of motion, see Sec.~\ref{sec:quantum}. Feedback cooling can also be used to ensure that the particle does not oscillate with large amplitudes, which would introduce nonlinear dynamics (as discussed in Secs.~\ref{sec:Basics}~\&~\ref{sec:thermodynamics}), limiting the force sensing capabilities of the levitated particle, see Ref.~\cite{Gieseler2013} and Sec.~\ref{sec:sensing}. Such cooling has proven invaluable for operating under high-vacuum conditions, where it is necessary to stabilize the motion of the particle to prevent loss from the optical potential, see Sec.~\ref{sec:thermodynamics}.

The most general modification to the dynamics of a levitated particle can be described by adding a feedback term $\fb$ to the Langevin equation \cite{Conangla2019}

\begin{equation}
\ddot{q}(t) + \Gam{CM}\dot{q}(t) + \w{q}^2q(t) = \frac{\sqrt{2\pi\Sf{ff}{}}}{\mass}\eta(t) + \fb,
\label{eqn:feedback}
\end{equation}
\noindent 
where $q$ is one degree-of-freedom, and all other terms are defined after eqn.~\eqref{eqn:EoM}. By engineering a feedback term proportional to the particle's velocity $\fb = G_{\dot{q}} \dot{q}(t)$, with some optimized gain $G_{\dot{q}}$ \cite{Conangla2019}, energy will be extracted from the particle's motion. This is referred to as ``cold damping'', since it damps the motion of the particle without introducing additional heating (dissipation without fluctuation). The addition of a term proportional to position $\fb = G_q q(t) + G_{\dot{q}} \dot{q}(t)$ leads to an increased cooling rate, but not a lower final temperature. 

Such an active feedback mechanism was used to stabilize the motion of levitated microparticles ($\Rad \approx 4\,\mu$m) as early at 1977 by Ashkin \cite{Ashkin1977b}. Feedback cooling came into prominence after work by Li \emph{et al.} \cite{Li2011}, who used a 3-beam radiation pressure scheme to cool the motion of a $\Rad = 1.5\,\mu$m sphere to 1.5\,mK. The radiation pressure force drops rapidly with size, so this protocol isn't suitable for cooling particles smaller than $\sim 1\,\mu$m. 

Recently, two groups have employed linear feedback cooling on 100-200\,nm diameter \emph{charged} particles using the Coulomb force \cite{Tebbenjohanns2019, Conangla2019} . An optical signal is used to generate an electrical feedback signal which is applied to electrodes in the vicinity of the optically levitated nanoparticle. This has been used to generate temperatures in a single degree-of-freedom as low as $100\,\mu$K, which represents less than 20 motional phonons \cite{Tebbenjohanns2019}.

\subsubsection{Nonlinear feedback cooling:}

In cases where it is desirable to work with both small and charge-neutral particles, it is most effective to utilize the gradient force. In 2012, Gieseler \emph{et al.} \cite{Gieseler2012} presented influential work where \emph{parametric feedback cooling} was used to reduce a $\Rad = 70\,$nm nanoparticle's centre-of-mass (c.o.m.) temperature to 50\,mK. This involved a parametric modulation of the trapping potential, with the feedback signal generated either by the trapping light (as in Fig.~\ref{fig:detection}b)), or by a probe beam.  The phase of the feedback signal is tuned such that as the particle travels away from trap-centre the potential is stiffened, and as it travels towards trap-centre the potential is relaxed. Such a scheme therefore modulates the potential at twice the frequency of oscillation, which is achieved via $\fb = G_{\rm nl}q(t)\dot{q}(t)$. 

This feedback process leads to damping on the particle motion $\delta\Gam{} \propto G_{\rm nl}\langle \nm\rangle$, where $\nm$ is the number of motional phonons. The power of the feedback cooling is a function of the oscillator amplitude, and hence it is referred to as \emph{nonlinear} feedback cooling. In linear damping schemes, as discussed above, the feedback damping rate $\delta\Gam{}$ is independent of the oscillator energy.

\begin{figure}[t]
\centering
\includegraphics[width=6in]{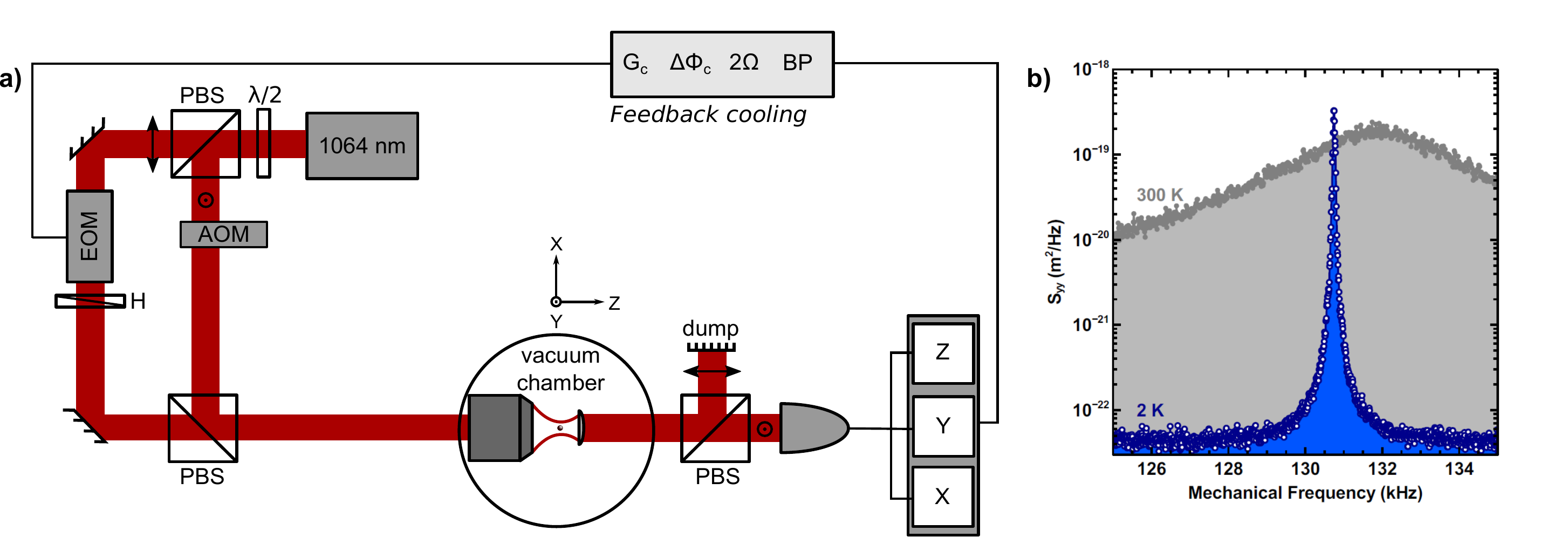}
\caption{\textbf{Feedback cooling} a) Typical experimental apparatus to feedback cool a trapped particle. In the feedback circuit, BP indicates bandpass filtering, 2$\w{}$ frequency doubling, $\Delta\Phi_{\rm c}$ phase shifting, and $G_{\rm c}$ electronic gain. The feedback signal is used to modulate an electro-optic modulator (EOM) to actuate the optical potential. In this example, a probe beam is used to monitor the particle's motion, and separately controlled by an acousto-optic modulator (AOM) b) Example of parametric feedback cooling. Blue open circles represent the spectrum of the $y$-coordinate with feedback cooling at a pressure of $2.5\times 10^{-4}$\,mbar; grey closed circles represent the same spectrum without feedback cooling at a pressure of 6\,mbar.}
\label{feedback_fig}
\end{figure}

Parametric feedback can be performed on all motional degrees-of-freedom simultaneously, since the gradient optical force always points towards the trap centre. Signals from all degrees-of-freedom are summed together and delivered to a device which actuates the trapping potential, such as an electro-optic modulator. Figure \ref{feedback_fig}a) outlines a typical experimental implementation of the feedback loop.

\subsection{Limits to feedback cooling}

\begin{figure}[t!]
\centering
\scriptsize
\tiny
\includegraphics[width=3.4in]{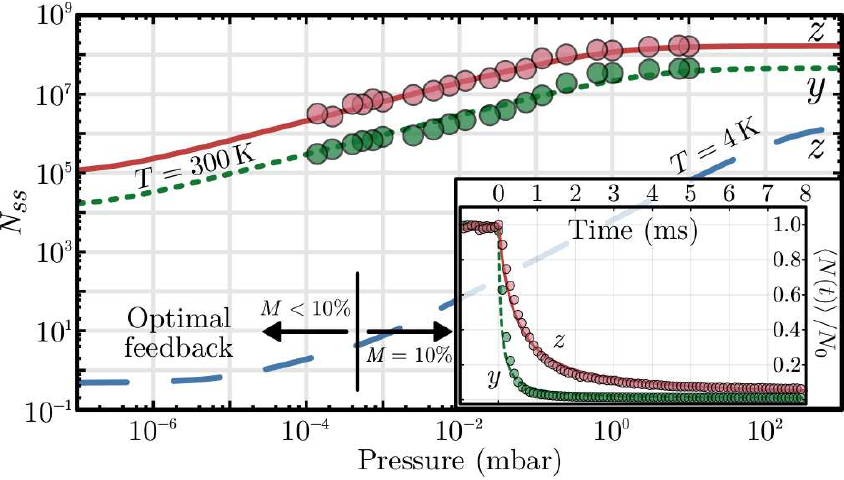}
\caption{\textbf{Nonlinear feedback cooling} Steady state phonon number ($N_\text{ss}$) versus pressure. Circles represent experimental data for axial ($z$) and transverse ($y$) oscillations at 300\,K. The red solid (green dashed) curve is a theoretical model \cite{Rodenburg2016}. The blue dashed curve represents the prediction for an experiment in a 4\,K environment. The feedback gain is varied continuously ($M$ is the trap intensity modulation).  The inset shows the phonon dynamics for the $z$ and $y$ modes. Reproduced from \cite{Rodenburg2016}.}
\label{fig:F2}
\end{figure}

With a feedback loop active, the damping constant in eqn.~\eqref{eqn:feedback} is modified to $\Gam{CM}\to\Gam{eff}=\Gam{CM}+\delta\Gam{}$, where $\delta\Gam{}$ represents the contribution of the feedback. The effective temperature of the c.o.m. motion is then modified to 

\begin{equation}
\T{eff}=\T{0}\frac{\Gam{CM}}{\Gam{CM}+\delta\Gam{}},
\label{eff_temp}
\end{equation}
\noindent
where $\T{0}$ is the c.o.m. temperature of the particle before the feedback is engaged. Thus, depending on the sign of $\delta\Gam{}$, which in turn depends on the phase-shift of the feedback loop, the effective temperature of the oscillator can be reduced (positive damping), or increased (negative damping). Figure \ref{feedback_fig}b) illustrates the effect of feedback cooling on the power spectral density of the particle's position coordinate, showing a reduced area with cooling activated (c.f. Sec.~\ref{sec:acf}), indicating a reduction in effective temperature from 300\,K to 2\,K.

It is natural to ask whether it is possible to reach the ground state via active feedback cooling. In the case of linear damping techniques, the possibilities for ground state cooling are analogous to those in the context of cavity optomechanics, which considers a linear damping parameter (the sum of a mechanical and an optomechanical damping) \cite{CavOptReview}. Explicitly, there is a trade-off between detection efficiency and shot-noise \cite{Conangla2019, Tebbenjohanns2019}, with the former a major limitation in levitated experiments. Methods to improve detection efficiency are discussed in Sec.~\ref{sec:det}, and it should be noted that combining linear feedback with passive cavity cooling (see Sec.~\ref{sec:cav_cool}) can further facilitate reaching the ground-state \cite{Genoni2015}.

It is not possible to take advantage of the standard theory of quantum cavity optomechanics \cite{CavOptReview} to address the nonlinear case, since the damping parameter is intrinsically related to the phonon occupation. Recent work addressed  the necessary conditions for mechanical ground state occupation via nonlinear cooling \cite{Rodenburg2016}. The difference between linear and nonlinear feedback schemes is highlighted by an equation of motion for the oscillator's phonon occupation 

\begin{equation}
\langle\dot{n}_{\rm m}\rangle=B\langle \nm\rangle^{2}-C\langle \nm\rangle+A.
\label{N_eqn}
\end{equation}
\noindent
The parameters $A,B,C$ depend upon the damping rates $\Gam{lin}, \Gam{nm}$ due to linear \& nonlinear feedback respectively, and the optical scattering damping rate $\Gam{rad}$, such that: $A=\Gam{rad}-6\Gam{nl}-\Gam{lin}$, $B=-24\Gam{nl}$ and $C=24\Gam{nl}+\Gam{lin}$ \cite{Rodenburg2016}. Gas damping has been neglected for simplicity, assuming operation in regimes dominated solely by optical scattering \cite{Jain2016}.  It is seen that the inclusion of non-linearity in the feedback induces dynamics that are nonlinear in the phonon occupation number, and therefore leads to non-exponential loss of oscillator energy, in contrast to linear feedback.

The results of this study  are presented in Fig.~\ref{fig:F2}, along with the predicted steady state phonon number when the experiment is placed in a cryostat at 4\,K (blue dashed curve). Starting at high pressures, the particle is cooled while continuously increasing the feedback gain to compensate for the $\langle \nm\rangle$ dependence on the feedback damping. Proceeding in this manner, it is predicted that below $\lesssim\ 10^{-5}$\,mbar nonlinear feedback could be used to cool to the ground state.\\

	\setcounter{footnote}{0}

\section{Sensing}
\label{sec:sensing}

A great driving force in the development of nanoscale oscillator devices has been their potential for detecting a wide range of forces. The minimum detectable force $\Fmin$ is usually limited by the thermal energy of the oscillator

\begin{equation}
\label{eqn:Fmin}
\Fmin = \sqrt{\frac{4 \ks{q} \kB \T{CM} b}{\w{q}\Qm}},
\end{equation} 
\noindent
where $b$ is the measurement bandwidth, $\ks{q}$ is the spring constant, $\T{CM}$ the centre-of-mass (c.o.m.) temperature of the oscillator, $\w{q}$ its resonance frequency, and $\Qm$ its mechanical quality factor. It is instantly clear that high quality factor oscillators and low temperatures enable high sensitivities. It has been possible to measure mass with yoctogram resolution \cite{Chaste2012} and sub-attonewton forces \cite{Ranjit2015a, Ranjit2016} with nanoscale devices. 

Levitated nanoparticles are seen as obvious candidates for high resolution force sensing, due to their low mass and high mechanical quality factors in vacuum\footnote{An early analysis by Libbrecht \& Black \cite{Libbrecht2004} recognized the potential for levitated microspheres to act as test masses with quantum-limited displacement readout due to their lack of thermal contact with the environment.}. Even with the amazing progress in creating standard high $\Qm$ nanomechanical devices, levitated systems offer several unique prospects, for example the potential to exploit macroscopically separated superposition states, see Secs.~\ref{sec:quantum} \& \ref{sec:conclusion}. In the specific case of a levitated particle, eqn.~\eqref{eqn:Fmin} can be rewritten as

\begin{equation}
\label{eqn:Fmin_lev}
\Fmin = \sqrt{4\kB\T{CM}\mass\Gam{CM}b},
\end{equation} 
\noindent
where $\Gam{CM}$ is the total c.o.m. damping, clearly illustrating that working at low pressures increases sensitivity. Note that if feedback or cavity cooling to a temperature $\T{eff}$ is applied to the motion of the levitated particle, a total damping rate $\Gam{eff}$ must be used to include the additional damping from the cooling mechanism, c.f. Sec.~\ref{sec:det_fb}. Since $\Gam{eff}\T{eff} = \T{CM}\Gam{CM}$, cooling the particle doesn't of itself increase sensitivity, yet this additional dissipation is necessary to operate in high-vacuum (Sec.~\ref{sec:thermodynamics}), and can yield advantages in terms of response-bandwidth.

In this chapter, sensitivities will be quoted in units of X\,Hz$^{-1/2}$, where X could be force, acceleration etc. Each quoted figure does not include the \emph{ultimate} sensitivity, or Allen minimum, since to the best of our knowledge this has not been extensively studied in the context of levitated optomechanics. We note that for practical applications, a careful study of long-term drifts, detector calibration stability, levitated particle mass stability etc. will have to be undertaken. Long-term stability is unlikely for particles levitated within an optical cavity, due to noise added through stabilization of the laser frequency to the cavity resonance, and the sensitivity of optical cavities to vibration and thermal effects.

\begin{figure}[t]
	 {\includegraphics{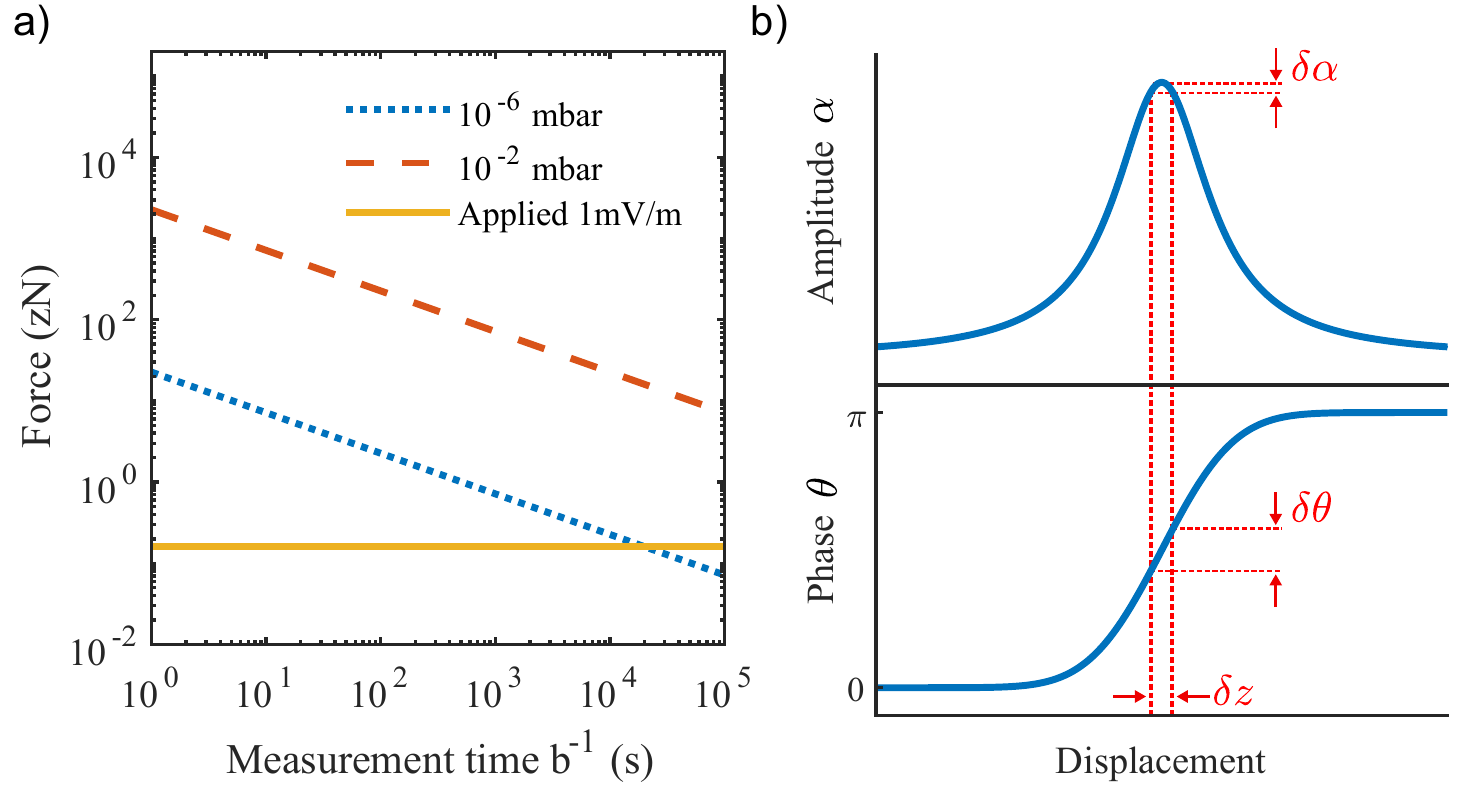}}	\centering
\caption{\label{fig:force} 
\textbf{Force sensing with levitated particles} a) By monitoring the position spectral density of a levitated particle, one can derive the forces acting on the particle, with an accuracy which improves with measurement time. The dotted/dashed lines show the thermal force acting on a 100\,nm radius sphere, at $10^{-6}\,$mbar and $10^{-2}\,$mbar respectively, which sets the minimum detectable force from any external source. The solid line, for example, is the force one would expect on a singly-charged sphere exposed to a field of 1\,mV/m, and would be detectable at $10^{-6}\,$mbar after $\sim 10^4$\,s of averaging. b) An optical cavity can be used to sensitively monitor the displacement $\delta z$ of a levitated nanoparticle, as its motion modulates the amplitude and phase of the light transmitted and reflected from the cavity. By pumping the cavity close to resonance, $\delta z$ produces a large phase shift $\delta \theta$, which can be sensitively measured, whilst avoiding a large shift in cavity amplitude $\delta \alpha$, which would impart a backaction force on the particle.
}
\end{figure}

Ranjit \emph{et al.} \cite{Ranjit2015a, Ranjit2016} exposed a charged levitated microsphere, which had been feedback-cooled to sub-Kelvin temperatures, to an electrical field oscillating at the particle's resonance frequency, and monitored the position spectral density, achieving a force sensitivity of 1.6\,aN\,Hz$^{-1/2}$, and after several hours of averaging measured at the 6\,zN level. See Hempston \emph{et al.} for a detailed analysis of force sensing with charged nanoparticles exposed to electric fields \cite{Hempston2017}. See Fig.~\ref{fig:force}a) for an illustration of force sensing with levitated particles.

The acceleration of levitated particles can be measured with high sensitivity. This is of interest, since the levitated particle system is potentially much more compact than existing commercial devices such as falling corner cubes or complex cold-atom systems. Monteiro \emph{et al.} \cite{Monteiro2017b} achieved a sensitivity of $0.4 \times 10^{-6}\,g$\,Hz$^{-1/2}$ using an optically levitated microsphere of diameter $\sim 10\,\mu$m, and after several hours of averaging achieved an ultimate sensitivity at the $10^{-9}\,g$ level. By monitoring the wave-packet momentum spread of a ground-state cooled nanoparticle, it is predicted that accelerations below $10^{-8}$\,m\,s$^{-2}$ could be measured on sub-second timescales \cite{Geraci2015}.

A quantum description of force sensing with optically levitated nanoparticles \cite{Rodenburg2016} finds that beyond the thermal limit, there is an optimum measurement strength, which represents a balance between minimizing shot noise (by using high optical power) and minimizing light scattering from the trapped particle (by using low optical power). This is nothing other than the Standard Quantum Limit (SQL) familiar from quantum measurement theory. For a thorough discussion of the quantum limits of measurement accuracy in optomechanical systems, and methods for circumventing them, see the following review articles \cite{CavOptReview, QuantNoiseReview}.

\subsubsection{Detection of surface forces}

Particles tightly confined in optical tweezers can be moved close to surfaces to search for short-range forces \cite{Geraci2010}, with a flexibility not provided by tethered oscillators. Rider \emph{et al.} \cite{Rider2016} trapped a charge-neutral microsphere, achieving a force sensitivity of $2 \times 10^{-17}\,$N\,Hz$^{-1/2}$ with 1000\,s interrogation time. By bringing the particle into close proximity (20\,$\mu$m) to an oscillating silicon cantilever they investigated novel screened interactions, such as those that could be provided by a speculated chameleon mechanism, ruling out some of the parameter space for the existence of such a mechanism. 

Diehl \emph{et al.} trapped and feedback-cooled a silica nanoparticle within 380\,nm of a SiN membrane \cite{Diehl2018}, and Winstone \emph{et al.} optically trapped a charged silica particle $4\,\mu$m from a SiO$_2$-coated Si wafer \cite{Winstone2018}. Both teams were able to reconstruct a distorted trapping potential for their particles, with the latter estimating a force sensitivity of $3 \times 10^{-7}\,$N\,Hz$^{-1/2}$. Magrini \emph{et al.} trapped a nanoparticle within 310\,nm of a photonic crystal cavity \cite{Magrini2018}.

\subsubsection{Sensing with levitated cavity optomechanics}
\label{sec:sens_cav}
Cavity optomechanical systems offer a route to extremely precise measurements of mechanical motion. This indeed motivated much of the early research into cavity optomechanics, culminating in the detection of gravitation waves by the LIGO interferometer \cite{LIGO}, which is a device capable of measuring displacements with an incredible sensitivity of $10^{-15}\,$m\,Hz$^{-1/2}$. The basis of displacement detection in cavity optomechanics is the shift in cavity resonance frequency $\w{cav}$ with oscillator displacement $z$, as encoded in the optomechanical coupling (also known in this context as the ``frequency pull'') $\go{} = \delta\w{cav}/\delta z$. Note that the geometry of the system is also encoded in the parameter $\go{}$, which for a moving-mirror Fabry-P{\'e}rot cavity of length $L$ is given by $\go{FP} = \w{cav}/L$, whereas for a levitated nanoparticle it's given in eqn.~\eqref{glin} in Sec.~\ref{sec:cav_cool}. 

In principle, one can monitor displacement by pumping the cavity off-resonance and observing the cavity amplitude fluctuations, but this leads to significant back-action onto the displacement. Hence, the standard protocol is to pump the optical cavity on-resonance and to measure the corresponding phase shift $\delta \theta$ of the light exiting the cavity, as illustrated in Fig.~\ref{fig:force}b). The phase shift is given by $\delta \theta \propto g \delta z/\kappa$, where $\kappa$ is the linewidth of the cavity, and the proportionality indicates that the exact shift depends on the phase-measurement technique which is employed, though we note that the minimum detectable phase shift is shot-noise limited $\delta \theta_{\rm min} = 1/2\sqrt{N}$, where $N$ is the number of photons which have passed through the cavity \cite{CavOptReview}. There are a wealth of highly-accurate phase monitoring techniques, and they can be particularly useful for quantum applications where particular quadratures of motion must be measured \cite{CavOptReview}.

For a nanosphere levitated within an optical cavity, displacement sensitivity at the $10^{-14}\,$m\,Hz$^{-1/2}$ level has been reported \cite{Delic2018, Windey2018}.  Geraci \emph{et al.} presented the first proposal for the detection of forces using a levitated cavity optomechanical system \cite{Geraci2010}, exploiting the potentially large values of $\Qm$.  A force gradient $\delta F / \delta z$ gives rise to a fractional shift in the cavity resonance of

\begin{equation}
\vert \delta \wc/\wc\vert = \frac{\vert \delta F / \delta z\vert }{2\ks{z}},
\end{equation} 
\noindent
where $\ks{z}$ is the spring constant in the $z$-direction. Geraci \emph{et al.} propose to trap a particle in the optical antinode of an optical cavity field \cite{Geraci2010} and with their parameters they expect to be able to detect fractional frequency shifts of $\vert \delta \wc/\wc\vert = 10^{-7}$ after 1\,s of averaging. By using an oscillating reflective substrate as one of the cavity mirrors, they aim to look for short-range forces between a trapped microparticle and the substrate, such as the Casimir force \cite{Canaguier-Durant2011} and short-range non-Newtonian gravity. Later work \cite{Geraci2015} extends this technique by introducing ground-state cooling and subsequent wave-packet expansion, and even further with additional matterwave interferometry (see Sec.~\ref{sec:MWI}). Very recent work \cite{Qvarfort2017} suggests that the fully quantum evolution of a levitated microparticle in an optical cavity could be used to achieve startling gravity-shift sensitivities of $10^{-16}\,g$\,Hz$^{-1/2}$.

\subsection{Other experimental configurations}

So far we have considered detecting forces using optically levitated particles. Various authors have suggested using other experimental configurations. The spin provided by \NV\-centres in levitated nanodiamonds (Sec.~\ref{sec:NV})	 can be used to generate a coupling between a magnetic field gradient and the mechanical motion of the particles. Kumar \& Bhattacharya \cite{Kumar2017} propose that such a coupling could be used to measure magnetic field gradients with 100\,mT\,m$^{-1}$\,Hz$^{-1/2}$ sensitivity under ambient conditions, and $1\,\mu$T\,m$^{-1}$\,Hz$^{-1/2}$ sensitivity if the oscillator is cooled to its ground-state under vacuum conditions.

Goldwater \emph{et al.} consider the case of a charged particle levitated in a Paul trap, whose motion induces a current in a nearby circuit \cite{Goldwater2018}. In this configuration, the dominant dissipation is a resistive coupling to the electrical circuit, which also determines the detection efficiency, which changes the nature of the force sensitivity as compared to an optomechanical system, and puts constraints on the measurement bandwidth. The authors predict minimum detectable forces below $10^{-19}\,$N after 1\,s of measurement in a room temperature environment, and below $10^{-21}\,$N in a 5\,mK environment. 

At the extreme cutting edge, Prat-Camps \emph{et al.} consider the magnetic levitation of magnetic particles above a superconducting surface \cite{Prat-Camps2017}. This system is predicted to be extremely low noise, and displacement readout is made via the magnetic flux induced in a nearby SQUID. By operating in UHV, and with an ambient temperature of 1\,K, the authors predict an impressive force sensitivity of $10^{-23}$\,N\,Hz$^{-1/2}$ with 100\,nm radius magnets, and an acceleration sensitivity of $10^{-15}\,g$\,Hz$^{-1/2}$ with 10\,mm radius magnets.

\subsubsection{Sensing via orientation:}

So far, we have considered the coupling of external forces to, and measurement via, the c.o.m. displacement of levitated particles. In this section we consider the orientational degrees-of-freedom, as discussed in greater detail in Sec.~\ref{sec:rotation}. For an anisotropic particle which is harmonically bound in its orientational degrees-of-freedom (librational motion), the motion is sensitive to externally applied torques $\torque$. In analogy to eqn.~\eqref{eqn:Fmin}, the thermal limit on the minimal detectable torque is \cite{Hoang2016}:

\begin{equation}
\torque^{\rm min} = \sqrt{\frac{4 \kB \T{\theta} I \w{\theta} b}{Q_{\theta}}},
\end{equation} 
\noindent
where $\T{\theta}, \w{\theta}, Q_{\theta}$ are the temperature, frequency and quality factor of the librational motion, respectively, and $I$ is the moment of inertia. Specifically for a levitated particle, this reduces to $\torque^{\rm min} = \sqrt{4 \kB \T{\theta} I \Gam{\theta} b}$ \cite{Ahn2018}, where $\Gam{\theta}$ is the damping on the librational motion (see Sec.~\ref{sec:rotation}). Although the sensitivity is strongly geometry dependent, it is suggested that torque sensitivities below $10^{-27}\,$Nm\,Hz$^{-1/2}$ are achievable in UHV \cite{Hoang2016, Ahn2018}, which is several orders of magnitude better than the state-of-the-art for non-levitating sensors. This would, for example, enable the measurement of Casimir torques \cite{Xu2017}.

\subsubsection{Detection of static forces:}

Most sensing schemes which exploit nano-oscillators consider the \emph{resonant} detection of forces. This means that the force under question must be varying at the same frequency as the resonant frequency of your detection oscillator. This is not always practical, since it requires one to add an external modulation to static forces. In addition, high quality factor sensors require long interrogation times to achieve high sensitivity, which can require demanding experimental stability. In recent work, levitated particles are emerging as detectors for truly static forces.  

Hebestreit \emph{et al.} use a technique only available to levitated oscillators: dropping \cite{Hebestreit2018c}. A particle is released from an optical potential after being cooled, exposed to a static force, and then recaptured. Since the force causes a DC shift in the particle's position, upon recapture the amplitude of its motion will be increased, which can be measured. This enabled both static gravitational and electrostatic force resolution at the 10\,aN level \cite{Hebestreit2018c}. 

Kuhn \emph{et al.} took a different approach \cite{Kuhn2017b}, by non-resonantly frequency-locking the rotation of a silicon nanorod to an external time-reference via pulses of circularly polarized light. The phase-lag between the optical drive and the response of the nanorod is effected by external forces, including scattering forces, and the authors inferred a torque sensitivity of $10^{-22}\,$Nm\,Hz$^{-1/2}$ under ambient conditions. This technique also offers MHz read-out rates \cite{Kuhn2017b}.

\subsection{Exotic sensing schemes}
\label{sec:exotic_sensing}

In this section, we will consider some of the novel forces that researchers have suggested a levitated sensor would be suitable to detect.

Moore \emph{et al.} undertook a detailed study of the response of charge-neutral microspheres to large electric fields, to look for the presence of anomalous milli-charges, which have been proposed as an extension to the standard model to partially explain the existence of dark matter \cite{Moore2014}. Their experiment was sensitive to charges above $10^{-5}\,e$, and concluded that such charges were limited to a maximum abundance of $2.5\times10^{-14}$ per nucleon.  

Arvanitaki \& Geraci propose a method for detecting high-frequency gravitational waves using a nanosphere (or microdisk) levitated within the field of an optical cavity \cite{Arvanitaki2013}. The scheme relies upon the fact that under the influence of a gravitational wave, the position of the particle relative to the cavity mirrors and the position of the trap-equilibrium are shifted by differing amounts, and this displacement of the sphere from the trap centre could be measured. This scheme requires a long (100\,m), low-finesse ($\mathcal F = 10$) cavity, which is still considerably more compact that existing gravitational wave detectors \cite{LIGO}. We note, that a recently proposed fibre-cavity configuration may be suitable for such an experiment \cite{Pontin2018}. The levitated nanoparticle gravitational wave sensor is particularly sensitive around the mechanical frequency of the trapped particle, and the authors state that this scheme increases sensitivity in the 100-300\,kHz band by up to 1000 as compared to Advanced LIGO, making it particularly suitable for detecting novel gravitational wave sources such as the annihilation of QCD axions \cite{Arvanitaki2013}.

Finally, Riedel has proposed that macroscopic superpositions could be used to detect low-mass dark matter \cite{Riedel2013, Riedel2015}. The core of this idea is that even if collisions with certain types of particle are classically undetectable, they could still cause detectable decoherence of a macroscopic superposition state in a matter-wave interferometer. As with any detection scheme which relies upon decoherence, a challenge is distinguishing the decoherence source of interest from other sources, though the author suggests that an anisotropic dark matter flux could be detected \cite{Bateman2015, Riedel2015}. Detection relies on a proposed ``coherent enhancement'' of scattering probability between low-mass dark matter and macroscopic particles \cite{Riedel2013, Bateman2015}.

\section{Levitated Cavity Optomechanics}
\label{sec:cav_cool}

\setcounter{footnote}{0}

\begin{figure}[ht]
{\includegraphics{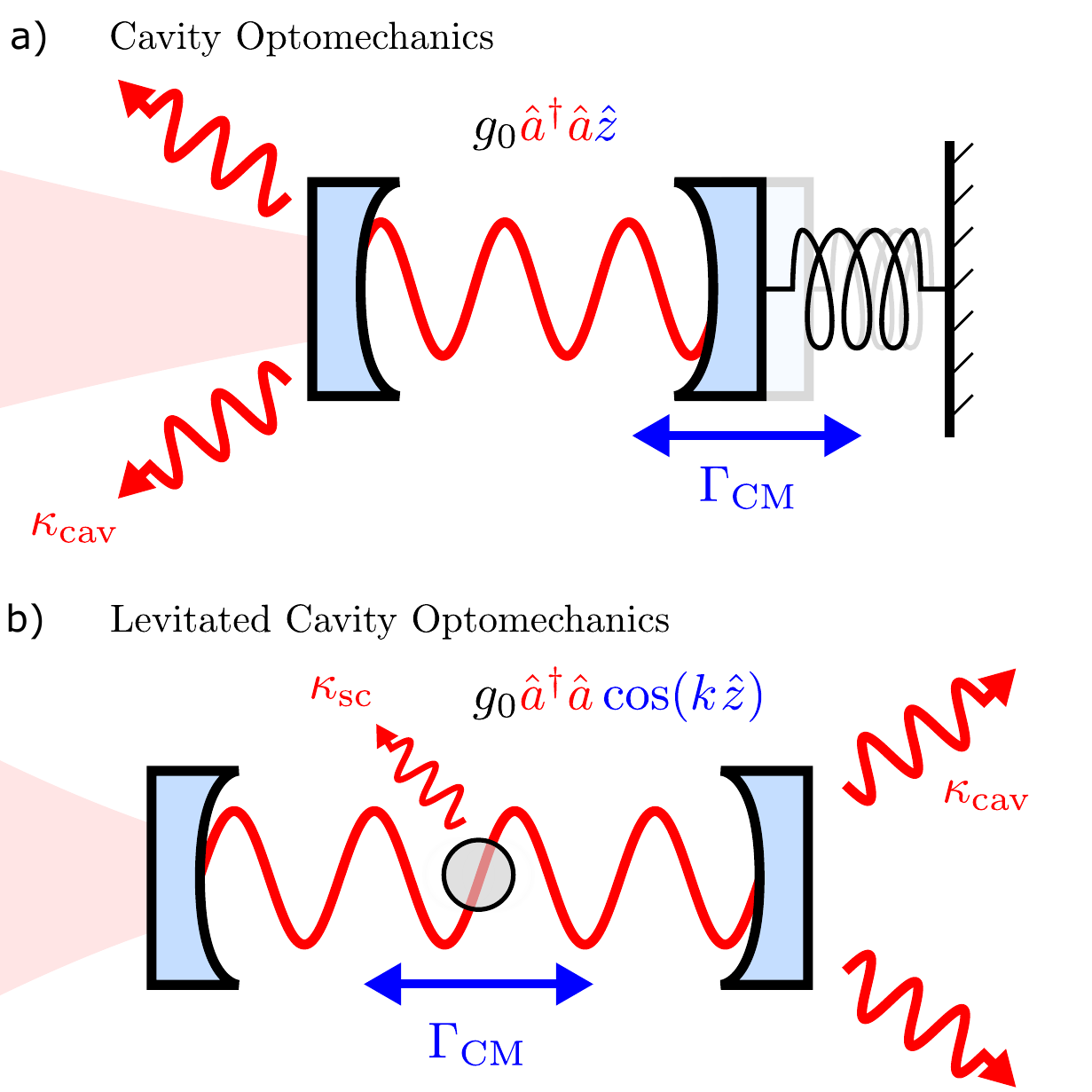}}	\centering
\caption{\label{levvsnonlev}
\textbf{Standard optomechanics versus levitated optomechanics} a)  In the canonical optomechanical set-up, a moving mirror of a Fabry-P{\'e}rot cavity oscillates about an equilibrium position at frequency $\w{q}$, altering the length of the cavity and hence its resonant frequency. If the cavity is driven off-resonance, with a detuning $\Det \simeq -\w{q}$, the result is the famous sideband cooling mechanism: radiation pressure forces from the cavity mode can cool the mirror's mechanical oscillation to its quantum ground state \cite{Bowenbook,CavOptReview}.  b) For levitated optomechanics, a nanoparticle is levitated within, and cooled by, the light field. In this case, the nanosphere couples to the cavity field via the dipole force. Motion of the nanosphere perturbs the intracavity electric field,  mimicking a change in the cavity length and in principle permitting ground state cooling.}

\end{figure}

The intense research interest - over the last decade or so - in cavity optomechanical systems consisting of an optical resonator with a movable mirror, arises from two unique capabilities of the Fabry-P{\'e}rot cavity:\\

 {\bf (1) Quantum limited-detection:} The exquisite sensitivity of an optical Fabry-P{\'e}rot cavity to small displacements of its mirror was underlined by the first detection of gravitational waves by the LIGO detector, which has attained displacement sensitivities as low as $10^{-18}$\,m\,Hz$^{-1/2}$ \cite{LIGO}, a small fraction of an atomic nucleus. But this success, the culmination of a decades-long project, involves very large scale science (many hundreds of millions of US Dollars), including the multinational VIRGO project. These efforts have stimulated far smaller-scale optomechanical set-ups whose potential for sensing quantum scale motion is widely recognized (see {\em The Economist } magazine, 28/01/2017). These experiments are motivated not only by foundational science but also metrology, accelerometry, navigation, and ultra-weak force sensing.\\

{\bf (2) Optical cooling:}  The well-known red-sideband cooling mechanism has enabled the cooling of small mechanical (tethered) oscillators to their quantum ground state.  There are excellent textbook treatments of the underlying mechanism \cite{Bowenbook, CavOptReview}. Cavity modes for a cavity of length $L$ are characterized by their frequency spacing (free spectral range) ${\rm FSR} = c/2L$, and a photon lifetime $\tau = 1/(2\pi\kap{cav})$, with $\kap{cav}= \frac{c\pi} {\F L}$. The finesse $\F$ encodes the mirror reflectivity, and ranges up to $\sim 10^6$ at the state-of-the-art. Each cavity mode corresponds to a Lorentzian resonance of FWHM $\kap{cav}$.

A key requirement for ground-state cooling is to attain the resolved-sideband regime, where the mechanical frequency $\w{q} \gtrsim \kap{cav}/2$. It is also necessary to minimize the coupling of the mechanical system to its environment, as quantified by the rate $\Gam{CM}$, which includes contributions due to collisions with the surrounding gas $\Gam{gas}$ and the scattering of photons $\Gam{rad}$.

Combining {\bf(1)} and {\bf(2)} in a \emph{levitated} cavity optomechanical system, see Fig.~\ref{levvsnonlev}, would open the door to the quantum control of levitated nanoparticles. Achieving  {\bf(1)}, sensing motion at the zero-point fluctuation $\zpf{x}$ level, seems achievable since $\zpf{x} = \sqrt{\frac{\hbar}{2\mass \w{q}} } \sim 10^{-12}$\,m for a levitated system, which is modest in comparison with the sensitivity of LIGO. However, achieving  {\bf(2)}, the stable trapping and strong cavity cooling of a levitated nanoparticle in high vacuum (where $\Gam{gas} \to 0$) remains a challenge, despite considerable progress.

\subsection{Levitated cavity optomechanics}

Arguably the most significant distinction between levitation within an optical cavity and levitation in an optical tweezer is that 
in the former the light field is naturally an \emph{active} participant in the dynamics. The conservative part of the linearized Hamiltonian of the light-matter system may be written

\begin{equation}
\frac{\hat{H}} {\hbar}= \frac{\w{opt}}{2} (\HPo^2+ \Hq^2)  + \frac{ \w{q} }{2}
 (\hat{p}^2+ \hat{q}^2) +   \Vint( \Hq, \hat{q}),
\label{Hamopto}
\end{equation} 
\noindent
where $\w{opt}$ is the frequency of the light field\footnote{Equation~\ref{Hamopto} is typically solved in a rotating frame of the optical field. Hence $\w{opt} \equiv \Det=\w{cav}-\w{L}$, which is the detuning between the cavity resonance frequency $\w{cav}$ and the laser frequency $\w{L}$.  From here on we use the standard optomechanical convention of using $\Det$ to characterize the optical frequencies.} described by conjugate variables $\HPo$ and $\Hq$, $\w{q}$ is the mechanical frequency of the levitated particle with conjugate variables $\hat{p}$ and $\hat{q}$, and $\Vint( \Hq, \hat{q})$ describes the interaction between the optics and mechanics. In other words there are two {\em coupled} quantum harmonic oscillators rather than, as for levitation without a cavity, a mechanical oscillator ($q\equiv \{x,y,z\}$) in a \emph{classical} external potential. The aim of many levitated set-ups is to achieve a light-matter coupling similar to the canonical optomechanical form

\begin{equation}
\Vint( \Hq, \hat{q})= \go{} \Hq \hat{q},
\label{Optoint}
\end{equation} 
\noindent
 where $\go{}$ is the optomechanical coupling.

All one requires to put the optical and mechanical degrees-of-freedom on the same footing as in eqn.~\eqref{Hamopto} are standard coordinate rescalings of the mechanical coordinates that eliminate the mass: the rescaling $ \hat{Q} = \sqrt{\hbar/(\mass\w{q})} \hat{q}$ and $\hat{P} = \sqrt{\hbar \w{q} \mass} \hat{p}$
transforms a mechanical harmonic oscillator Hamiltonian 
$\frac{\hat{P}^2}{2\mass}+ \frac{1}{2}m\omega^2 \hat{Q}^2$ to the form shown in eqn.~\eqref{Hamopto} $\frac{ \w{q} }{2}
 (\hat{p}^2+ \hat{q}^2)$. 

This symmetrized form of the Hamiltonian has underpinned a raft of standard cavity optomechanical phenomena, including hybridization of photon and phonon modes associated with 
$ \Hq, \hat{q}$ respectively \cite{Bowenbook,CavOptReview}.  We can also write the linearized limit of eqn.~\eqref{Hamopto} using optical field operators, $\Hq=\frac{1}{\sqrt{2}} (\hat{a}^\dagger + \hat{a})$  such that
$\frac{\hat{H}} {\hbar}=  \Det \hat{a}^\dagger\hat{a}+ \frac{ \w{q} }{2}
 (\hat{p}^2+ \hat{q}^2) + \go{} (\hat{a}^\dagger + \hat{a} ) \hat{q}$. Similarly, we can define field operators for the mechanics $\zpf{Q} \hat{q}=\sqrt{\hbar/(2\mass\w{q})}(\hat{b}^\dagger + \hat{b})$,
where $\zpf{Q} $ characterizes the quantum zero-point fluctuations of the mechanical oscillator. 

The most important process in near-resonant cavity optomechanics is the well-known red-sideband cooling mechanism that occurs for $\Det \approx  -\w{q}$. This yields an optomechanical cooling rate $\Gam{opt}$, which in the sideband resolved regime ($\w{q} \gtrsim \kap{cav}/2$) can be large enough to enable ground-state cooling, assuming $\Gam{opt} \gg \Gam{gas}$\footnote{The thermal heating rate $\Gam{gas}$ is minimized in standard optomechanics by working in a cryogenically cooled environment. Levitated systems have the advantage of good isolation from the thermal environment if operating at low ambient pressures.}.

The mechanical and optical harmonic oscillators in eqn.~\eqref{Hamopto} each give rise to a Langevin equation subject to uncorrelated Gaussian noises, in the uncoupled $\go{}=0$ case. In the regime where the dominant source of dissipation is due to collisions with gas molecules ($\Gam{CM}\simeq\Gam{gas}$), the mechanical mode $\dot {\hat{p}}+\w{q} {\hat{q}}+\Gam{gas} {\hat{p}}=\zeta(t)$,  experiences thermal noise forces \cite{Bowenbook}

\begin{equation}
\langle \zeta(t')\zeta(t) \rangle \simeq \Gam{gas} (\nm+\frac{1}{2})  \delta(t-t'),
\end{equation} 
\noindent
where $\nm=\kB \T{env}/(\hbar \w{q})$ is the thermal occupancy when coupled to the environment. The effect of optical scattering is considered in Sec.~\ref{sec:challenges}.

Optical photons have energy scales much larger than the thermal environment  $\hbar \w{L} \gg \kB \T{env}$, and so optical cavity modes can be considered to have zero thermal occupation $\nopt \simeq 0$ for a shot-noise limited laser, even under ambient conditions. Hence, the optical modes experience amplitude fluctuations
\begin{equation}
\langle \zeta_{\rm opt}(t')\zeta_{\rm opt}(t) \rangle \simeq \kap{\rm cav} (\nopt+\frac{1}{2})  \delta(t-t')\equiv \frac{\kap{cav}}{2} \delta(t-t'),
\end{equation} 
\noindent
 corresponding to a zero temperature bath, $\T{opt}=0$. In a non-idealized case, $\zeta_{\rm opt}$ would also parametrize classical noise arising from laser amplitude or frequency fluctuations.

In the scenario where background gas dominates environmental noise on the mechanical degree-of-freedom $\Gam{CM} \approx \Gam{gas}$, and in the presence of optomechanical coupling and hence optical damping $\Gam{opt}$, the equilibrium temperature of the mechanical mode is thus \cite{Bowenbook, CavOptReview}
\begin{equation}
\T{CM}= \frac{\Gam{gas}\T{env}+ \Gam{opt}\T{opt}}{\Gam{gas}+ \Gam{opt}}\simeq \frac{\Gam{gas}}{\Gam{opt}}\T{env}.
\end{equation}
\noindent
Hence, for a levitated system in high vacuum where $\Gam{gas}/\Gam{opt} \to 0$, the ground state is approached\footnote{Other noise sources, including photon scattering (see below), make an increasing contribution to $\Gam{CM}$ in ultra-high vacuum, and the above description should be modified accordingly.}. 

\subsubsection{The quantum back action (QBA) regime:}
For both clamped and levitated cavity optomechanics, the coupling to the cavity generates the well-known quantum back-action (QBA) force $F_{\rm QBA} \propto \go{}^2$. The QBA regime occurs when $F_{\rm QBA}$ exceed the Langevin thermal noise forces and is associated with two of the most interesting and significant quantum signatures of cavity optomechanics: quantum sideband asymmetry and ponderomotive squeezing 
\cite{Bowenbook,CavOptReview}.

The figure of merit for achieving the QBA regime is the quantum cooperativity

\begin{equation}
\Coop=\frac{4\go{}^2}{\kap{cav} \Gam{CM} \nm} \gtrsim 1.
\label{Cooperativity}
\end{equation} 
\noindent 
The total mechanical damping rate $\Gam{CM}$ is the sum of all noise sources acting on the particle's centre-of-mass (c.o.m.), and $\Gam{CM}\nm$ is the mechanical heating rate. This regime has been achieved with levitated ultracold atoms \cite{Brooks2012}, and is being approached with levitated nanoparticles \cite{Windey2018}. 

One should make a clear distinction between QBA in cavity optomechanics and the backaction recently investigated in non-cavity levitated optomechanics \cite{Jain2016}, where recoil heating on the levitated particle due to shot noise was measured. In QBA, one has a two stage process: firstly the amplitude of the shot noise fluctuations disturbs the mechanical oscillator; secondly the resultant mechanical fluctuations in turn are imprinted on the phase of the intracavity field, resulting in noise amplitude-phase correlations in the cavity output; for example these can manifest in noise floor levels in the measured output signal which are {\em lower} over some ranges than the incoming shot noise levels.

\subsection{Levitated cavity optomechanics:  four challenges}
\label{sec:challenges}
 In 2010 three independent proposals \cite{Chang2010,RomeroIsart2010,Barker2010a} for the optomechanical cooling of levitated nanoparticles opened a new sub-field of quantum levitated cavity optomechanics. There has been enormous progress since, and many ingenious developments and important steps towards the ultimate goal of attaining ground-state cooling or the quantum back-action regime.

We focus on four particular road-blocks  which specifically hinder progress in levitated cavity optomechanical systems. A difficulty is in the interdependencies: optimizing one of these road-blocks typically translates into degraded performance in one of the others. Solving these challenges simultaneously approximately coincides with optimizing the quantum cooperativity $\Coop$:

\begin{enumerate}
\item maximizing $\w{q}$ to minimize the mechanical occupation $\nm$ for a given cooling rate $\Gam{opt}$,
\item maximizing the optomechanical coupling $\go{}$, 
\item minimizing $\Gam{rad}$, the optical scattering contribution to $\kap{cav}$, and hence c.o.m. heating $\Gam{CM}$,
\item stable trapping at high vacuum. 
\end{enumerate}

\subsubsection*{1. Maximizing the mechanical frequency:}

Levitated nanoparticles, unlike clamped oscillators, have no intrinsic natural frequency so this must be set by some auxiliary classical (typically optical) levitating field which generates the trapping frequency $\w{q}$.  Hence the coordinate rescaling  of the conservative Hamiltonian used in 
 eqn.~\eqref{Hamopto} is not immediately possible. The departure point is the  nonlinear Hamiltonian
\begin{equation}\label{Hamclass2}
\frac{\hat{H}}{\hbar}=  \Det  \hat{a}^\dagger \hat{a} + \frac{\hat{P}^2}{2\mass}+
 U_{\rm class}( \hat{Q})+\Vint(\hat{a}^\dagger \hat{a}, \hat{Q}),
\end{equation} 
\noindent
where the mechanical  coordinates are unscaled values, in ordinary laboratory units. The auxiliary classical field, or fields, $U_{\rm class}$ now provide the trapping potential and the trapping frequency $\w{q}$ is an emergent property from analysis of the form of the trap.

Both $U_{\rm class}$ and $\Vint$ fields can originate from the \emph{same} optical field i.e. the total electric field in the optical cavity. This may arise from multiple cavity modes as well as external fields such as an optical tweezer used to hold the particle within the cavity.  One exception \cite{Millen2015,Fonseca2016}  involves an additional electrical trapping potential for charged nanoparticles, which contributes to $\Upot{\rm class}$ but not $\Vint$. For the fully optical cases, we distinguish two important scenarios:

\noindent{\bf I: Self-trapping}
In this case, first proposed in \cite{Chang2010, Barker2010a}, the standing wave of a cavity mode (or multiple cavity modes) determines the mechanical frequency  $\w{q}$.  In particular,
 $\Upot{class} \equiv \Upot{class}(\nopt, \hat{Q})$ which depends on the mean photon number $\nopt=\langle \hat{a}^\dagger \hat{a}\rangle$  in the cavity.
We consider the cavity axis to be along $z$, thus $\hat{Q}\equiv z$. The dipole force potential for a single cavity mode is 
 \begin{equation}\label{Dip}
\Upot{d}^{\rm cav}=  -A\hbar  \hat{a}^\dagger \hat{a} \cos^2 {\kl\hat{z}} \env(\hat{x},\hat{y}),
\end{equation} 
\noindent
where $\env$ is the transverse envelope of optical beam (set to 1 for now) and $A = \frac{3\V{s}}{2\V{m}}\frac{\epsilon_\mathrm{r}-1}{\epsilon_\mathrm{r}+2} \w{L}$. The potential $\Upot{d}^{\rm cav}$ has a minimum at the antinode $\hat{z}=0$, and there are multiple optical wells separated by $\lambda/2$. A classical potential is obtained by considering small quantum fluctuations of the cavity field about a classical steady state value $\overline{a} =\sqrt{\nopt}$, hence we replace $\hat{a}\to \overline{a} +\hat{a}$, so 
$\Upot{class}/\hbar=  -A \nopt \cos^2 {\kl\hat{z}} $ and $\nopt=|\overline{a}|^2$.

The classical well depth is related to the parameter $A$ which depends on the sphere volume $\V{s}=4/3\pi \Rad^3$, the cavity mode volume $\V{m}=\pi \waist^2 L$ (with $\waist$ the waist of the cavity field and $L$ the length of the cavity) and the laser frequency $\w{L}$. Considering small oscillations about the equilibrium position $z_0$ one can also replace  $\hat{z} \to z_{0} +\hat{z}$ and obtain an effective trap frequency
 \begin{equation}\label{Dip}
\w{z}^2 \approx  2\hbar  \frac{A}{\mass} \nopt  \kl^2 \cos 2\kl z_0 \equiv 2\hbar  \frac{A}{\mass} \nopt  \kl^2,
\end{equation} \noindent
for small oscillations about an antinode $\kl z_0=N\pi$, for $N=0,\pm 1,\pm 2...$. 
In this case, the interaction potential:
 \begin{equation}\label{Vint}
\Vint(\hat{a},\hat{z})=  -\frac{A}{2} \sqrt{\nopt}  [(\hat{a}^\dagger+ \hat{a}) \cos {2\kl(z_0+\hat{z})}],
\end{equation} 
\noindent
where we assume the mean field $\overline{a}$ to be real. \\

\noindent{\bf II: External trapping}
This scheme was introduced and considered in \cite{RomeroIsart2010} and considers a nanoparticle levitated within the cavity field by an optical tweezer. For these set-ups, the  potential $\Upot{class} \equiv \Upot{d}^{\rm cav} (\nopt, \hat{z})+ \Upot{d}^{\rm tw}(\Io,\hat{z}) $
combines the cavity standing wave potential with an additional optical tweezer trap potential of intensity $\Io$;  but  $\Upot{d}^{\rm tw} \gg \Upot{d}^{\rm cav}$ so the dominant contribution is from the external tweezer potential, which is approximately harmonic. However the interaction term is still given by $\Vint(\hat{a},\hat{z})$ in eqn.~\eqref{Vint}. 

\begin{figure*}[ht]
\begin{center}
{\includegraphics{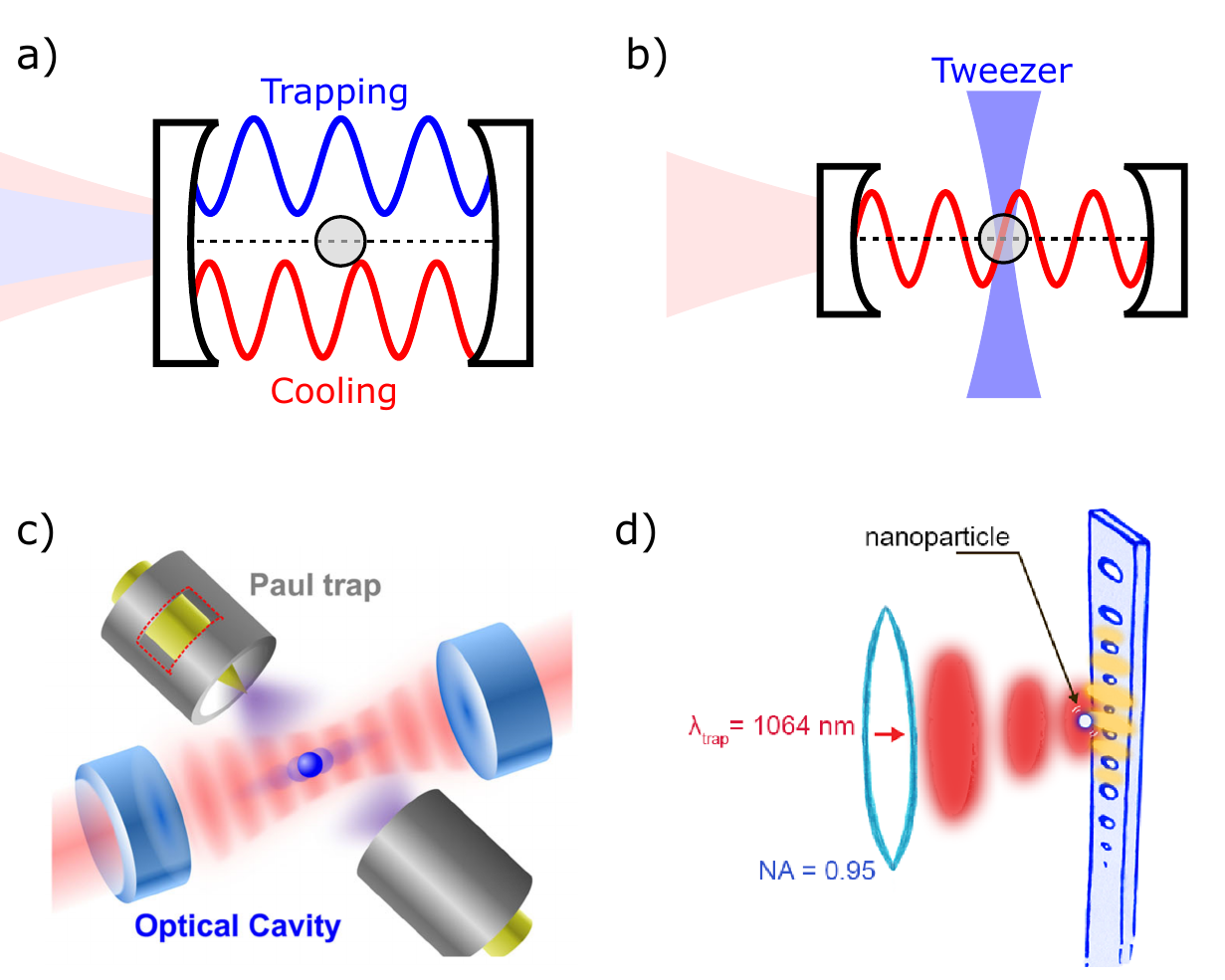}}	\centering
\end{center}
\caption{{\bf Levitated cavity optomechanics set-ups} a)  Two-mode cooling. A self-trapping system with separate modes for trapping and cooling. Proposed and investigated in \cite{Chang2010,Pender2012,Monteiro2013}, and implemented in \cite{Kiesel2013} (fields displaced for clarity). b) Set-up with external optical tweezer for trapping and a cavity mode for optomechanical cooling and potentially read-out. Introduced in \cite{RomeroIsart2010} with a realization in \cite{Mestres2015}. It is also possible to achieve cooling by coherently scattering light from the tweezer into the cavity mode \cite{Windey2018, Delic2018, GonzalezBallestero2019} c) Hybrid system consisting of a Paul trap and optical cavity, introduced and realized in \cite{Millen2015,Fonseca2016}, image taken from \cite{Millen2015}. Self-trapping as $\w{q}$ is set by the cavity mode, with the Paul trap enabling stable trapping at high vacuum. d) A standing-wave optical tweezer enables coupling to the near-field of a photonic crystal cavity \cite{Magrini2018}, achieving high single-photon optomechanical coupling, image taken from \cite{Magrini2018}.} 
\label{Cavitysetups}
\end{figure*}

The external trapping field can be useful for isolating the nanoparticle loading stage from the delicate optical cavity \cite{Mestres2015}, for pre-cooling, and for bringing the particle close to near-field cavity structures \cite{Magrini2018}. There are also significant advantages in decoupling the control of the mechanical frequency from control of the interaction and optomechanical coupling.\\

For both self-trapped and externally trapped cases, frequencies in the range $\w{q}/2\pi \approx 20-200$\,kHz have been achieved, which is significantly lower than the majority of tethered systems which achieve frequencies in the MHz-GHz range, facilitating quantum operation. Levitated oscillators in vacuum or near-vacuum must cool from ambient temperatures $\T{env}\approx 300$\,K and $\nm\sim 10^7$, several orders of magnitude higher than for the most favourable clamped set-ups for which the initial $\nm \sim 10^2-10^4$, which when aided by cryogenic cooling only require a modest degree of red-sideband cooling to reach the quantum regime. In addition, higher frequencies facilitate the side-band resolved condition $\w{q} \simeq \kap{cav}/2$, which for dispersive cooling enables the quantum ground state to be reached. Typical optical cavities \cite{Kiesel2013,Asenbaum2013,Millen2015,Fonseca2016} have $\kap{cav} \sim 20-300$\,kHz.

To increase $\w{q}$ one must increase the optical power through $\nopt$ or $\Io$\footnote{Recent experiments \cite{Kuhn2017a} have demonstrated that you can increase $\w{q}$ by increasing $\pol{}$ through both using higher refractive index materials, and non-spherical geometries, see Sec.~\ref{sec:rotation}. Such particles are also suitable for cavity optomechanics \cite{Stickler2016a}, including in their rotational degree-of-freedom.}; the drawback is that the coupling of $\Upot{class}$ to free photon modes through $\Gam{rad}$ (as opposed to cavity modes) leads not only to loss of photons from the cavity (and larger $\kap{cav}$) but also to recoil heating, currently a severe limiting factor (see below, challenge 3). Trapping with an electric field such as \cite{Millen2015,Fonseca2016} would entirely avoid the recoil heating problem; the drawback is that currently these attain $\w{q} \lesssim 1$ kHz so such set-ups still rely on optical self-trapping.

\subsubsection*{2. Maximizing the optomechanical coupling:}

Like standard optical trapping, the optical-cavity field standing-wave can provide an effective harmonic potential $\propto \hat{z}^2$ of comparable $\omega_{z}$. However, in  cavity
levitation,  this  now becomes a two-way process:  the particle itself results in a position dependent perturbation to the cavity electric field and hence the cavity modes $\hat{a}$ described by eqn.~\eqref{Vint}.

The field should localize the nanoparticle about an equilibrium position $z_0$ where there is an appreciable linear component of the interaction. For small
displacements about $z_0$ we write

 \begin{equation}\label{Vint2}
\Vint(\hat{a},\hat{z}) \approx  - \kl A\sqrt{\nopt}  (\hat{a}^\dagger+ \hat{a}) \sin {2\kl z_0} \hat{z}= \go{} (\hat{a}^\dagger+ \hat{a})(\hat{b}^\dagger+ \hat{b}),
\end{equation} 
\noindent
where we have introduced $\hat{z}=\zpf{z}(\hat{b}^\dagger+ \hat{b})$ position field operators. The optomechanical coupling strength is thus
 \begin{equation}\label{glin}
\go{}= \kl A\sqrt{\nopt}  \zpf{z} \sin {2\kl z_0} =\go{0}\sqrt{\nopt}
\end{equation} 
\noindent
where $\go{0}$ is the single-photon coupling. Since $\Gam{opt} \propto \go{}^2$ it is important to maximize $\go{}$; this can be done by increasing $\nopt$, at the cost of increased photon recoil heating. 

To increase the optomechanical coupling, one can also decrease the mode volume \cite{Kuhn2017c, Magrini2018}, since $A\propto \V{s}/\V{m}$, which involves significant engineering challenges. One can also increase the nanoparticle size, since $\go{}^2\propto \Gam{opt} \propto \Rad^6$. For a self-trapping system, the dependencies can be more complicated as  $\zpf{z}$ also depends on $A$ via $\omega_{z}$, with a more detailed consideration given in e.g. \cite{Monteiro2013}. Recoil heating rates also increase with particle size as $\Gam{rad}\sim \Rad^3$ for modest sized nanoparticles, but as $\Gam{rad}/\Gam{opt}  \ll 1$ is required for ground-state cooling, larger nanospheres are still more advantageous under this particular consideration. However, in quantum regimes at ultra-high vacuum, where the heating is dominated by photon scattering, the typical number of coherent oscillations of the mechanical oscillator $N_{\rm osc} \propto \Gam{rad}^{-1} $  \cite{Chang2010} so minimizing $\Gam{rad}$, for example by working with smaller particles, would be necessary to observe quantum effects.

 There is a limit to the size where the point-dipole approximation holds $\Rad/\lambda  \lesssim 0.2$, hence $\Rad\approx 200$\,nm is close to the upper limit for the commonly used laser wavelength $\lambda =1064$\,nm. Even there, the sphere diameter of 400\,nm is surprisingly large relative to typical optical well widths of $\lambda/2=532$\,nm. There are further challenges to observing quantum effects with such large spheres, see Sec.~\ref{sec:quantum}.\

A critical parameter is the location of the particle's equilibrium position $z_0$ relative to the cavity field structure. For a $\Rad \approx 200$\,nm silica nanosphere with $\w{z} \sim 10-100 \times 2\pi$\,kHz the term $\kl A \zpf{z} \sim 1$\,rad/s, hence the one-photon coupling rate $\go{0} \sim  \sin {2\kl z_0}$, which ranges from zero at the antinode up to a maximal value of 1\,rad/s.  

Displacing the nanoparticle from the antinode can be done in a variety of ways. Using a second cavity mode was investigated in \cite{Chang2010,Pender2012,Monteiro2013} and implemented experimentally in \cite{Kiesel2013}, as illustrated in Fig.~\ref{Cavitysetups}a). A separate optical trap can position the nanoparticle relative to a cavity mode used only for cooling \cite{Mestres2015}, as illustrated in Fig.~\ref{Cavitysetups}b) \& d). In \cite{Millen2015,Fonseca2016} a Paul trap was used alongside the optical cavity field, as shown in Fig.~\ref{Cavitysetups}c), which both stabilizes the particle and slowly drives the particle away from the optical antinode at  $z_0=0$.  


\subsubsection*{3. Minimizing optical heating:}

This challenge of minimizing recoil heating \cite{Jain2016} is pertinent to all optical trapping experiments aiming to cool the motion of levitated particles. Specific to cavity cooling, however, is the additional requirement to attain  $\kap{cav} \gg \Gam{rad}$, such that the dynamics are controlled by photon decay through the mirrors rather than by scattering from the particle. This aim is, to a degree, in conflict with efforts to maximize $\go{}$ and $\Gam{opt}$, since scattering increases with increasing $\Rad$ and intra-cavity power.

The optical cavity cooling rate $\Gam{opt} \propto \Rad^6$, whilst $\Gam{rad} \propto \Rad^3$, so using larger nanospheres $\Rad \simeq 200$\,nm is comparatively advantageous for achieving ground-state cooling. However, $\Rad \gtrsim 200$\,nm runs against the point-dipole requirement $\Rad \ll \lambda/2$. Ultra-high finesse cavities $ \F \gtrsim 100,000$  are advantageous for sideband resolved cooling, yet make the $\kap{cav} \gg \Gam{rad}$ condition harder to achieve as linewidths of $\kap{cav} \approx  10$\,kHz are typical. It is also important to minimize $\Gam{rad}$ to maintain coherence of the mechanical oscillations, since $N_{\rm osc} \propto \Gam{rad}^{-1} $ as indicated in the previous section.

Overcoming the challenge of minimizing optical heating is likely to be the biggest stumbling-block in achieving quantum-control of levitated nanoparticles, in any experimental format, and may require the use of non-optical fields. 

\subsubsection*{4. Minimizing gas heating; stable trapping at high vacuum:}
At sub-millibar pressures (with some variation across experiments), optically levitated nanoparticles become unstable. The exact mechanism is not well understood, but thought to be a combination of optical absorption \cite{Millen2014, Hebestreit2018}, optical scattering \cite{Jain2016}, and potentially even transfer of  angular momentum from the light to the particles \cite{Svak2018}. Operation in high vacuum is essential to minimize $\Gam{CM}$, and eventually to avoid decoherence, see Sec.~\ref{sec:decoherence}.

An additional challenge in an optical cavity rather than an optical tweezer, is the relatively weak transverse confinement of the particle, exacerbating particle loss. For the same reason, cooling rates in an optical cavity are greatly reduced in the transverse directions. Presumably, suitably optimized two-mode cavity cooling would stabilize the particle enough to operate stably at low pressures, but this is yet to be demonstrated.

By using an external optical trapping field, as in Fig.~\ref{Cavitysetups}b), optical feedback \cite{Gieseler2012} can be used to stably trap a particle within the cavity field, as demonstrated by \cite{Mestres2015}. The interplay of active feedback and passive cavity cooling is a complex process, which can lead to  interesting dynamical effects \cite{Genoni2015}. Due to the vastly deeper trapping potential, a Paul trap can stably trap a charged nanoparticle in high vacuum, where it can then be cooled by a weaker optical cavity field \cite{Millen2015, Fonseca2016}. It may be extremely challenging to perform quantum experiments with a charged particle, due to strong decoherence mechanisms, though well-established particle neutralization techniques exist \cite{Moore2014, Frimmer2017, Hempston2017}. Otherwise, there are proposals to work with magnetically levitated neutral particles \cite{RomeroIsart2012, Via2015}. 

Another potential solution is to avoid the necessity for stable trapping at all. Asenbaum \emph{et al.} \cite{Asenbaum2013} demonstrated the cooling of nanoparticles \emph{in transit} through an optical cavity in high-vacuum. As the particles transit, they become trapped in one-dimension (along the cavity axis), and are cooled. This is a suitable geometry if your final goal is matterwave interferometry, see Sec.~\ref{sec:MWI}. Further work \cite{Stickler2016a} has predicted that it is possible to cool all degrees-of-freedom of an anisotropic nanoparticle that transits through an optical cavity. This same group has demonstrated the detection of nanoparticles in transit through a silicon microcavity \cite{Kuhn2017c}.

\subsection{Levitated cavity optomechanics: state of play }

Relatively few research groups have published experimental work claiming optical cavity cooling of levitated nanoparticles. Key parameters for existing experiments are given in Table~\ref{table:cav}. 

This does not include the work by Mestres \emph{et al.} \cite{Mestres2015}, in which a feedback-cooled particle is transferred to an optical cavity field, since the trapping is for a short time and they report no cooling or coupling to the cavity field. Nevertheless, this work points towards a promising direction in the field: a ``best of both worlds'' scenario which would combine the strong cooling and read-out offered by cavity sideband cooling and phase sensitivity to displacement, with the stable trapping and 3D cooling offered by tweezers with active feedback cooling. Two recent works \cite{Windey2018} and \cite{Delic2018} are in this vein, and currently achieve mK temperatures. These systems trap within a cavity using a tweezer, as in Fig.~\ref{Cavitysetups}(b) but the cavity is not driven; it is coherently populated by the scattered tweezer light field \cite{GonzalezBallestero2019}. As the cavity resonance is blue shifted relative to the tweezer laser, optomechanical cooling is demonstrated. One remarkable feature of these set-ups is that they offer 3D cooling, whereas most cavity schemes only offer strong axial cooling.

Table~\ref{table:cav} includes the work of Magrini \emph{et al.}, where a nanoparticle is trapped in a tweezer in the near field of a photonic crystal cavity \cite{Magrini2018}. Of particular note is the very-high single-photon optomechanical coupling $\go{0} = 9.3\,$kHz. The authors do not use the cavity for cooling in this case, rather for read-out, suggesting that this system will enable feedback-cooling to the quantum regime. We note the quantum cooperativity $\Coop \sim 10^{-9}$, which may seem surprising given the particularly large optomechanical coupling rate in this experiment, however this is balanced by the large cavity decay rate $\kap{cav}/2\pi \sim 5\,$GHz.

The highest quantum cooperativities achieved to date are the recent coherent scattering experiments discussed above, with \cite{Delic2018} operating in the sideband resolved regime but limited by operating pressure, and \cite{Windey2018} operating at low pressures but not in the sideband resolved regime. As the optomechanical coupling arises from a
cross-term between the strong tweezer field and the cavity field, extremely strong values of effective $g \gtrsim 10$ kHz result, potentially opening the way to quantum levitated cavity optomechanics in the very near future.

\begin{table}
	\centering
		\begin{tabular}{ |l|c|c|c|c|c|c|c|c| }
		\hline
			 & $\F$ & $\kap{cav}/2\pi$ & $\w{q}/2\pi$ & $\Gam{CM}/2\pi$ (Hz) & $\go{0}/2\pi $ &  $\T{fin}$ & $\Coop$  \\
			 &			& (Hz)             & (Hz)         & $\Pg^{\rm min}$ (mbar)         &  (Hz)          &  (K)  &  \\
			\hline
			Ref.~\protect\cite{Kiesel2013}  & $7.6\times10^4$ & $1.8\times10^5$ & $1.7\times10^5$ & $7.2\times10^3$ & 1.2 &  64 & $10^{-6}$ \\
			Figure~\protect\ref{Cavitysetups}a)  & & & & 4.0 & & &  \\ 
			\hline
			Ref.~\protect\cite{Asenbaum2013}  & $3.0\times10^5$ & $8.3\times10^4$ & $1.5\times10^5$ & $3.0\times10^{-4}$ & n/a$^{+}$ &  n/a$^{++}$ & n/a$^{+}$ \\
																						& & & & $1.0\times10^{-8}$ & & & \\ 
			\hline
			Ref.~\protect\cite{Delic2018}  & $7.3\times10^{4}$ & $1.9\times10^5$ & $1.9\times10^{5}$ & $1\times10^{2}$ & n/a$^{+}$ &  1 & $10^{-2}$ \\
			Figure~\protect\ref{Cavitysetups}b)  & & & &$6\times10^{-2}$ & & &  \\ 
			\hline
			Ref.~\protect\cite{Windey2018}  & $2.2\times10^{4}$ & $1.1\times10^6$ & $1.4\times10^{5}$ & $4\times10^{-2}$ & n/a$^{+}$ &  $3\times10^{-3}$ & $10^{-2}$\\
			Figure~\protect\ref{Cavitysetups}b)  & & & &$2\times10^{-5}$ & & & \\ 
			\hline
			Ref.~\protect\cite{Fonseca2016}  & $5.0\times10^4$ & $1.0\times10^5$ & $4.0\times10^4$ & $1.1\times10^{-3}$ & 0.26 &  $3\times10^{-2}$ & $10^{-3}$\\
			Figure~\protect\ref{Cavitysetups}c)  & & & &$5.0\times10^{-6}$ & & &  \\ 
			\hline
			Ref.~\protect\cite{Magrini2018}  & $-^{\$}$ & $5.0\times10^9$ & $4.5\times10^{5}$ & $>10^3$ & 9300 &  n/a$^{\$\$}$ & $10^{-9}$\\
			Figure~\protect\ref{Cavitysetups}d)  & & & & $1.5$ & & & \\ 
			\hline
					
		\end{tabular}
		\caption{\label{table:cav} \emph{Levitated cavity optomechanics state-of-the-art:} This table lists the key cavity-cooling parameters for existing experiments in the field: $\Pg^{\rm min}$ is the minimal operating pressure, and $\T{fin}$ is the ultimate particle temperature after cooling.\newline
		$^{+}$ This parameter is not well defined in this experiment.\newline
		$^{++}$ The kinetic energy of the transiting particle was reduced by a factor of 30.\newline
		$^{\$}$ The authors do not provide this information.\newline
		$^{\$\$}$ In this work, the cavity was used for readout, but not cooling.}
\end{table}

\subsection{Further physical studies possible with levitated cavity optomechanics}

Although levitated cavity optomechanics is attracting great interest in terms of its future potential as an experimental probe of fundamental physics, such as the detection of gravitational waves \cite{Arvanitaki2013}, it already offers a rich playground of nonlinear, stochastic and coupled light-matter physics.
 
In particular, it offers an easily tunable Hamiltonian as dynamical parameters such as the mechanical frequency, optomechanical coupling strength and damping may be varied and even modulated in time. Sideband structures from temporal modulation were experimentally investigated in Fonseca \emph{et al.} \cite{Fonseca2016} and theoretically in Aranas \emph{et al.} \cite{Aranas2016}. In addition, spectral signatures of nonlinearities are easily detectable.  Tunable position-squared coupling has been detected \cite{Fonseca2016}, where the effects of an optomechanical coupling of the form $\go{} (\hat{a}+\hat{a}^\dagger) q^2$ was manifested in optical sidebands at twice the mechanical frequency. Such nonlinearities are distinct from behaviours resulting from anharmonicities in the classical optical potential \cite{Gieseler2013}. 
 

The cavity system in addition allows for studies of dissipative optomechanics, which differs in significant ways from dispersive optomechanical systems as it allows, in principle, ground state cooling away from the good cavity limit \cite{Bowenbook}.
To date there have been few experimental studies in dissipative optomechanical regimes \cite{Sawadsky2015}.  Whether or not levitated particles offer an experimental arena remains to be seen; however larger particles lead to high position dependent scattering rates, which may make a significant contribution to optical losses from the cavity. More studies may follow once experimental challenges in trapping and control are overcome, allowing for studies involving collective particle dynamics \cite{Habraken2013} (with multiple particles, either interacting directly or coupled indirectly via the cavity mode) as well as set-ups with multiple cavities. Dissipative coupling many also be an important component in hybrid set-ups coupling different physical modes, including clouds of atoms and quantum spins.

\section{Tests of Quantum Physics}
\label{sec:quantum}

\setcounter{footnote}{0}

The development of quantum theory in the 20\textsuperscript{th} century drastically changed our view of the world, and unsettled our notion of realism. This theory was essential in understanding new technologies of the time, such as transistors and lasers. For many years, the philosophical implication of quantum theory was \emph{the} hot topic in the field \cite{WhitakerBook}. Towards the end of the 20\textsuperscript{th} century, and into the 21\textsuperscript{st}, these concerns have been somewhat sidelined by what is often referred to as the ``second quantum revolution'', where aspects of quantum physics are exploited to create fundamentally new technologies. We now see an explosion of research into quantum enhanced sensing, quantum communication and quantum computing.

The conceptual difficulties of quantum physics have not gone away. Leggett summarizes the problem succinctly \cite{Leggett1985}:  

\begin{quotation}
...most physicists have a very non-quantum mechanical notion of reality at the macroscopic level [and make the following assumption.]  Macroscopic realism: A macroscopic system with two or more macroscopically distinct states available to it will at all times be in one or the other of these states... A direct extrapolation of quantum mechanics to the macroscopic level denies this.
\end{quotation}

In other words, although quantum theory tells us that the position of any object is undefined until measurement, somehow it's hard to believe that this is true for macroscopic objects. This (perhaps unsettling) notion is not conceptually solved either by decoherence theory, which expertly explains why it is difficult to maintain coherence in a macroscopic system but does not solve the measurement problem, nor by noting that for macroscopic objects the position uncertainty is in general immeasurably small.

The originators of quantum theory attempted to solve this conceptual issue by drawing a dividing line between the quantum and classical worlds, but locating this ``shifty split'' \cite{Bell1990} is a notoriously difficult problem. People searching for this split usually consider systems of increasing numbers of particles, or total mass\footnote{It is also suggested that complexity \cite{Home1996} or even consciousness \cite{Zeh1970} could define the split between quantum and classical systems.}. In 2010, a mechanical resonator tens-of-microns in size was cooled to its quantum ground-state of motion \cite{O'Connell2010}, and since then similar sized objects have been used for coherent state transfer \cite{Palomaki2013a}, optical-mechanical entanglement \cite{Palomaki2013b}, mechanical-mechanical entanglement \cite{Riedinger2017} and phonon interferometry \cite{Hong2017}. 

Does this mean that quantum theory is the correct description up to at least the micron-scale? The general consensus is no, since in the above mentioned systems, the ``size'' of the produced quantum states is very small\footnote{In the above examples, the ground-state uncertainty is of the order femtometres.}, and we do not learn whether quantum physics holds on macroscopic scales \cite{Nimmrichter2013}. To test quantum theory on larger scales, we must produce a quantum state which is spatially separated by more than the ground-state uncertainty. Marshall \emph{et al.} proposed such an experiment, involving a movable mirror inside a Michelson interferometer \cite{Marshall2003}, though the realization of this protocol remains elusive, since it would require extremely large single-photon coupling rates, and simultaneously low mechanical frequencies. 

In this context, levitated particles are seen as a paradigmatic system for realizing a macroscopically separated quantum state, due to their potential to undergo free coherent evolution. This is possible since, unlike other optomechanical systems, levitated particles are not tethered or clamped, so once the levitating field is switched off the particles fall freely under gravity. 

This section will cover various aspects of quantum physics that can be uniquely tested using levitated particles.

\subsection{Interferometry}
\label{sec:MWI}

A dramatic and clear demonstration of a quantum phenomenon is the creation of a position superposition state. As discussed above, this provides a test of macroscopic quantum physics when the position states are separated by more than the ground-state uncertainty, and the test is even more convincing when the separation is as large as the object in question. However, proving the existence of a superposition state is not trivial, since upon measurement the result is often identical to a classical statistical mixture. 

Matterwave interferometry with single particles is a clear route to proving that a position-superposition state was created \cite{Arndt2014}. A coherent source of particles is incident upon a diffractive element, which may be a physical grating or a standing light wave. The separated superposition state is produced at the diffractive element, and consequent evolution over space and time enables a measurable interference pattern to build up. This technique has been used to provide the most stringent test on the macroscopicity of quantum theory so far observed, via the diffraction of $10^4$\,amu molecules \cite{Eibenberger2013}. 

The first step in matterwave interferometry is to prepare a source which can coherently illuminate multiple slits of the diffractive element. The coherence width $\sigma_{\rm c}$ at a distance $L$ after a source of width $a$ is $\sigma_{\rm c} = 2L\ldB/a$ \cite{Hornberger2012}, where $\ldB = h/p$ is the de Broglie wavelength of a particle with momentum $p$. Hence, $\sigma_{\rm c}$ must be larger than the grating period, and it is immediately clear that for massive particles this means one either needs long interferometers, or cold particles (which both reduces $\ldB$ and the effective source width $a_{\rm eff}$, see below). It is proposed to use either cavity \cite{RomeroIsart2011a, Kuhn2017c} (Sec.~\ref{sec:cav_cool}) or feedback \cite{Bateman2014} (Sec.~\ref{sec:det_fb}) cooling to realize a cold and narrow nanoparticle source, yielding an \emph{effective} slit width $a_{\rm eff}$. This can be estimated, for a particle cooled to a 1-D temperature along the $x$ direction $T_x$ in a harmonic potential, as $a_{\rm eff} \approx \sqrt{\kB T_x/\mass\omega_x^2}$, for a harmonic frequency $\omega_x$. Note, that for objects as massive as nanoparticles it is required that $a_{\rm eff} \ll \Rad$, which is not realizable with a physical slit.

Romero-Isart \emph{et al.} \cite{RomeroIsart2011a} then propose to implement an ``optomechanical double slit''. A position superposition state is formed by dropping a cavity-cooled nanoparticle through the antinode of a second optical cavity, whereby a pulsed measurement of $x^2$ is made (enhanced by the initial coherent expansion) to generate the superposition. At some time later, the position is measured, and repeated realizations will generate an interference pattern. The authors propose that a particle of diameter 40\,nm would spend $\sim128\,$ms in the interferometer, require an initial particle phonon occupancy of $\bar{n} = 0.1$, an operating pressure of $<10^{-16}\,$mbar and an environmental temperature of 4.5\,K. These are challenging operating conditions, required to minimize decoherence, which is discussed below.

\begin{figure}[t]
	 {\includegraphics{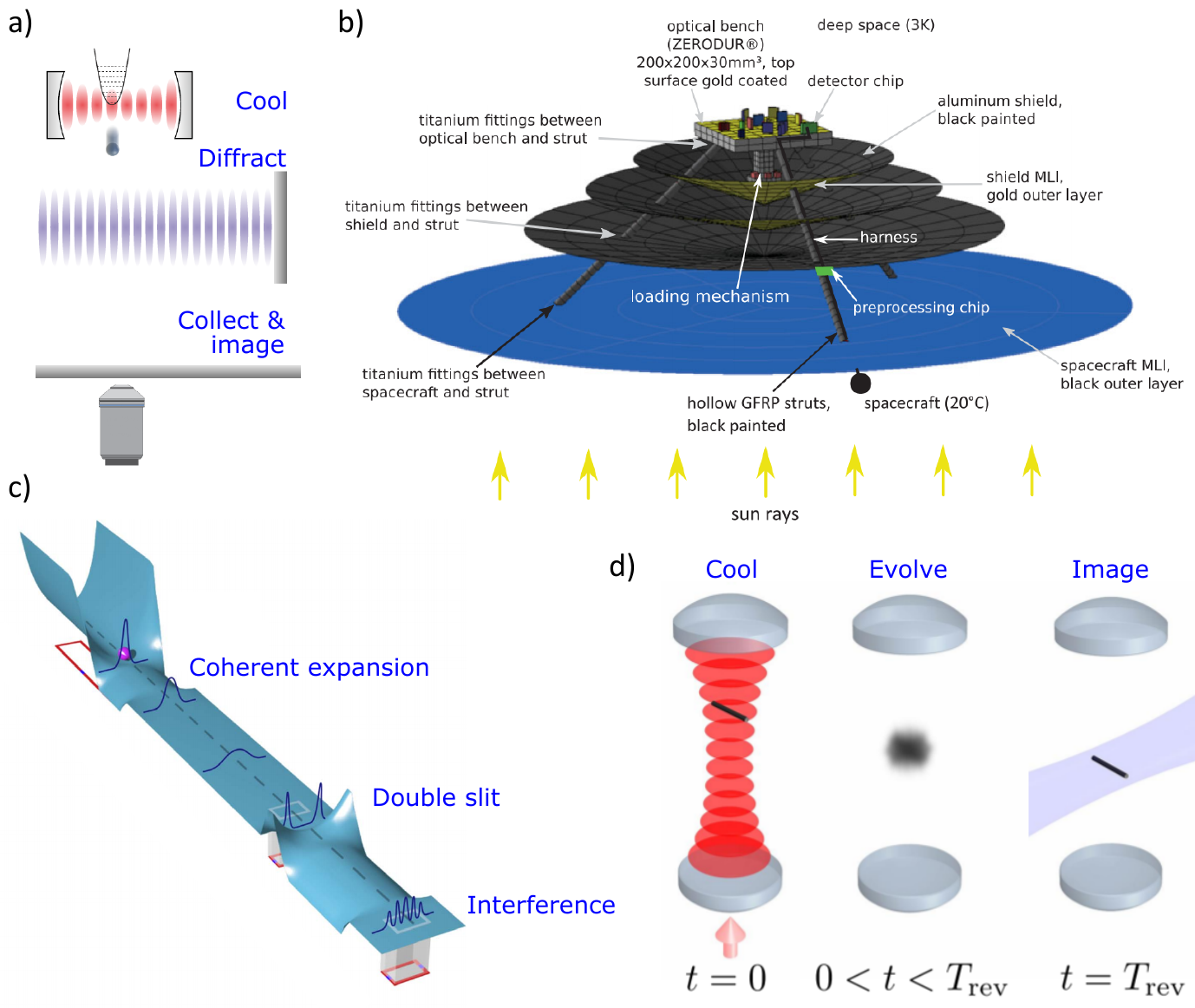}}	\centering
\caption{\label{fig:interferometry} 
\textbf{High mass interferometry} a) By cooling the centre-of-mass of a nanoparticle, for example via cavity cooling, an optical diffraction grating can be coherently illuminated, forming a near-field matterwave interference pattern with $10^{6-7}\,$amu particles, figure from \cite{KuhnThesis}. b) To overcome the path-length and decoherence limitations of matterwave interferometers, it is proposed to perform a satellite based experiment to push the mass limit to $10^{11}\,$amu, figure from \cite{Kaltenbaek2016}. c) Cavity magnetomechanics could be used to perform a magnetomechanical double-slit experiment on a chip, in a cryogenic environment, pushing the mass to the $10^{13}\,$amu level, image taken from \cite{Pino2018}. d) A different approach is to look for quantized angular momentum via observation of orientational quantum revivals of a $10^5\,$amu nanorod, figure adapted from \cite{Stickler2018}.
}
\end{figure}

Another approach is to follow the coherent source by an optical diffraction grating, as illustrated in Fig.~\ref{fig:interferometry}a). An optical standing wave can either act as a phase grating, imprinting a position-dependent phase, or an absorptive grating which removes particles that pass through the optical anti-nodes (for example via ionization using UV light) \cite{KuhnThesis}. The length of the interferometer can then be reduced by observing the near-field interference pattern \cite{Bateman2014}. Bateman \emph{et al.} suggest a particle $\sim 10\,$nm diameter particle falling for a total of $\sim 240\,$ms, with an initial particle temperature of 20\,mK achieved through feedback cooling, an operating pressure of $10^{-10}\,$mbar, and in a room temperature environment. These conditions are less challenging than the previous proposal, though feedback or cavity cooling such a small particle would be difficult. 

The maximum particle size in these experiments is limited by two factors: more massive particles having shorter de Broglie wavelengths, and hence requiring longer interferometers\footnote{Recent work proposes that an inverted potential can be used to coherently accelerate wavefunction expansion \cite{RomeroIsart2017}.}; and more massive particles having in general higher decoherence rates (see below), which is compounded by the requirement for longer evolution times in the interferometer. However, it is not trivial to work with dielectric particles below a few 10s nm in diameter. For feedback cooling, the trap depth and the feedback signal drop with the particle volume, and for cavity cooling the optomechanical coupling drops with particle volume. In the latter case, a solution is to work with cavities with small mode volumes \cite{Kuhn2017c}, and a thorough discussion of the limits of matterwave interferometry with dielectric particles can be found in Ref.~\cite{KuhnThesis}. 

One method to minimize decoherence and the requirement for large interferometers is to perform the experiment in free-fall in zero-gravity, i.e. in space. The MAQRO mission proposes to do just this \cite{Hechenblaikner2014, Zanoni2016, Kaltenbaek2016}, as illustrated in Fig.~\ref{fig:interferometry}b), by putting a optical nanoparticle interferometer on a satellite which is thermally shielded from the Sun. Of course a satellite based experiment comes with a whole new set of experimental and financial challenges.

\subsubsection{Other interferometric schemes:}

The matterwave interferometry schemes described above require strict initial cooling and many iterations of the experiment with an ensemble of particles to build up an interference pattern. In this section, we discuss some interferometric schemes that relax these requirements.

Scala \emph{et al.} propose a form of Ramsay interferometry to probe high-mass superpositions \cite{Scala2013}. A levitated nanodiamond with a single \NV\ centre (Sec.~\ref{sec:NV}) is exposed to a magnetic field, which exerts a spin-state dependent force on the centre-of-mass (c.o.m.). The particle is optically probed to produce an \NV\ spin-state superposition, which in turn produces a c.o.m. superposition, since the magnetic field displaces the equilibrium position of the nanodiamond in a spin-dependent manner. By operating in a coordinate system which is tilted with respect to gravity, one of the spin states picks up a gravitationally induced phase relative to the other, which can be measured using Ramsay interferometry to infer the macroscopic c.o.m. superposition. 

This experiment can be repeated with the same particle to build up statistics. The authors propose that this experiment would be possible with 200\,nm diameter diamond spheres, without c.o.m. cooling, and in an ambient environment. However, in this experiment the superpositions that are produced are not particularly macroscopic $\sim 1\,$pm. A refined Ramsey scheme suggests a free-fall experiment \cite{Wan2016a}, where a spin-dependent force is used to split and merge the matter wave-packet, which for the same experimental requirements as just mentioned would produce superposition states of $\sim 100\,$nm. Ramsey interferometry with levitated nanodiamonds seems to be, in theory, remarkably robust against initial conditions \cite{Wan2016b}, yet requires stringent control over magnetic fields and their alignment relative to gravity.

Pino \emph{et al.} propose to push the mass to the $10^{13}\,$amu scale using a $2\,\mu$m diameter magnetically levitated superconducting Niobium sphere \cite{Pino2018}. The authors suggest a superconducting ``skate-park'' circuit, as illustrated in Fig.~\ref{fig:interferometry}c), which recycles the particle to allow multiple realizations, in a cryogenic ($50\,$mK) ultra-high vacuum ($10^{-17}$\,mbar) environment, where coherent inflation \cite{RomeroIsart2017} is used to accelerate wavefunction expansion. The protocol is similar to the optomechanical double slit mentioned above \cite{RomeroIsart2011a}, using cavity magnetomechanics \cite{RomeroIsart2012} instead of cavity optomechanics. This scheme requires cutting edge technology, and the development of new techniques, but offers the potential to test quantum mass limits several orders of magnitude above any other suggestion.

Finally, Stickler \emph{et al.} have proposed a different method for testing quantum physics with high mass objects, by looking for quantization of the angular momentum of a nanorod \cite{Stickler2018}, as illustrated in Fig.~\ref{fig:interferometry}d). In this scheme, an optically levitated nanorod is prepared in an alignment state of modest temperature ($\sim 1\,$K) via feedback or rotational cavity cooling \cite{Stickler2016a}, and then dropped so that its alignment state freely evolves into a superposition of all orientations. After a characteristic revival time $T_{\rm rev}$, the particle will return to its initial alignment, which is evidence of interference between orientation states. Though the proposed mass in these experiments is low (for example, $10^5\,$amu carbon nanotubes), the experimental requirements are modest: moderate cooling, $10^{-8}\,$mbar pressures, millisecond experimental periods with particle recycling, and a room temperature environment.

\subsection{Decoherence}
\label{sec:decoherence}

Decoherence describes a loss of coherence due to an interaction between a quantum system and the environment. We stress that this is different from wavefunction collapse, which is discussed below. Larger objects are typically more susceptible to environmental decoherence, which explains why observing quantum effects with massive objects is experimentally challenging. One motivation for working with levitated particles is the potential to minimize thermal decoherence as compared to many optomechanical systems. 

We will consider 1D position localization decoherence, which is easiest described via a master-equation, following Ref.~\cite{Romero-Isart2011_3}, which considers individual scattering events for an object delocalized over a distance $|x - x'|$ with a characteristic decoherence rate $\Gdec$ such that $\langle x | \rho(t) | x' \rangle \propto e^{-\Gdec t}\langle x | \rho(0) | x' \rangle$. The decoherence rate is defined as

\begin{equation}
\Gdec(|x - x'|) = \gdec\left(1-e^{-\frac{|x - x'|^2}{4\adec^2}}\right),
\end{equation}
\noindent

where $\gdec, \adec$ are a localization strength and distance respectively\footnote{Note, parameters in this section are labelled with a tilde, to distinguish them from other symbols in this manuscript, whilst being consistent with the literature.}, which depend upon the exact form of decoherence. The localization distance, for our purposes, is related to the thermal wavelength of the scattering particles $2\adec = \lambda_{\rm th}$ (e.g. the mean de Broglie wavelength of air molecules in gas-collision decoherence). The size of this parameter relative to the extent of the position-coherence (i.e. the range over which $\langle x | \rho(t) | x' \rangle$ is finite) is important for determining the scaling behaviour of the decoherence process

\begin{equation}
\Gdec(|x - x'|) = 
\begin{cases}
      \Ldec |x - x'|^2, & |x - x'| \ll 2\adec, \\
      \gdec, & |x - x'| \gg 2\adec,
    \end{cases}
  \end{equation} 
	\noindent
which introduces the final important parameter, the localization parameter $\Ldec \equiv \gdec/(4\adec^2)$. When $|x - x'| \ll 2\adec$ the position correlations decay as $\langle x | \rho(t) | x' \rangle \propto e^{-\Ldec |x - x'|^2 t}\langle x | \rho(0) | x' \rangle$, illustrating that a large localization parameter leads to faster decoherence, and that larger delocalized states also decohere faster.  When $|x - x'| \gg 2\adec$ the position correlations decay as $\langle x | \rho(t) | x' \rangle \propto e^{-\gdec t}\langle x | \rho(0) | x' \rangle$, which tells us that in this short thermal wavelength regime the decoherence rate doesn't depend upon $|x - x'|$, and a single scattering event will cause full decoherence. 

Now we will state the relevant parameters for various decoherence mechanisms. For collisions between a sphere and air molecules we have \cite{Romero-Isart2011_3}

\begin{align}
\lambda_{\rm th}^{\rm air} &= \frac{2\pi\hbar}{\sqrt{2\pi\mg\kB\T{env}}}, \\
\Ldec_{\rm air} &= \frac{8\sqrt{2\pi}\mg v_{\rm th}\Pg \Rad^2}{3\sqrt{3}\hbar^2}, \nonumber
\end{align}
\noindent
where $v_{\rm th}$ is the thermal velocity of the gas molecules of mass $\mg$. By inspection one can see that decoherence rates drop with pressure and gas temperature. For the interaction with blackbody radiation, we consider three parts: absorption, emission and scattering (abs, emis, scat) by the particle

\begin{align}
\lambda_{\rm th}^{\rm bb} &= \frac{\pi^{2/3}\hbar c}{\kB\T{env}}, \\
\Ldec_{\rm abs(emis)}^{\rm bb} &= \frac{4\pi^4c}{189\eo}\left[\frac{\kB\T{env(int)}}{\hbar c}\right]^6\pol{bb}'', \nonumber \\
\Ldec_{\rm scat}^{\rm bb} &= \frac{8!\xi(9)c}{18\pi^3\eo^2}\left[\frac{\kB\T{env}}{\hbar c}\right]^9(\pol{bb}')^2, \nonumber
\end{align}
\noindent
where all terms are defined in Sec.~\ref{sec:int_temp}. Here it is clear that decoherence rates drop rapidly with both the environmental and particle-internal temperature. 

The final contribution we will consider is scattering of monochromatic light, which is trapping or otherwise incident upon a spherical, sub-wavelength particle. In this case the thermal wavelength is just the wavelength of the relevant light source.  Below are the expressions for the localization parameter for a particle suspended both via a tweezer (tw) and an optical cavity (cav) \cite{RomeroIsart2011b}

\begin{align}
\Ldec_{\rm scat}^{\rm tw} &= \frac{\pol{}^2\w{tw}^5\V{S}\Po}{6\pi^2\eo^2c^6\rho\waist^2\w{q}}\frac{1}{\lambda_{\rm tw}^2},\\
\Ldec_{\rm scat}^{\rm cav} &= \frac{\pol{}^2\w{cav}^6\V{S}\nopt}{12\pi\eo^2c^5\rho \V{m} \w{q}}\frac{1}{\lambda_{\rm cav}^2}, 
\end{align}
\noindent
where $\pol{}, \V{S}, \rho$ are the polarizability, volume and density of the particle respectively; $\Po, \waist$ the optical power and beam waist of the tweezer respectively; $\nopt, \V{m}$ are the photon occupancy and mode volume of the cavity respectively; and $\w{q}$ is the mechanical trapping frequency. Here it is clear that decoherence rates drop rapidly with the size of the particle.

We refer the reader to Ref.~\cite{RomeroIsart2011b} for a thorough discussion of other potential decoherence mechanisms. As an example of the role of these terms in a more experimental language, consider the coherent expansion of a wavepacket initially cooled to a thermal state of mean occupancy ${\bar n}$ in a potential with harmonic frequency $\w{q}$, for example as the initial stage in a matterwave interferometer. The coherence length will reach a maximum value ${\tilde \xi}_{\rm max}$ in the short wavelength limit given by \cite{Romero-Isart2011_3}

\begin{equation}
{\tilde \xi}_{\rm max} = \sqrt{2}\left[\frac{2\hbar\w{q}}{3\mass\Ldec^2(2{\bar n} + 1)}\right]^{1/6}.
\end{equation} 
\noindent

In an interferometer, this value would have to be larger than the slit separation to see interference. Note that the maximum coherence length depends critically upon the localization parameter. As a sense of scale \cite{Romero-Isart2011_3} a $\Rad = 50\,$nm silica sphere cooled to ${\bar n} = 0.1$ phonons with $\w{q} = 2\pi\times100$\,kHz, in a cryogenic environment (5\,K) and with a bulk temperature of 200\,K would reach a maximum coherence length of 20\,nm, making traditional matter-wave interferometry extremely challenging for particles this large.

\subsection{Wavefunction collapse models}

There is a theoretical programme to introduce a physical mechanism to objectively explain wavefunction collapse, which operates in a mass- and size-dependent manner, such that large objects cannot be in a macroscopically-distinct quantum superposition state for an extended period of time. There are various different formulations of these phenomenological ``wavefuntion collapse models'', though they are tightly constrained so that they do not counter existing observations or allow super-luminal signalling. The models can be roughly split into two: those which invoke a classical noise field as a nonlinear extension to the Schr{\"o}dinger equation to cause wavefunction collapse, so-called dynamical reduction models; and those that invoke gravitational interactions\footnote{There are other proposed wavefunction collapse mechanisms \cite{BassiReview, Romero-Isart2011_3}, which are rarely discussed in the context of levitated particles and will not be covered here.}.

\subsubsection{Dynamical reduction models:}

The most developed dynamical reduction model is known as ``Continuous Spontaneous Localization'' \cite{Ghirardi1990}, or CSL, and the reader is directed to Bassi \emph{et al.} \cite{BassiReview} for a thorough review of such processes, and to Toro{\v s} \emph{et al.} \cite{Toros2017} for more recent extensions to the model. Such a model must be \emph{stochastic} to maintain the probabilistic nature of quantum mechanics, and \emph{nonlinear} to explain irreversible wavefunction collapse. For example, a potential modification of the Schr{\"o}dinger equation for a single particle in one spatial dimension $q$ is \cite{BassiReview}

\begin{equation} \label{eqn:collapse}
\dif \psi(t) = \left [ -\frac{i}{\hbar}H\dif t + \sqrt{\lambda}(q - \langle q \rangle)\dif W(t) - \frac{\lambda_c}{2}(q - \langle q \rangle)^2\dif t \right] \psi(t),
\end{equation} 
\noindent
where $H$ is the standard Hamiltonian of the system, $\langle q \rangle \equiv \langle \psi(t) | q | \psi(t) \rangle$ is the position expectation, and a constant $\lambda_c$ determines the strength of the collapse process. Note that the second term on the right hand side is stochastic, where $W(t)$ is a Wiener process, and the third term is nonlinear, as required. 

The CSL model is more sophisticated that the simple case presented in eqn.~\eqref{eqn:collapse}, in that it considers mass distributions rather than single particles, and the stochastic process is random in space and time. In this case we have two phenomenological collapse constants: a length $\rc$, which for a single particle approximates the separation above which superpositions are suppressed, or a dividing line between the micro- and macro-scopic realms; and a localization rate $\lcsl$ which approximates the collapse rate for a single nucleon. Current models and experiments suggest \cite{Diosi2015}: $\rc \approx 100\,$nm and $2.2 \times 10^{-17} {\rm Hz} \leq \lcsl \leq 2.2 \times 10^{-8\pm2}\,$Hz. 

One method to test these models would be to perform matterwave interferometry, as discussed above, and look for a loss of interference contrast \cite{Romero-Isart2011_3, KuhnThesis}. Since the collapse processes are predicted to be both mass and superposition-size dependent, matterwave interferometry with massive objects is the natural test-bed, rather than, for example, simply interrogating the evolution of ground-state cooled oscillators \cite{Romero-Isart2011_3}. For a rigid massive object, the wavefunction collapse rate assuming the CSL model would be \cite{Toros2017}

\begin{equation} \label{eqn:CSLrate}
\Lcsl = \frac{n_a}{n(\rc)}\left (\frac{m_an(\rc)}{m_0}\right )^2\lcsl,
\end{equation} 
\noindent
where $n_a$ is the number of atoms in the object, $m_a$ the atomic mass, $m_0$ the proton mass and $n(\rc)$ the number of atoms contained within a spherical volume of radius $\rc$. For a rigid object of mass $\mass$ in a superposition state that is larger than $\rc$ this simplifies to $\Lcsl = (\mass/m_0)^2\lcsl$, illustrating the suppression of macroscopic superpositions with mass.

The challenge in this approach would be distinguishing collapse induced decoherence from environmental sources, since standard decoherence theory also predicts the practical difficulty of delocalizing large objects \cite{Romero-Isart2011_3}, see Sec.~\ref{sec:decoherence}. Proposed experiments to test collapse models are based on a thorough minimization and understanding of environmental decoherence, combined with protocols varying experimental parameters such as object mass to look for scaling behaviour unique to CSL \cite{Romero-Isart2011_3}.

The stochastic noise term in eqn.~\eqref{eqn:collapse} implies that there is a classical fluctuating field that imparts energy and momentum onto particles, even in the classical regime. Collett \& Pearle \cite{Collett2003} noted that this means particles will undergo a form of random walk due to the existence of a collapse process\footnote{This may or may not lead to a violation of energy conservation \cite{Toros2017}.}), which could be detected. In comparison to eqn.~\eqref{eqn:EoM}, CSL would have a force density \cite{Nimmrichter2014}

\begin{equation} \label{eqn:CSLforce}
\Sf{ff}{CSL} = \lcsl\left (\frac{\hbar}{\rc}\right )^2\alpha_{\rm CSL},
\end{equation} 
\noindent
where $\alpha_{\rm CSL}$ is an object geometry dependent term, common examples can be found in Nimmrichter \emph{et al.} \cite{Nimmrichter2014}. As a reminder, the force spectral density is related to the fluctuating force $\mathcal{F}$ through $\langle \mathcal{F}(t)\mathcal{F}(t')\rangle = 2\pi S_{\rm ff}\delta(t - t')$. Note, that this fluctuating force does not have a dissipative component. Bera \emph{et al.} \cite{Bera2015} suggest that pressures below $10^{-12}\,$mbar, motional temperatures below $1\,$mK and cryogenic environmental temperatures are required to observe such force noise.

For a cooled harmonically bound particle of mass $\mass$, CSL leads to an increase in the mean phonon number of \cite{Goldwater2016} 

\begin{equation}
\langle \dot{n}(t) \rangle = \frac{\hbar \lcsl\alpha_{\rm CSL}}{\mass\w{q}\rc^2}.
\end{equation} 
\noindent
Goldwater \emph{et al.} \cite{Goldwater2016} propose to look for this heating with a charged particle levitated in a Paul trap, which requires initial cooling to a few tens of motional quanta via optical cavity cooling. They suggest pressures of $10^{-12}\,$mbar, environmental temperatures of 4\,K and particle bulk temperatures of 65\,K are sufficient to see significant heating due to CSL. We briefly note that it is also possible to look for the effects of collapse-induced noise on the rotational degrees-of-freedom of disks \cite{Collett2003, Bera2015} and rods \cite{Stickler2017a}.

A final method for experimentally discerning dynamical reduction processes is to look for a change in the frequency response of a massive oscillator \cite{Bahrami2014b}, explicitly an additional broadening of the position power spectral density above that due to thermal effects. However, it is predicted that this would lead to only micro-Hertz increases in spectral linewidth and require months of continuous measurement to resolve \cite{Goldwater2016}.

\subsubsection{Gravitational collapse models:}

It is one of the biggest open problems in physics; how to marry the theories of general relativity and quantum physics (and one not to be considered here!). The Schr{\" o}dinger-Newton equation involves the addition of a classical gravitational self-interaction to the Schr{\" o}dinger equation, and describes how a quantum superposition responds to classical gravity

\begin{align}
i\hbar\dif \psi(t,{\bf r}) = \left [ H\dif t + \Upot{g}(\psi) \right ] \psi(t,{\bf r}), \\
\Upot{g}(t,{\bf r}) = -G \int\dif^3r'|\psi(t,{\bf r'})|^2I_{\rho}({\bf r} - {\bf r'}),  \nonumber 
\end{align}
\noindent
where $H$ is the standard Schr{\" o}dinger equation Hamiltonian, $G$ is the gravitational constant, $I_{\rho}$ is a term which depends on the mass distribution and can be found for example in Ref.~\cite{Grossardt2016}, and the dependence of the gravitational interaction term $\Upot{g}(\psi)$ on the wavefunction has been made explicit in the first line.

Gro\ss{}ardt \emph{et al.} \cite{Grossardt2016} show that this nonlinear term would lead to an energy-level dependent shift in the otherwise harmonic frequency spectrum of a levitated object; a shift which increases with mass. They propose levitating $5\,\mu$m diameter osmium disks in a cryo-Paul trap, and predict a frequency shift of order 0.1\,mHz. We note this proposal requires the production of mechanical Fock states. A somewhat similar proposal suggested that the dynamics of massive mechanical squeezed states may also validate the Schr{\" o}dinger-Newton equation \cite{Yang2013}.

While such tests would suggest that we understand the interaction between a quantum state and classical gravity, we don't have an understanding of the gravitational field \emph{produced by} a superposition (i.e. does it lead to a superposition of space-time?), let alone whether the gravitational field should be quantized\footnote{Recent proposals have suggested levitated optomechanical experiments to explore this problem \cite{Bose2017, Marletto2017}.}. This lack of understanding has motivated some to suggest that gravity is the ultimate cause of wavefunction collapse. 

This could occur via various mechanisms, and we refer the reader to other sources for a thorough discussion on this topic \cite{Romero-Isart2011_3, BassiReview, Bassi2017}. We will briefly discuss one commonly cited mechanism here. Di{\'o}si proposed \cite{Diosi1987} that the noise-term found in dynamical reduction models (eqn.~\eqref{eqn:collapse}) is due to fluctuations in the classical gravitational field. Independently, Penrose considered the inherently nonlinear gravitational self-interaction between disparate parts of the same macroscopic superposition \cite{Penrose1996}. This approach leads to an energy-uncertainty that makes the superposition unstable, in quantitatively the same way as Di{\'o}si's proposal; hence this framework is called the Di{\'o}si-Penrose (DP) model.

The DP model predicts a collapse rate for macroscopic superpositions of a spherical object \cite{Romero-Isart2011_3}

\begin{equation}
\LDP = \frac{G\mass^2}{2\Rad\hbar},
\end{equation} 
\noindent 
where $G$ is the gravitational constant, and $\Rad$ is both the radius of the object and the critical superposition size in this model (somewhat equivalent to $\rc$ in the CSL model). Note that this model doesn't contain free phenomenological parameters, which is attractive. This collapse rate is much weaker than the CSL prediction, with coherence times only dropping below a millisecond once micrometer sized objects are prepared in a macroscopic spatial superposition. For this reason, tests of the DP model are considered extremely experimentally challenging, in any physical system \cite{Wezel2008}. 

As with CSL, a collapse process of gravitational origin would cause c.o.m. diffusion of a levitated object \cite{Collett2003, Bera2015}, but this effect would be extremely weak as compared to CSL, requiring background pressures of $10^{-18}\,$mbar to observe \cite{Bera2015}. 

\subsection{Preparing mechanical quantum states}

We have discussed in detail the potential to cool the c.o.m. motion of levitated objects to the ground-state. In this section we will briefly consider the potential to generate novel mechanical quantum states with levitated particles.

When cavity cooling of levitated nanospheres was first considered, several authors noted that there is potential to create novel quantum states once the particle is prepared in its ground state and strong coupling has been achieved, due to the state transfer possible between the optical field and mechanical motion (photon-to-phonon transfer). Romero-Isart \emph{et al.} \cite{RomeroIsart2010} showed that it would in principle be possible to produce a mechanical superposition state if the cavity was pumped with a single photon, and a subsequent homodyne measurement was performed on the cavity output. This would be a mechanical superposition of the first two harmonic energy levels in the confining potential.  

Chang \emph{et al.} \cite{Chang2010} showed entanglement between two light fields could be transferred to the motion of two separate nanospheres in different optical cavities. In a different, non-cavity approach, Ralph \emph{et al.} \cite{Ralph2016} showed you could entangle the position and orientation of a levitated nanosphere through measurement of a point on the surface of the sphere, a route not available when working with atomic or molecular systems. 

Modulation of the trapping potential, or optomechanical spring-constant, is a route to generating squeezed motional states. In the quantum regime, squeezing enables one to push the fundamental quantum uncertainty below the standard quantum limit, with applications in sensing. A thermal state can be squeezed to reduce the uncertainty in one of the quadratures at the expense of anti-squeezing the other. Chang \emph{et al.} \cite{Chang2010} showed that a modulation of the trapping field would lead to time-dependent quantum squeezing, which can also be transferred to the optical field. The authors point out that a linear coupling to the light field should yield significant squeezing of up to 30\,dB for the mechanics and 15\,dB for the optical field. Modulation of the trapping potential was used to generate a classical squeezed thermal state with a levitated nanoparticle \cite{Rashid2016}, achieving 2.7\,dB of squeezing. In a different approach, Genoni \emph{et al.} \cite{Genoni2015} propose to generate a non-classical squeezed \emph{steady state} of $\sim 2\,$dB, through continuous measurement of both scattered and cavity fields, a method which could also enhance cooling rates. Rashid \emph{et al.} demonstrate how continuous measurement can be used to generate quantum states without the presence of an optical cavity \cite{Rashid2017}.

Various groups \cite{Ranjit2015b, Chen2017} have noted that it is possible to couple levitated nanoparticles to the spin degrees-of-freedom of atomic systems, which are cooled deep into the quantum regime, opening up the full toolbox of cavity QED in analogy to various other optomechanical proposals (see references within \cite{Ranjit2015b}). This could involve atoms co-trapped in the same or distant optical cavities.

\section{Spin Systems (Nitrogen-Vacancy Centers in Diamond)}
\label{sec:NV}

\setcounter{footnote}{0}

A hybrid quantum system can be formed when multiple distinct quantum systems are coupled together \cite{Wallquist2009, Treutlein2014}. For optomechanical experiments, this is a natural description since system dynamics involve degrees-of-freedom in both the optical field (photons) and mechanical motion of the oscillator (phonons). Such hybridization may prove useful in the fields of quantum information and measurement \cite{Fiore2011}, or for transduction between otherwise incompatible quantum systems \cite{Gavartin2012}. 

Another way to hybridize an optomechanical system is to use a mechanical oscillator that has both a motional degree-of-freedom and an intrinsically quantum degree-of-freedom. For example, the oscillator could contain a defect centre with atom-like properties, or be formed entirely from a quantum dot. Access to such quantum systems allows different kinds of physics, such as an enhanced ability to interact with electromagnetic fields or temperature \cite{Mamin2013, Rondin2014, Maze2008, Dolde2011, Kucsko2013, Toyli2012}. In this section, we will discuss the hybrid optomechanical system formed using a levitated nanodiamond as the mechanical oscillator, with a defect centre as the quantum spin degree-of-freedom. Explicitly, we consider the nitrogen-vacancy centre in diamond, since this has been the focus of the vast majority of work in the field. 

\subsubsection{The NV centre:}

\begin{figure}[ht]
\centering
\includegraphics[width=4.5in]{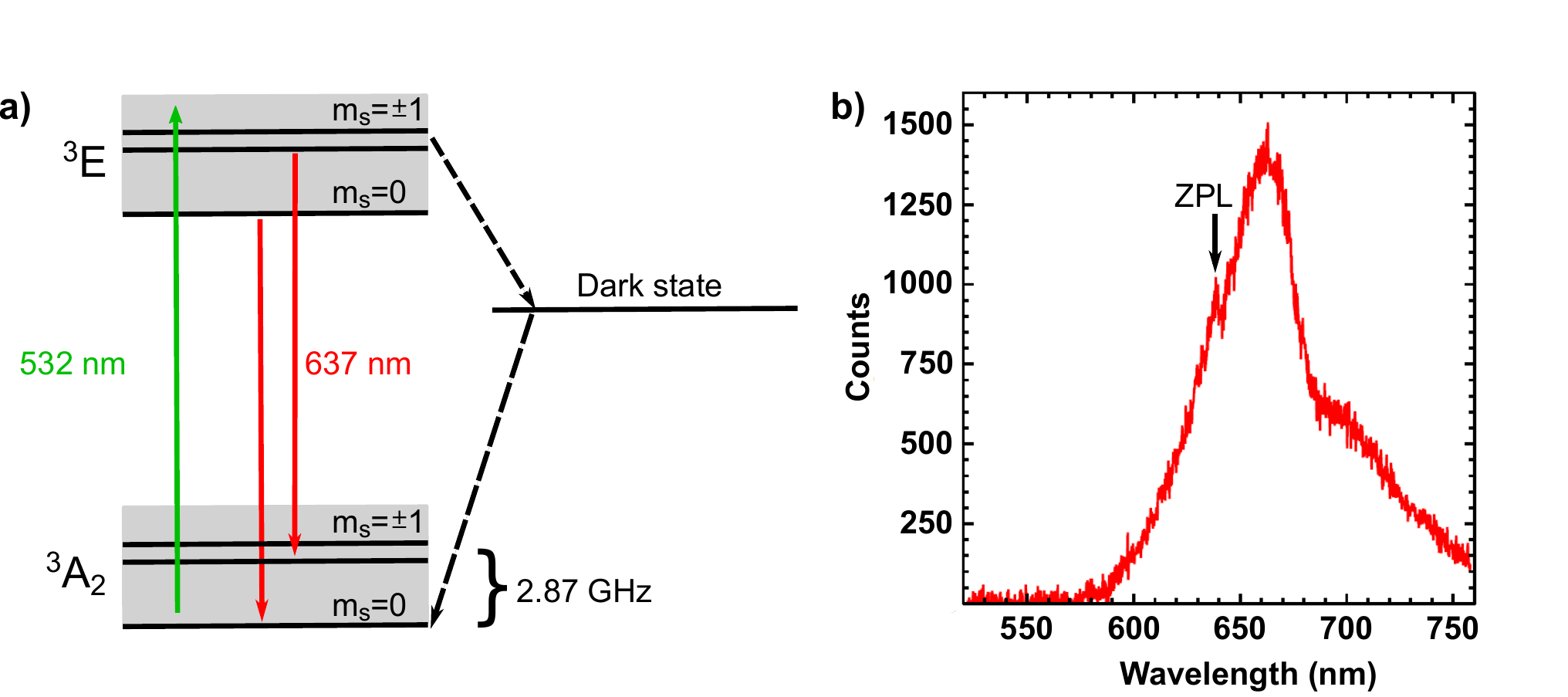}
\caption{\textbf{The Nitrogen vacancy centre.} a) Energy level diagram of the \NV\ centre showing the spin-selective inter-system crossing out of the excited state (\Es). Solid arrows indicate optical excitation at 532\,nm and radiative decay at the zero-phonon line (ZPL) at 637\,nm. Dashed arrows represent non-radiative transitions through the inter-system crossing. The spin resonance within the ground state (\gs), centred around 2.87\,GHz, is shown. c) Typical photoluminescence spectrum at room temperature. The ZPL at 637\,nm is highlighted. Modified from \cite{Pettit2017}.}
\label{fig:NVfig}
\end{figure}

The nitrogen-vacancy (NV) centre is formed by the coalescence of a lattice vacancy adjacent to a nitrogen impurity atom within the diamond lattice, forming a defect with trigonal $C_{3V}$ symmetry. There are two primary charge states of the defect centre, a neutral state (NV$^{0}$), and a negatively charged state (\NV). Both centres are optically active, but it is the \NV\ centre that is the basis for this section \cite{Doherty2013}, owing to its compatibility with optical manipulation at a wavelength of 532\,nm (see Fig.~\ref{fig:NVfig}a)), and its versatility as a quantum system in temperatures ranging from 600\,K down to cryogenic temperatures \cite{Toyli2012}.

Photoluminescence from the centre is broadband at room temperature, ranging from the zero-phonon line at 637\,nm, out to 800\,nm, through phonon broadened transitions within the ground state (\gs) and excited state (\Es) triplets, see Fig.~\ref{fig:NVfig}b). Polarization and readout of the spin state within the ground-state triplet is possible through optical techniques by exploiting the spin-selective inter-system crossing out of the \NV\ excited state triplet. Coupling to the inter-system crossing favours the excited state $m_{s}=\pm1$ spin projections over the $m_{s}=0$ counterpart. Decay through the inter-system crossing is non-radiative, and preferentially re-populates the $m_{s}=0$ ground state spin sub-level. Thus, a reduction in the emitted photoluminescence results if the \NV spin has been prepared in either of its $m_{s}=\pm1$ projections prior to excitation above the zero-phonon line at 637\,nm. Polarization of the spin is possible through repetition of this cycle.

The \NV\ centre in diamond has emerged as a versatile tool in quantum information studies \cite{Fuchs2011, Neumann2010, Togan2010, Wrachtrup2006} and nanoscale sensing \cite{Mamin2013, Rondin2014, Maze2008, Dolde2011, Kucsko2013, Toyli2012, Beams2013}. Efforts have been made to couple the spin of the \NV\ centre to mechanical resonators, creating hybrid systems with both mechanical and spin degrees-of-freedom \cite{Kolkowitz2012, Arcizet2011}. 

\subsubsection{Levitated nanodiamond:}
Recently, these efforts have melded with the field of levitated optomechanics \cite{Neukirch2014a}. In particular, there is now interest in coupling the \NV\ electron spin to the centre-of-mass (c.o.m.) motion of a levitating nanodiamond in a vacuum environment. This type of hybrid optomechanical system could provide a platform to perform fundamental tests on the limits of quantum superposition \cite{Scala2013, Yin2013, RomeroIsart2011a} (see Sec.~\ref{sec:MWI}), test theories of quantum gravity \cite{Albrecht2014, Marletto2017, Bose2017}, explore novel methods of mechanical squeezing \cite{Ge2016}, and even enhance mass spectrometry \cite{Zhao2014}.

Progress in this field has been developing quickly, and thus far experiments have shown the ability to optomechanically control the fluorescence of single \NV\ centres in low vacuum \cite{Neukirch2015}, as well as detect the electron spin resonance (ESR) of single \NV\ centres at atmospheric pressure \cite{Neukirch2015}, and of ensembles of \NV\ centres in vacuum \cite{Hoang2016, Delord2017c}. Ion traps have been shown to be suitable for trapping nanodiamonds and manipulating \NV\ centre spins in high vacuum \cite{Delord2018}, including the observation of coupling between the electron spin and rotational motion of the nanodiamond \cite{Delord2017b, Delord2019}.

\begin{figure}[h!]
\centering
\includegraphics[width=5.5in]{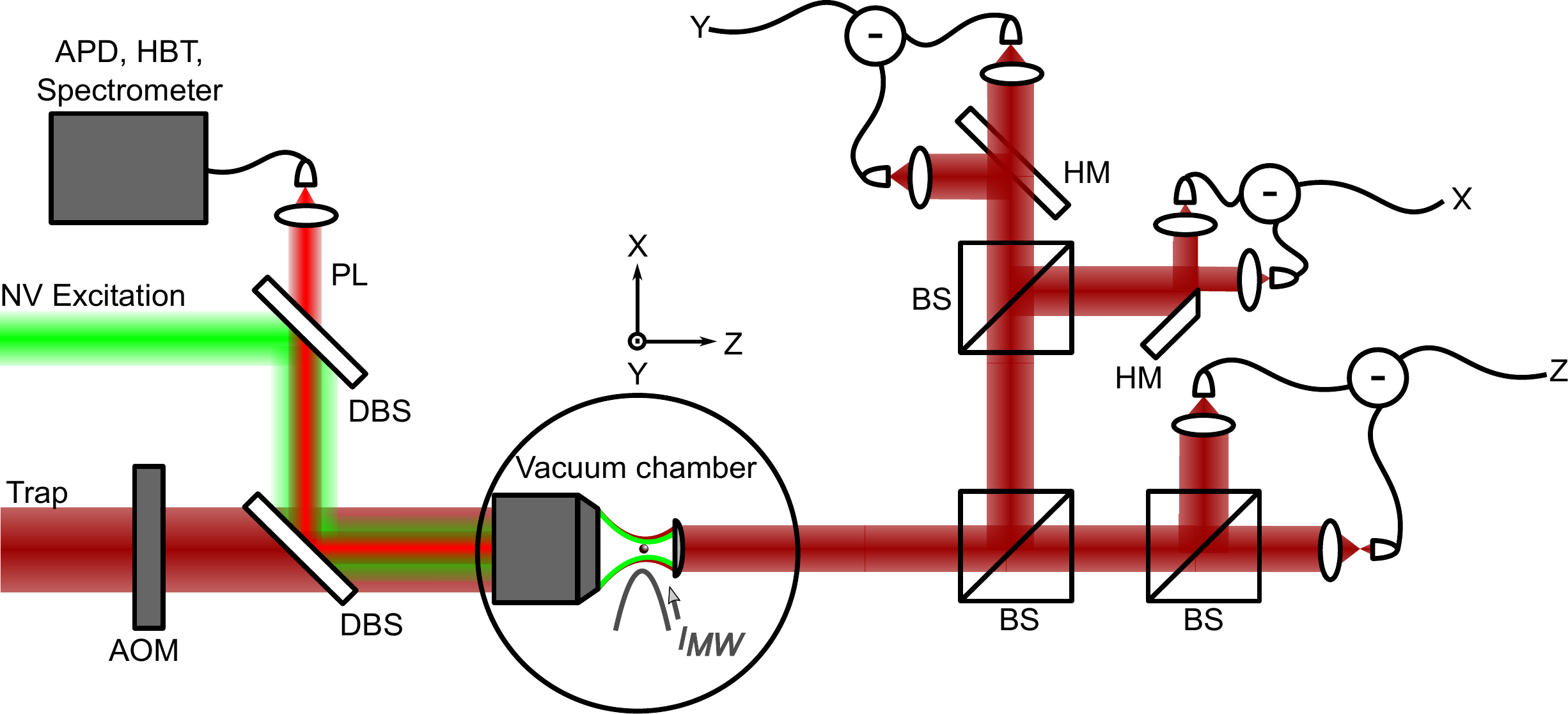}
\caption{Example set-up for levitating nanodiamonds and collecting the \NV\ fluorescence.  DBS - dichroic beam splitter, BS - beam splitter, HM - half mirror. Reproduced from \cite{Pettit2017}.}
\label{NVsetup}
\end{figure}

An example nanodiamond optical levitation set-up is shown in Fig.~\ref{NVsetup}. Nanodiamonds are trapped by a tightly focussed laser beam with a wavelength of 1064\,nm. Dichroic beam splitters allow the confocal alignment of an \NV\ excitation beam of wavelength 532\,nm and the collection of photoluminescence from the nanodiamond. Spin manipulation in the \NV\ centre is conducted by driving a bare-wire loop placed in the vicinity of the focal volume with microwave frequency currents, allowing the transient magnetic field to couple with the dipole moment of the spin.

\begin{figure}[ht]
\centering
\includegraphics[width=4.5in]{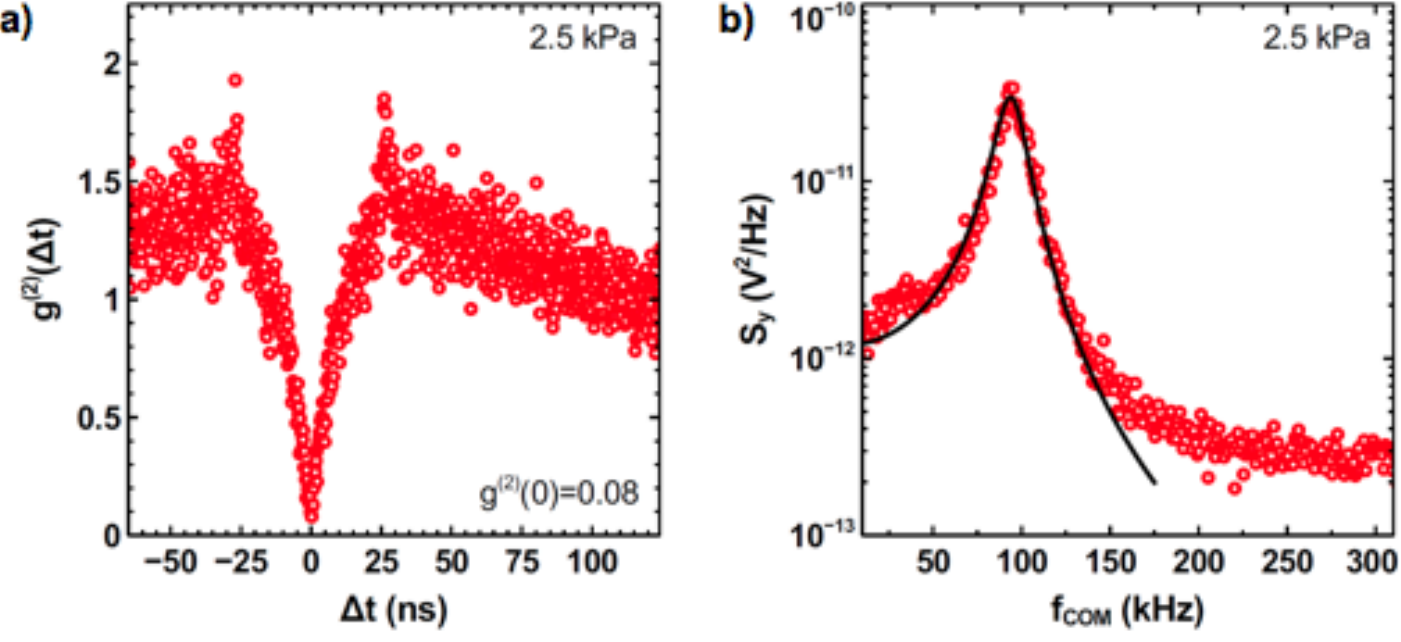}
\caption{a) Photon autocorrelation showing anti-bunching at zero delay, indicating a single \NV\ centre. b) Power spectral density of a nanodiamond's motion in the $y$-direction, from the set-up in Fig.~\ref{NVsetup}. Modified from \cite{Pettit2017}.}
\label{fig3}
\end{figure}

To confirm that the collected photoluminescence comes from a single \NV\ centre, a photon autocorrelation measurement can be conducted with a Hanbury-Brown and Twiss style correlator, and the second order correlation function, $g^{(2)}(\Delta t)$, measured. Figure~\ref{fig3}a shows a typical result, displaying an anti-bunching dip with $g^{(2)}(0)=0.08$. The measured autocorrelation also shows photon bunching at finite delay times, which is a characteristic of systems containing non-radiative decay pathways out of the excited state, as found in the \NV\ centre \cite{Berthel2015}.
 
Figure \ref{fig3}b) shows an example single-sided power spectral density of a time trace of a nanodiamond's c.o.m. position, recorded along the $y$-axis of the trap at 2.5\,kPa, illustrating that the same techniques are available for nanodiamonds as for other nanoparticles discussed in this review. Since the shapes of nanodiamonds are highly irregular, spectral analysis of the trapped particle's c.o.m. motion cannot provide an exact measure of the particle's size, as is done in experiments involving levitated nano- and micro- spheres. This difficulty has been addressed in previous experiments by coating individual nanodiamonds in spherical shells \cite{Neukirch2015}. The irregular shape of nanodiamonds has also been exploited to induce librational vibrations \cite{Hoang2016b} and rotational motion \cite{Delord2017b}.

\begin{figure}[ht]
 \centering
 \includegraphics[width=4.5in]{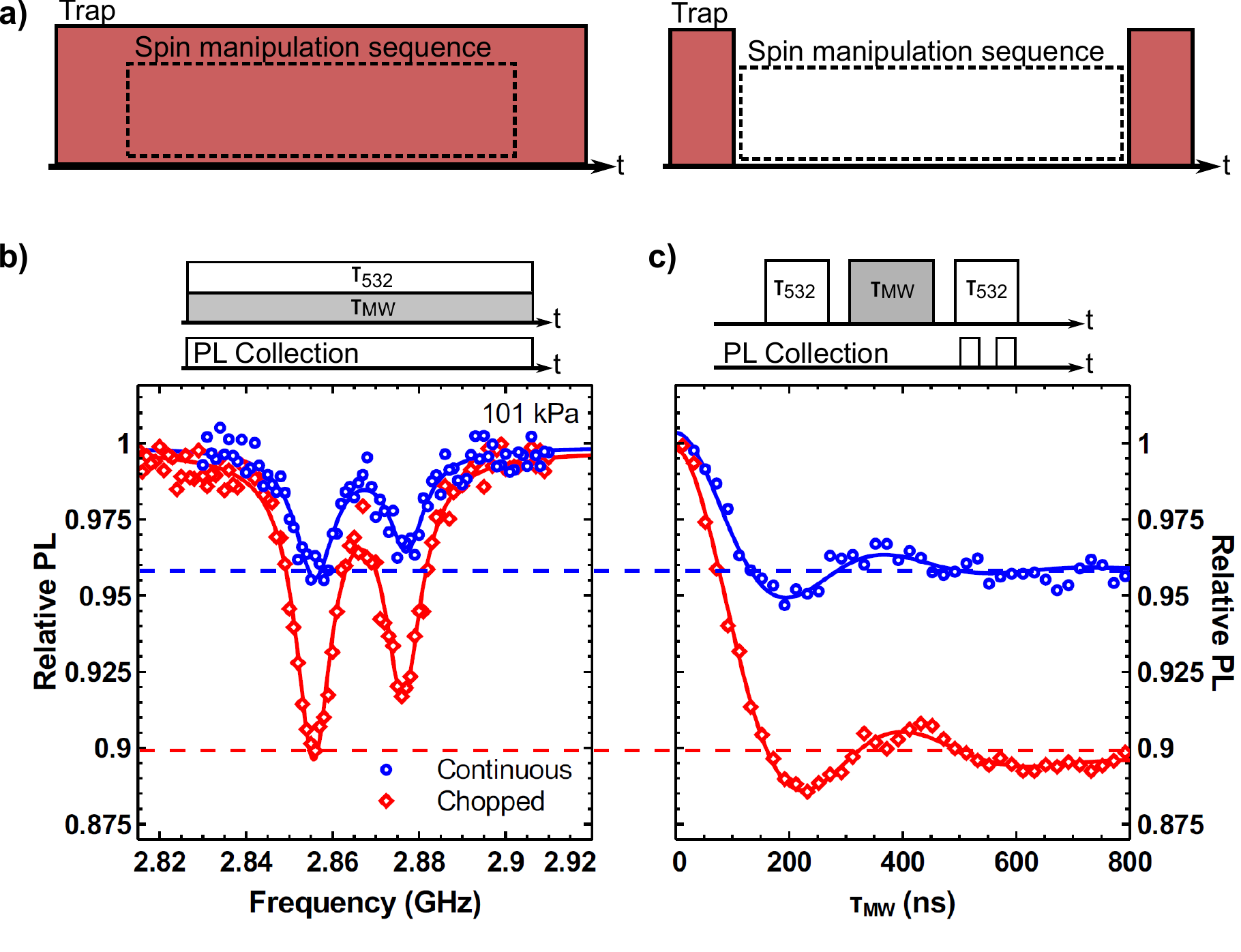}
 \caption{\textbf{\NV\ spin manipulation} a) Illustration of spin-manipulation protocols with levitated nanodiamonds, with the trapping beam on continuously, or chopped such that the trapping beam is off during the protocol. b) Comparison of continuous microwave ODMR scans in both continuous and chopped traps at atmospheric pressure. c) Comparison of time-resolved electron spin transients in both continuous and chopped traps at atmospheric pressure. Reproduced from \cite{Pettit2017}.}
 \label{protocols}
 \end{figure}

The \NV\ centre ground state is a spin triplet, and it is possible to drive transitions within this triplet with microwave frequency radiation, while reading out the spin state optically using the technique of optically detected magnetic resonance (ODMR) \cite{Doherty2013}. At atmospheric pressure, it is possible to maintain stably trapped particles while chopping the trap with duty cycles as high as 50\% with sufficient trapping power. This presents an opportunity to study spin manipulations of the \NV centre in the presence and absence of the trapping beam, as illustrated in Figure~\ref{protocols}a). For the experiments shown in this study, the trapping duty cycle was 40\%.

Figure~\ref{protocols}b) shows two continuous ODMR scans, one each with the trapping beam on or off. It is seen that the contrast is reduced in the presence of the trapping beam as reported in previous studies \cite{Neukirch2015, Hoang2016}. Figure~\ref{protocols}c) shows the corresponding time-resolved ODMR scans for a single transition, where the spin was rotated by a microwave pulse of varying duration. In both cases highly damped spin oscillations are observed, and the steady-state values of the transients approach the observed contrast in the continuous ODMR measurement. Fits to the time-resolved data enable one to extract the transverse spin coherence time, $T_2$. This result suggests that while exposure to 1064\,nm radiation reduces the emitted fluorescence from the \NV\ centre, the coherence properties of its ground state spin remain unaffected.

Working with nanodiamonds, coherent control of single spins from the \NV\ defect centre has been demonstrated, opening the door for these advances to merge into a hybridized optomechanical system. Challenges on this front remain, however. Optically levitated nanodiamonds do not typically survive excursions beyond low vacuum. Thermal degradation of the internal spins, and even graphetisation of the diamond has been observed in optical traps at pressures below $\sim 50\,$mbar \cite{Rahman2015, Neukirch2015}. This is due to absorption by amorphous carbon on the surface, and nitrogen defects in the bulk. Internal heating can be mitigated by using particles milled from low nitrogen chemical vapour deposition grown bulk diamond \cite{Frangeskou2018} and via laser refrigeration using rare-earth doped glasses \cite{Rahman2017, Kern2017}.

\section{Further Topics}
\label{sec:further}

\setcounter{footnote}{0}

\subsection{Librational and rotational optomechanics}
\label{sec:rotation}

Until now, we have only considered the centre-of-mass (c.o.m.) motion of levitated particles. In this section we will consider their alignment degrees-of-freedom. These are accessible when the particle's polarizability can be described by a tensor due to its anisotropic geometry (i.e. it is not spherical), when the particle is birefringent, or if the particle is highly absorbing, facilitating transfer of angular momentum from a light field to the particle.

This section will be split into two. First, we consider librational motion, where the alignment of the particle is harmonically trapped in an optical potential. Secondly, we consider rotational motion, where the alignment is free or driven to rotate about an axis. A review on librational and rotational optomechanics was written by Shi \& Bhattacharya \cite{Shi2016}.

\subsubsection{Librational optomechanics:}

Torsional resonators have a long history in making sensitive measurements, and in analogy to vibrational optomechanics, there is a desire to miniaturize torsional devices to further improve sensitivity \cite{Kim2013}. Torsional motion is distinct from vibrational motion, with the difference largely due to the fact that vibrational motion depends on mass $M$, whereas torsional motion depends on the \emph{distribution} of mass, i.e. the moment of inertia $I$. Considering a sphere of radius $\Rad$: $M \propto \Rad^3$, whereas $I \propto \Rad^5$, one consequence being that small objects are easier to move in angular rather then linear directions. Explicitly, for a harmonically bound object, the vibrational frequency $\w{q} = \sqrt{\ks{q}/M}$ for vibrational spring constant $\ks{q}$, whereas the torsional frequency $\w{\theta} = \sqrt{\ks{\theta}/I}$ for torsional spring constant $\ks{\theta}$. This means that for nanoscale objects, the torsional modes tend to have higher frequencies than the vibrational modes, as directly observed with levitated particles \cite{Kuhn2017a, Hoang2016}, which is desirable for quantum optomechanics, as it raises the ground-state energy. Following the same argument, optomechanical cavity coupling also scales favourably for the torsional degrees-of-freedom: for vibration the coupling rate $\go{q} = \frac{\hbar}{2M\w{q}}\frac{\partial\wc}{\partial q}$, and for torsion $\go{\theta} = \frac{\hbar}{I\w{\theta}}\frac{\partial\wc}{\partial \theta}$ (for cavity resonance frequency $\wc$) \cite{Shi2016}.

Let us explicitly consider levitated particles. Firstly, we will drop the term ``torsion'', since the particles do not experiment any stress, and instead use the correct term ``libration''. In a linearly polarized light field, an anisotropic particle aligns with, and is harmonically bound about, the polarization vector of the light. To date, librational motion has been observed in anisotropic diamond crystals \cite{Hoang2016} and nanofabricated silicon cylinders \cite{Kuhn2017a}, and Yb$^{3+}$:YLF crystals have been aligned using linearly polarized light \cite{Rahman2017}. In analogy to other proposed levitated sensors (Sec.~\ref{sec:sensing}), having access to the librational modes is predicted to lead to exceptional torque sensitivity \cite{Hoang2016}. 

Understanding the optical forces acting on a non-spherical particle requires knowledge of the polarization field inside the object \cite{Barnett2006}, which in general requires numerical methods to calculate. However, if one works in the Rayleigh-Gans approximation, analytic solutions are possible for some shapes, such as disks and cylinders \cite{Stickler2016a}, and ellipsoids \cite{Hoang2016}. This approximation requires that at least one dimension is much smaller than the wavelength of the light \cite{Schiffer1979}. 

\begin{figure}[ht]
	 {\includegraphics{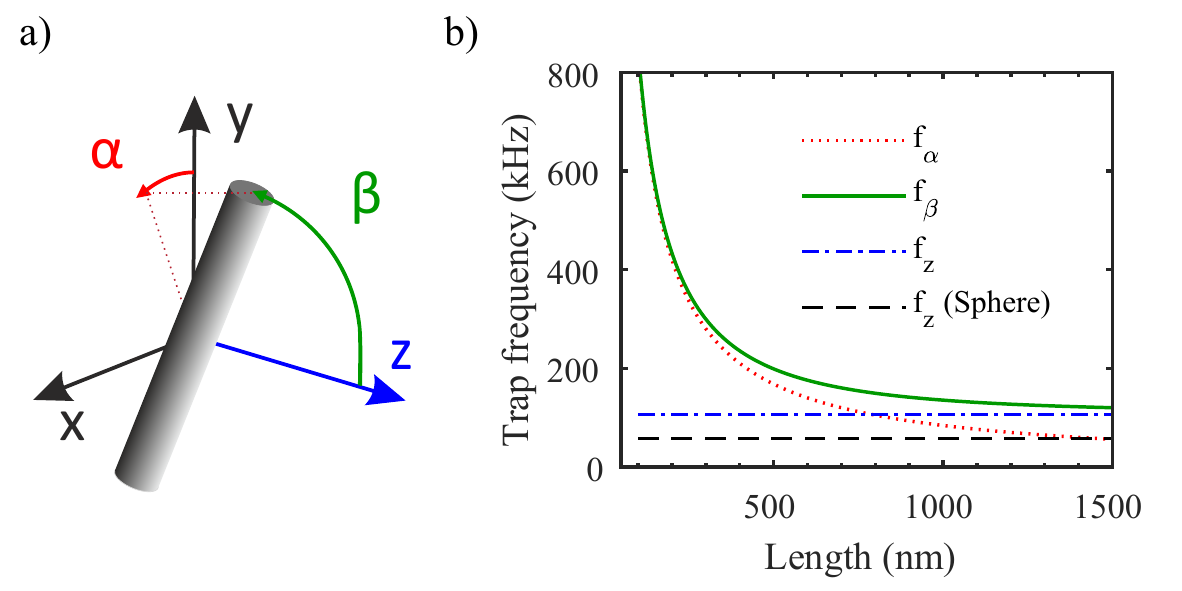}}	\centering
\caption{\label{fig:torsional_freqs} 
\textbf{Librational motion} a) Relevant degrees of freedom for an anisotropic particle. The linear degrees of freedom are labelled $x,y,z$, and the alignment $\alpha, \beta$. b) Variation in trapping frequencies for a cylinder of various lengths in the Rayleigh-Gans approximation, and comparison to a sphere of equivalent volume. The trapping light propagates along the $z$-axis, and is linearly polarized along the $x$-axis.
 Calculated for the same experimental parameters as in Kuhn \emph{et al.} \cite{Kuhn2017a}.
}
\end{figure}

We now consider the illustrative example of cylinders that obey the Rayleigh-Gans approximation. The theory of the interaction of such particles with light can be found in Stickler \emph{et al.} \cite{Stickler2016a}, and a thorough experimental study in Kuhn \emph{et al.} \cite{Kuhn2017a}. An illustration of the relevant degrees-of-freedom is shown in Fig.~\ref{fig:torsional_freqs}a). The cylinder has length $\ell$, diameter $d$ and volume $\V{rod}$. We define the translational degrees of freedom $q \in \left\{x,y,z\right\}$, as for a sphere, and define the alignment via the angles $\{\alpha, \beta\}$, where $\alpha$ is the angle between the $x$-axis and the projection onto the $x-y$ plane, and $\beta$ is the angle between the rod's symmetry axis and $z$. The symmetry axis of the rod is aligned along the vector $\mathbf{m}$. The optical potential for such an object depends upon the alignment relative to the polarization vector of the light, which we chose to be in the $x$-direction, with unit vector $\mathbf{e}_x$, and is given by

\begin{equation}
\Upot{opt}^{\rm cyl} = \Upot{0}^{\rm cyl} f(\mathbf{r})\left [ \frac{\chi_{\perp}}{\chi_{\|}} + \frac{\Delta \chi}{\chi_{\|}} (\mathbf{m}\cdot\mathbf{e}_x)^2 \right ],
\end{equation}
\noindent

with $\Upot{0}^{\rm cyl} = -\pol{cyl}'E_0^2/4$, where $\pol{cyl}'$ is the real part of the polarizability of a cylinder $\pol{cyl} = \eo \V{cyl} \chi_{\|}$. The susceptibilities $\chi$ are defined as $\Delta \chi = \chi_{\|} - \chi_{\perp}$ (the susceptibility anisotropy) and $\chi_{\|} = \er - 1 (= \chi_{\rm max})$, $\chi_{\perp} = 2(\er - 1)/(\er +1)$. The harmonic frequencies of such a nanorod trapped along the polarization axis ($x$) of a standing light wave propagating along the $z$-axis, have the following values 

\begin{eqnarray}\label{eqn:trapped}
\omega_{x,y} &= &\sqrt{\frac{8 \Po\chi_{\|}}{ \pi \rho c \waist^4}}, \quad \omega_z = \sqrt{\frac{4 \Po \chi_{\|} \kl^2}{\pi \rho c \waist^2 }}, \\
\w{\beta} & = & \sqrt{\frac{48 \Po\chi_{\|}}{ \pi \rho c \waist^2 \ell^2} \left( \frac{\Delta\chi}{\chi_{\|}} + \frac{(\kl \ell)^2}{12}\right)}, \\
\w{\alpha} &= &\sqrt{\frac{48 \Po \Delta\chi}{ \pi \rho c \waist^2 \ell^2}},
\end{eqnarray} 
\noindent
where $\waist$ is the beam waist radius, $\kl$ the wavevector of the light, and $\ell$ the length of the cylinder. The moment of inertia of a cylinder is $I = M\ell^2/12$.  The first important thing to note is that the alignment degrees-of-freedom $\w{\alpha, \beta}$ depend upon the moment of inertia, whereas the centre-of-mass modes $\w{q}$ are both mass and moment of inertia independent, as with spherical particles. That means $\w{\alpha, \beta}$ can be significantly increased by working with short rods, as illustrated in Fig.~\ref{fig:torsional_freqs}b). The second thing to note is that $\chi_{\|} = \er-1$ is greater than that of a sphere $\chi_{\mathrm{sphere}} = 3(\er-1)/(\er_r+2)$, which means that the c.o.m. frequencies (and optical couplings in general) are higher for an anisotropic particle than for a sphere of equivalent volume. For the full description of the translational damping rate of the motion of a cylinder, see ref.~\cite{Martinetz2018}.

From the above discussion it is clear that the susceptibility of an anisotropic particle depends upon its alignment in a linearly polarized light field\footnote{When averaged over all rotations, an anisotropic particle still has a susceptibility twice that of a sphere.}. For this reason, the coupling between an anisotropic particle and the field of an optical cavity is alignment dependent, the consequence being that cavity cooling of both the alignment and translational degrees-of-freedom is possible, with ground-state cooling predicted \cite{Stickler2016a}\footnote{An earlier proposal suggested cavity cooling the alignment degree-of-freedom using two counter-propagating Laguerre-Gauss modes \cite{RomeroIsart2010}.}, with higher cooling rates for all degrees-of-freedom than a sphere of equivalent volume. The coupling between an optical cavity and a nanorod has been observed \cite{Kuhn2015}. 

Recently, the theoretical toolbox required to quantum mechanically describe the motion of levitated and aligned nanomagnets has been developed \cite{Rusconi2016}, with the same team developing the mechanism to ground-state cool the librational motion of such objects.

\subsubsection{Rotational optomechanics:}

The previous section considered the dynamics of the alignment of anisotropic particles which are trapped about the polarization vector of a light field. In this section, we consider either the free or driven rotational dynamics of levitated nanoparticles. The free rotation of nanorods transiting an optical cavity under high vacuum ($<10^{-8}\,$mbar) has been studied \cite{Kuhn2015}, where rotational frequencies of 50\,MHz were observed. 

Rotation can also be driven by transferring angular momentum from the light field to the particle, with each photon transferring $\hbar \kl$ of momentum. First, we consider the spin angular momentum of circularly polarized light. Spin angular momentum can be transferred to absorbing particles, such as levitated dust grains \cite{Abbas2004}, graphene flakes \cite{Kane2010, Coppock2016, Coppock2017} or silica nanospheres \cite{Monteiro2017a, Reimann2018}. Rotation rates in excess of 50\,MHz have been obtained with electrically levitated graphene \cite{Coppock2016}, and of GHz with optically levitated nanospheres \cite{Reimann2018}. Spin angular momentum transfers to birefringent crystals, such as spheres of Vaterite, with few-micron diameter particles rotating at 10\,MHz \cite{Arita2013}. Arita \emph{et al.} report the synthesis and rotation of sub-micron diameter Vaterite particles \cite{Arita2016}. Spin angular momentum transfers to particles with anisotropic susceptibility (i.e. non-spherical particles), for example silicon nanorods \cite{Kuhn2017a}, via optical scattering. The optically induced torque $\mathbf{N_{\rm opt}}$ depends upon the induced polarization $\mathbf{P} = \mathbf{\alpha E}$ (where $\mathbf{\alpha}$ is the tensor polarizability) through $\mathbf{N_{\rm opt}} = \langle \mathbf{P} \times \mathbf{E}^*\rangle$. 

The maximum rotation rate $\w{\rm max}^{\rm rot}$ depends upon the balance between the optically induced torque $\mathbf{N_{\rm opt}}$, and the damping rate for the rotational motion $\Gam{rot}$, for example in the $\alpha$ direction $\w{\rm max}^{\rm rot} = N_\alpha/(I\Gam{\alpha})$. Ahn \emph{et al.} have observed rotation rates above 1\,GHz for nano-dumbbells \cite{Ahn2018}. The rotation rate can be tuned through changing $N_\alpha$, which increases with intensity and decreases as the ellipticity of the light polarization deviates from perfectly circular, by changing the birefringence/anisotropy of the particle, and can be increased by reducing the pressure of the surrounding gas \cite{Arita2013, Kuhn2017a}. Gyroscopic stabilization of translational motion has been observed with rotating particles \cite{Arita2013, Kuhn2017a}. 

For a sphere, the rotational damping rate is 

\begin{equation}
\label{eqn:sphere_rot_damp}
 \Gam{rot}^{\rm sph} = \frac{10\acc \Pg \Rad^2 }{3\mass}\sqrt{\frac{2\pi \mg}{\kB \T{env}}},
\end{equation}
\noindent
where $\acc$ is the accomodation coefficient, and all terms are defined in Sec.~\ref{sec:thermodynamics}. For a cylinder, the rotational damping rate is

\begin{equation}
\label{eqn:cylinder_rot_damp}
 \Gam{rot}^{\rm cyl} = \frac{\Pg d \ell }{\mass}\sqrt{\frac{2\pi \mg}{\kB \T{gas}}}\left (2 - \frac{1}{2}\acc + \frac{\pi}{4}\acc \right ).
\end{equation}
\noindent
The optical torque acting on a cylinder is

\begin{equation}
N_\alpha = \frac{\Pg\Delta\chi \ell^2 d^4 \kl^3}{48 c  \waist{}^2}  \left [\Delta \chi \eta_1(\kl \ell) +\chi_\perp \eta_2(\kl \ell) \right ],
\end{equation}  
\noindent
where the functions $\eta_{1,2}(\kl \ell)$ are given by

\begin{eqnarray}
\eta_1(\kl \ell) & = & \frac{3}{4} \int_{-1}^1 \mathrm{d}\xi~(1 - \xi^2) \mathrm{sinc}^2 \left ( \frac{\kl \ell \xi}{2} \right )\\
\eta_2(\kl \ell) & = & \frac{3}{8} \int_{-1}^1 \mathrm{d}\xi~(1 - 3\xi^2) \mathrm{sinc}^2 \left ( \frac{\kl \ell \xi}{2} \right ).
\end{eqnarray}
\noindent
For short rods, $\kl \ell \ll 1$, one has $\eta_1 \simeq 1$ while $\eta_2 \simeq 0$. The torque acting on a birefringent particle depends on the exact material properties.

Optical orbital angular momentum (OAM) can also be transferred to particles, which is potentially interesting since each photon can carry many quanta $\hbar \kl$ of angular momentum, with the value referred to as the topological charge. There is, in principle, no limit to the value of the topological charge, and as of now the record value is 10,010 quanta \cite{Fickler2016}. OAM beams often have interesting intensity profiles, such as doughnut beams, which may be useful to avoid optical absorption and heating \cite{Zhou2017}. Transfer of OAM does not require a polarization anisotropy, rather the c.o.m. of a particle undergoes a (potentially complex) orbit. This has only recently been observed for particles levitated under vacuum conditions \cite{Mazilu2016, Zhou2017, Arita2017}. The rotation rate increases with increasing topological charge, but at high topological charge the motion may become extremely complex, and potentially unstable \cite{Arita2017, Svak2018}.

We now briefly consider why rotational motion may be of interest, and for a more detailed discussion see Shi \& Bhattacharya \cite{Shi2016}. Rotational motion is fundamentally different from translational or harmonically bound motion. For example, the equations of motion are nonlinear with respect to the dynamic variable \cite{Shi2016}. The nonlinear dynamics of a levitated rotor have been exploited to stabilize the rotational frequency of a silicon nanorod to better than one part in $10^{11}$ \cite{Kuhn2017b}. The ground-state energy of free rotation is zero, so cooling the free rotation would bring the rotational motion to an absolute halt, with the prospect of producing a mechanical rotational superposition state \cite{Shi2016, Stickler2018}. 

Shi \& Bhattacharya discuss the dynamics of a dielectric sphere confined by OAM carrying Laguerre-Gauss (LG) modes of an optical cavity \cite{Shi2016}. By pumping the cavity with a single LG mode, at the centre of the cavity the particle would be confined to rotate about a ring. The authors suggest that by pumping the cavity with multiple LG modes the system could act as a sensitive rotation detector. The quantum description of this rotational optomechanical system is currently lacking.

In this direction, the theoretical framework required to understand the rotational decoherence, friction, diffusion, and thermalization of a quantum rotor has recently been developed \cite{Stickler2016b, Papendell2017, Stickler2017}. It has been suggested that rotation can be non-destructively measured in levitated fluid droplets, since rotation distorts the shape of the fluid, which can be inferred by optical means \cite{Childress2017}.  

Finally, one application of rapidly rotating particles might be a study of quantum and vacuum friction. Quantum friction considers charges produced on surfaces via quantum fluctuations. If two neutral plates move parallel to each other, these virtual charges induce a (somewhat controversial \cite{Philbin2009, Pendry2010}) drag, or friction, against their motion \cite{Pendry1997}. The same argument can be extended to a sphere rotating above a surface \cite{Zhao2012}, and the authors suggest the effect should be observable for a graphite particle a few nm in size spinning a few 10s nm above a room temperature graphite surface, though it requires intimidating rotation frequencies on the order of THz. Vacuum friction does not require a surface, and considers the interaction between a moving surface and the vacuum EM field, which should induce drag \cite {Kardar1999}. The same effect should occur for a neutral particle rotating in vacuum \cite{Manjavacas2010a, Manjavacas2010b}, which will dissipate energy through emission of radiation at the rotational frequency, with the effect predicted to be observable for 10\,nm graphite particles rotating at 10s of GHz frequencies.

\subsection{Novel cooling mechanisms}
\label{sec:other_cool}
\begin{figure}[t]
	 {\includegraphics{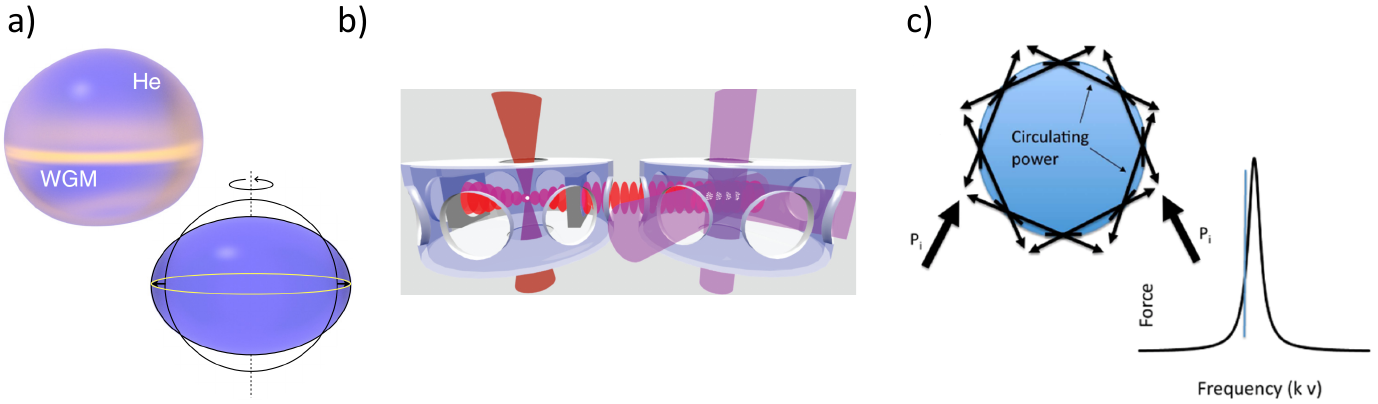}}	\centering
\caption{\label{fig:cool_meth} 
\textbf{Novel cooling methods} a) Droplets of superfluid Helium can support whispering gallery modes, which are coupled to both surface vibrational modes (upper) and droplet rotation through deformation (lower), figure adapted from \cite{Childress2017} b) Ultracold atoms can be used as a coolant for levitated particles via optical lattice coupling, figure taken from \cite{Ranjit2015b} c) By exciting the whispering gallery modes of a microsphere, a frequency-dependent scattering force can be engineered, which can be used to Doppler-cool the resonator, figure adapted from \cite{Barker2010b}.
}
\end{figure}

So far in this manuscript, we have considered two cooling methods: optical cavity cooling (Sec.~\ref{sec:cav_cool}) and optical feedback cooling (Sec.~\ref{sec:det_fb}). In this section we discuss some other cooling methods.

Optical cavity cooling requires high finesse cavities (or equivalently, operation in the good cavity limit), and efficient feedback cooling requires high measurement precision. Ranjit \emph{et al.} propose a method to overcome both of these limitations, by sympathetically cooling levitated nanospheres via coupling to a cloud of cold atoms \cite{Ranjit2015b}. A 1D optical lattice confines a cloud of cooled $^{87}$Rb, and is used to pump a medium finesse ($\mathcal{F} \sim 400$) optical cavity.  A nanosphere is trapped at the linear slope of the cavity field by an optical tweezer. Motion of the sphere changes the phase of the light, which imparts a force on the atoms. Conversely, motion of the atoms modulates the intensity of the light field, imparting a force to the trapped nanosphere. This optomechanical coupling is predicted to have a strength of a few kHz for spheres a few hundred nm in size, and would enable ground-state cooling, with the potential to couple to the internal states of the cold atoms to add an extra degree of quantum control. Coupling the motion of a ground-state cooled nanoparticle to a quantum bath of ultracold atoms may enable studies of quantum thermodynamics \cite{Millen2016, Millen2018}. Practically, it is desirable to work with lower-finesse cavities, as they are less sensitive to contamination from stray nanoparticles.

When considering levitated cavity optomechanics, it is clear that the diameter of the object must be smaller than the optical lattice spacing $\lambda/2$ \cite{Monteiro2013}, realistically limiting cavity cooling to radii $\Rad \leq 400\, $nm. It was observed by Barker \& Schneider that instead of using an external optical cavity, the \emph{intrinsic} optical cavity provided by Whispering Gallery Modes (WGMs) of micron-scale spheres could be used to cool their motion \cite{Barker2010b}, as illustrated in Fig.~\ref{fig:cool_meth}c). In the Mie scattering regime, the optical scattering force on a levitated microsphere is modified in the proximity of WGMs \cite{Ashkin1977}. The presence of a frequency dependent scattering force enabled Barker \& Schneider to construct a cooling scheme analogous to the Doppler cooling of atoms and molecules. This work suggests mK temperatures are possible for 10\,$\mu$m radii silica spheres. In-situ tuning of WGMs by several nm in levitated silica microparticles has been demonstrated \cite{Minowa2017}. WGMs have been used to cool the c.o.m. motion of a $180\,\mu$m diameter microsphere tethered to a cantilever \cite{Li2016}. 

On the subject of WGMs, a recent proposal by Childress \emph{et al.} discusses using a levitated superfluid droplet as a cavity optomechanical system \cite{Childress2017}. In this scenario, a mm radius droplet of superfluid $^3$He or $^4$He is magnetically levitated \cite{Weilert1996, Weilert1997}, and the cavity provided by a WGM of the droplet couples to surface vibrations, as illustrated in Fig.~\ref{fig:cool_meth}a), since these change the optical path length of the WGM. The authors expect a single photon optomechanical coupling rate $\go{0} \sim 200\,$Hz, which exceeds the frequency of the mechanical mode $f \sim 20\,$Hz, making strong-coupling easy to achieve. Furthermore, the internal temperature of levitated droplets of superfluid helium self-cools through evaporation \cite{Childress2017}. 

Cirio \emph{et al.} suggest levitating a cluster of superconducting loops, of radius $5\,\mu$m, in an inhomogeneous magnetic field \cite{Cirio2012}. By coupling such a system to a flux qubit, ground-state cooling is predicted. Romero-Isart \emph{et al.} propose a similar scheme, where a superconducting sphere of a few $\mu$m diameter (made out of, e.g. Pb), cooled into its Meissner state, is magnetically levitated and coupled to superconducting quantum circuits to achieve ground-state cooling \cite{RomeroIsart2012}, with coupling rates of $\sim 2\pi\times1\,$kHz when coupled to a flux qubit, and $\sim 2\pi\times100\,$kHz when coupled to an LC circuit. These magnetically levitated superconductors acting as ``magnetomechanical'' systems are interesting as compared to optomechanical systems for several reasons. There is no heating or dissipation due to scattering of photons, which limits the quality factor in optical systems \cite{Jain2016}\footnote{There is dissipation due to hysteresis effects in the superconducting pick-up circuit, but the quality factor is still expected to be far in excess of $10^{10}$ \cite{RomeroIsart2012}.}. The superconductors are pre-cooled to reach their Meissner states, so the internal temperature is already low, and there is no heat source to increase their internal temperature, facilitating many quantum applications. This cooling method also applies to few-micrometer sized objects of mass $10^{14}\,$amu, not possible in optical cavity cooling. Recent work has developed sophisticated experimental protocols for testing high mass quantum physics with such magnetically levitated microspheres \cite{RomeroIsart2017, Pino2018}.

A proposal by Artusio-Glimpse \emph{et al.} offers a method for slowing and manipulating small clouds of sub-wavelength particles, inspired by the application to remotely control debris in space \cite{ArtusioGlimpse2016}. In analogy to gravitational assist spacecraft manoeuvre, a focused beam swept past a dielectric particle can cause significant deceleration, being able to bring single particles to a halt, and being able to manipulate clouds of many thousands of particles to a distribution of velocities where 90\% are slowed by 50\%. 

Goldwater \emph{et al.} show that charged nanoparticles levitated in a Paul trap can be manipulated via the current they induce in nearby electrodes, allowing cooling to the temperature of the trapping circuitry, or feedback cooling to sub-Kelvin temperatures with room-temperature circuitry. State of the art refrigeration would allow ground-state cooling. This technique would work for sub-nanometre sized particles, and objects $> 10\,\mu$m in size. Also considering small particles, plasmonically trapped few-nm metal particles experience optomechanical backaction in plasmonic cavities \cite{Mestres2016}. To the best of our knowledge there are no proposals suggesting cooling by this method, and plasmonically trapped particles experience extreme internal heating, limiting application to particles trapped in liquids or levitated in ambient conditions \cite{Jauffred2015}.

Finally, we discuss recent work on cooling the \emph{internal} temperature of levitated particles, which would be useful to minimize blackbody radiation and decoherence, for example in matterwave interferometry (see Sec.~\ref{sec:MWI}), and could mitigate optical absorption which leads to evaporation \cite{Millen2014} or burning \cite{Rahman2015}. Optical absorption by solids can lead to the emission of blue shifted photons, with the excess energy coming from bulk phonons, hence leading to internal cooling. Bulk Yb$^{3+}$-doped YLiF$_4$ (YLF) was cooled to 90\,K from ambient temperatures upon irradiation by 1020\,nm laser light \cite{Melgaard2016}. Rahman \& Barker report the cooling of optically levitated $1\,\mu$m diameter crystals of YLF to 130\,K using the less optimal wavelength of 1031\,nm \cite{Rahman2017}. Kern \emph{et al.} propose bulk cooling of levitated nanodiamonds by about 60\,K, either through excitation of NV defects at $\sim 780\,$nm or SiV defects at $\sim 745\,$nm \cite{Kern2017}.

\section{Conclusion}
\label{sec:conclusion}

\setcounter{footnote}{0}

In this review we have covered a multitude of applications for the optomechanical system formed by nano-scale particles levitated in optical fields. This optomechanical platform is unique, since the resonance frequencies are not determined by fixed mechanical properties, but rather by the levitating field, giving a great deal of experimental flexibility. In addition, we observe the centre-of-mass (c.o.m.) mechanical mode, rather than the flexural modes seen in standard optomechanics, which yields lower energy dissipation and the possibility to create macroscopically separated quantum states. The research community has made excellent progress towards reaching the quantum regime, using active feedback to reach a few tens of motional quanta, achieving strong coupling to optical cavities, and demonstrating the control of spins embedded within levitated material.

In addition, levitated particles have proven to be a paradigmatic platform for studying stochastic thermodynamics at the nanoscale, in particular due to the ability to dynamically vary the coupling between a particle and its thermodynamic bath. The ability to effectively isolate a levitated object from the surrounding environment has enabled impressive proof-of-principle force sensing.

\subsection{Discussion}
In this section we raise some topical discussion points.

\subsubsection{Comparison of levitated oscillators and state-of-the-art tethered oscillators:} One often quoted motivation in the field of levitated optomechanics is the ability to engineer an oscillator with extremely high mechanical quality factors $\Qm$. As illustrated in Fig.~\ref{Qfactor_fig}, it is predicted that an optically levitated nanoparticle suspended in high vacuum can have quality factors $\Qm > 10^{11}$, limited by photon scattering. Practically, $\Qm \leq 10^8$ has been observed, due to the instability of levitated particles at low pressures, see Sec.~\ref{sec:int_temp}.

Recent advances in phononic shielding and resonator strain engineering has produced tethered oscillators operating in the range $10^8<\Qm<10^{11}$ \cite{Tsaturyan2017, Ghadimi2018, Maccabe2019}. This challenges the assumption that the levitated platform can offer the lowest energy dissipation rates. Nonetheless, levitated optomechanical systems have some unique advantages. 

When considering the exploitation of high $\Qm$ oscillators for force sensing (Sec.~\ref{sec:sensing}), lower masses enable the detection of smaller forces, and levitated systems are amongst the lowest-mass optomechanical systems. Mechanical-oscillator force sensing relies on a resonant interaction between the force and the oscillator, which is greatly simplified when the resonance frequency can be tuned as in levitated optomechanics. In addition to this point, recent work has demonstrated the \emph{non-resonant} detection of forces using levitated particles \cite{Kuhn2017b, Hebestreit2018c}. The geometry of optically trapped particles, in particular their tiny volume, enables them to be placed extremely close to surfaces \cite{Rider2016, Diehl2018, Winstone2018, Magrini2018}, for the detection of short-range forces \cite{Geraci2010}. 

Since the mode of oscillation for levitated particles is the c.o.m., they offer the best route to producing a macroscopically-separated superposition state of a massive object, as discussed extensively in Sec.~\ref{sec:quantum}. Not only would this shed light on fundamental physics, but would open up a vista of quantum-enhanced sensing (Sec.~\ref{sec:exotic_sensing}).

\subsubsection{Comparison of feedback and cavity cooling:}
A major motivation for building optomechanical experiments with levitated particles is the potential for studying fundamental quantum physics (Sec.~\ref{sec:quantum}). To do this, the motion of an optically trapped particle needs to be cooled to, or close to, the ground state. There are two main routes to cooling\footnote{Though other cooling methods are proposed, see Sec.~\ref{sec:other_cool}.}, active feedback (Sec.~\ref{sec:det_fb}) and passive cavity cooling (Sec.~\ref{sec:cav_cool}). Initially, the latter was thought of as the optimal platform for reaching the ground state \cite{Chang2010,RomeroIsart2010,Barker2010a}, but active feedback has emerged as the dominant tool to reach low temperatures \cite{Gieseler2012, Jain2016, Tebbenjohanns2019}. 

On one level, this is purely due to experimental practicalities. High-finesse optical cavities are challenging to work with, and the finesse is easily degraded when nanoparticles hit the surfaces of the mirrors, often requiring separate trapping and cooling regions within the experimental apparatus \cite{Mestres2015}. In comparison, much of the challenge when performing active feedback cooling is in the feedback electronics, where commercial solutions are available. There are now firm predictions that ground-state cooling can be achieved using active feedback methods \cite{Rodenburg2016, Tebbenjohanns2019, Conangla2019}.

The unique advantage offered by levitated cavity optomechanics is that a hybrid optomechanical system is formed between the mechanical motion of the levitated particle and the optical field of the cavity, as discussed in detail in Sec.~\ref{sec:cav_cool}. This would enable the transfer of coherent information between the optics and mechanics, or between different optical fields as mediated by the mechanics \cite{Monteiro2013}. In this sense, cavity cooling offers a clearer route to the integration of levitated particles into a quantum network. 

Of course, the real answer is that feedback and cavity cooling are complimentary, with a host of experiments exploiting both technologies \cite{Mestres2015, Delic2018, Windey2018}, and the combination of both cooling methods is predicted to lead to deeper cooling and coherent control \cite{Genoni2015}. It also seems likely that the recent addition of coherent scattering to the toolbox of levitated cavity optomechanics \cite{Delic2018, Windey2018, GonzalezBallestero2019} will lead to ground-state cooling in the very near future. 

\subsection{Outlook}

The field of optomechanics with levitated particles has seen considerable growth in the last 10 years, largely motivated by the tantalizing prospect of operating in the quantum regime. And indeed, experiments are extremely close to reaching this goal \cite{Windey2018, Delic2018, Conangla2019, Tebbenjohanns2019}. This is, of course, a two way street, and there have been many developments in the theory of macroscopic quantum physics inspired by levitated optomechanics \cite{RomeroIsart2011a, Bateman2014, Wan2016a, Stickler2017, RomeroIsart2017, Bose2017, Marletto2017, Stickler2018} .  However, the field is notable for the wide range of applications which have been uncovered. 

Levitated particles have been used to elucidate a wide range of nanothermodynamic processes \cite{Millen2018, Gieseler2018}, and this is an application which will continue to flourish as research teams develop complex spatial \cite{Rondin2017} and temporal optical potentials. Further development in this direction will illuminate mechanical processes in biological systems, acting as a flexible experimental analogue. 

Force sensing has also been a motivation in the field, especially due to high-performance in a room-temperature environment. It will be interesting to see the field produce miniaturized technologies, with first steps already taken \cite{Alda2016, Kuhn2017c, Wachter2019}. At the other extreme, the coming years will see the further development of a space-based platform for levitated particle experiments \cite{Kaltenbaek2016}.

Perhaps the most exciting coming developments in the field will involve moving away from \emph{opto}-mechanics, with the use of electrical \cite{Millen2015, Alda2016, Delord2017, Goldwater2018} and magnetic \cite{Cirio2012, RomeroIsart2017, Pino2018, Slezak2018} traps, overcoming some of the limitations present when particles are exposed to optical fields \cite{Jain2016}. This direction opens new vistas for control, cooling and technological integration, whilst pushing the field into unprecedented regimes of mass (both large \cite{Pino2018, Prat-Camps2017} and small \cite{Goldwater2018}) whilst enabling the manipulation of exotic materials, perhaps even extending to living objects \cite{RomeroIsart2010}.

\section*{Acknowledgments}
The authors would like to thank U. Deli\'{c}, T. Delord, A. A. Geraci, S. Kuhn, M. Rashid, R. Reimann, B. A. Stickler, D. Windey and G. P. Winstone for useful discussions.

JM is supported by the European Research Council (ERC) under the European Union's Horizon 2020 research and innovation programme (Grant agreement No. 803277), and by EPSRC New Investigator Award EP/S004777/1. R.P. and A.N.V. acknowledge generous support from The Institute of  Optics and the Department of Physics and Astronomy at the University of Rochester and Office of Naval Research awards N00014-17-1-2285 and N00014-18-1-2370.

\bibliography{Lev_opto_review_arxiv}

\end{document}